\documentclass[final,5p,times,twocolumn]{elsarticle}

\usepackage{lineno,hyperref}

\usepackage{graphics} 
\usepackage{epsfig} 
\usepackage{amsmath} 
\usepackage{amssymb}  

\usepackage{algpseudocode}
\usepackage{algorithm}
\usepackage{epstopdf}

\usepackage{url}

\usepackage{verbatim}

\usepackage{amsthm}
\usepackage{pgfplots}

\usepackage{booktabs}
\usepackage{color}
\usepackage{cancel}

\usepackage{enumitem}

\usepackage{tikz}

\usepgflibrary{arrows}

\newtheorem{assumption}{\bf Assumption}

\newtheorem{theorem}{\bf Theorem}
\newtheorem{proposition}{\bf Proposition}
\newtheorem{lemma}{\bf Lemma}

\journal{Annual Reviews in Control}
\date{\copyright   2024 The Author(s).}
\begin{document}
\begin{frontmatter}

\title{Analysis and design of model predictive control frameworks for dynamic operation\\ An overview
\tnoteref{mytitlenote}}
\tnotetext[mytitlenote]{
Johannes K\"ohler was supported by the Swiss National Science Foundation
under NCCR Automation (grant agreement 51NF40 180545)} 

\author{Johannes K\"{o}hler$^1$, Matthias A. M\"uller$^2$, Frank Allg\"{o}wer$^3$} 

\address{$^1$Institute for Dynamic Systems and Control, ETH Z\"urich, Z\"urich CH-8092, Switzerland.\\
$^2$ Leibniz University Hannover, Institute of Automatic Control, 30167 Hannover, Germany.\\
$^3$Institute for Systems Theory and Automatic Control, University of Stuttgart, 70550, Stuttgart, Germany. 
} 

\begin{abstract} 
This article provides an overview of model predictive control (MPC) frameworks for dynamic operation of nonlinear constrained systems.
Dynamic operation is often an integral part of the control objective, ranging from tracking of reference signals to the general economic operation of a plant under online changing time-varying operating conditions.
We focus on the particular challenges that arise when dealing with such more general control goals and present methods that have emerged in the literature to address these issues. 
The goal of this article is to present an overview of the state-of-the-art techniques, providing a diverse toolkit to apply and further develop MPC formulations that can handle the challenges intrinsic to dynamic operation. 
We also critically assess the applicability of the different research directions, discussing limitations and opportunities for further research. 
\end{abstract}
\begin{keyword}
Model predictive control (MPC), 
Tracking MPC, 
Economic MPC, 
MPC without stabilizing terminal cost, 
Closed-loop stability,
Dynamic system operation
\end{keyword}
\end{frontmatter}
 
\section{Introduction}
\label{sec:intro}
Model predictive control (MPC), also called receding horizon control, is a modern optimization-based control method. 
The underlying principle is to repeatedly solve finite horizon open-loop optimal control problems online. 
Feedback is generated implicitly by only implementing the initial part of the optimized input trajectory and repeating the online optimization in the next time step. 
MPC is widely used in practice (cf. the surveys by \cite{qin2003survey} and \cite{samad2020industry}) and actively researched in academia~\citep{mayne2014model}. 
This success of MPC is primarily due to some intrinsic advantages: (i) direct consideration of state and input constraints; (ii) applicability to general nonlinear MIMO systems; (iii) optimization of general performance criteria.

There have been significant advances in academia over the last decades, resulting in a mature stability theory for MPC~\citep{mayne2000constrained}. 
Much ongoing research in MPC is focused on deriving efficient implementations~\citep{verschueren2022acados}, accounting for model errors ~\citep{kouvaritakis2016model,mayne2016robust}, or learning the model online~\citep{hewing2020learning}. 
In contrast, this article focuses on the design and analysis of MPC framework that can be applied to \textit{dynamic operation}. 

\subsection{Dynamic operation}
\label{sec:intro_dynamic}
The motivation of the present article comes from many emerging control applications in which the control goal is not accurately reflected by the setpoint stabilization problem, which is classically studied in MPC. 
We specifically consider the challenges intrinsic in \textit{dynamic operation}, which has received less attention in the MPC community and is rarely studied in a unified fashion. 
By \textit{dynamic operation}, we primarily consider the following three challenges related to the control goal: 
\begin{enumerate}[label=(C.\arabic*)]
\item Stationary operation is not desired;
\label{enum_motivation_1}
\item Desired mode of operation changes online in an unpredictable fashion;
\label{enum_motivation_2}
\item Desired mode of operation cannot be directly specified in terms of a given setpoint/trajectory of the system state.
\label{enum_motivation_3}
\end{enumerate}
The desired mode of operation captures the control goal, which is encoded in a cost function to be minimized. 
Examples of desired mode of operation include: staying at a setpoint, following a time-varying trajectory, or operating on some general set. 
\ref{enum_motivation_1} implies that an optimal controller does not drive the system to a steady-state, which, e.g., arises naturally if a time-varying cost function is chosen. 
\ref{enum_motivation_2} considers that this cost function is not only time-varying, but also the future evolution cannot be predicted. 
Lastly, \ref{enum_motivation_3} reflects that determining an optimal state trajectory for a given cost function is non-trivial, i.e., the cost function does not simply minimize the distance to some (feasible) setpoint/trajectory/set. 
Next, we illustrate these abstract challenges~\ref{enum_motivation_1}--\ref{enum_motivation_3} using applications. 
Consider a motion planning or trajectory optimization problem, as encountered in robotics, aerospace, or autonomous driving. 
\ref{enum_motivation_1}: The primary goal is to track/follow some time-varying dynamic trajectory/path. 
 \ref{enum_motivation_2}: This reference is often generated online by a separate unit (e.g., using artificial intelligence and visual feedback) independent of the controller (and hence unpredictably). 
 \ref{enum_motivation_3}: The reference is primarily specified in terms of a low dimensional output of the system, which is often not physically realizable due to the dynamics or constraints on the system. 
As a completely different application, consider a heating, ventilation, and air conditioning (HVAC) system to regulate temperature in a building. 
\ref{enum_motivation_3}: The control goal naturally revolves around \textit{economic criteria} such as the energy consumption. 
\ref{enum_motivation_2}: The optimal mode of operation depends on external factors, such as temperature, occupancy, and price/demand, which fluctuate in an unpredictable fashion. 
\ref{enum_motivation_1}: The same external variables are subject to significant changes over time, e.g., due to the periodic day--night cycle, making the optimal mode of operation non-stationary. 
Analogous considerations apply to water distribution networks, power networks, power generation using kites and many more. 
Overall, in many applications the desired mode of operation is dynamic; the controller needs to change the mode of operation based on external, online changing, variables; and the optimal mode of operation is not explicitly specified in terms of a known feasible setpoint/trajectory of the system state. 

\subsection{Contribution}
\label{sec:intro_contribution}
We provide an overview of recent advances in the design and analysis of MPC formulations that can accommodate and address the challenges intrinsic to dynamic operation~\ref{enum_motivation_1}--\ref{enum_motivation_3}. 
We are specifically interested in classical system theoretic properties, such guaranteeing recursive feasibility, constraint satisfaction, and stability/performance for general nonlinear systems. 
We provide a set of tools and methods to design MPC schemes for such challenging control applications with guaranteed closed-loop properties. 
We discuss the existing work and methods in a broad context, highlighting gaps in the existing literature and discussing different strategies in a unified fashion. 
In general, the challenges~\ref{enum_motivation_1}--\ref{enum_motivation_3} and the studied methods touch on a number of different research fields in MPC, e.g.: 
trajectory tracking~\citep{faulwasser2012optimization}, output regulation~\citep{kohler2021constrained}, 
artificial references in MPC~\citep{limon2018nonlinear}, economic MPC~\citep{faulwasser2018economic}, and MPC without terminal constraints~\citep{grune2017nonlinear}. 
We study these formulations in a unified way, focusing on how these frameworks address the challenges outlined above.

\subsection{Outline}
First, we introduce \textit{preliminaries} regarding the design of a stabilizing MPC scheme and explain the challenges in applying this design to dynamic operation (Sec.~\ref{sec:prelim}). 
Moreover, we present designs for the \textit{stabilizing terminal ingredients}, primarily the local control Lyapunov function (CLF), for general tracking problems (Sec.~\ref{sec:terminal}). 
Then, we show how infeasible and online changing references can be accommodated using \textit{artificial reference trajectories} (Sec.~\ref{sec:artificial}). 
Furthermore, we discuss
how economic performance criteria can be directly optimized using \textit{economic} MPC formulations (Sec.~\ref{sec:economic}). 
Finally, we present an alternative framework, focusing on the analysis of simpler MPC schemes \textit{without stabilizing terminal ingredients} (Sec.~\ref{sec:UCON}). 
The paper concludes with a \textit{discussion} regarding the different provided tools, their benefits, limitations, and open issues in the literature (Sec.~\ref{sec:discuss}). 
We note that Section~\ref{sec:challenges} explains how these different developments are motivated by the challenges~\ref{enum_motivation_1}--\ref{enum_motivation_3} and Section~\ref{sec:discussion_summary} summarizes how the different designs address the challenges~\ref{enum_motivation_1}--\ref{enum_motivation_3}.

\subsection{Notation}
We denote the set of integers in an interval $[a,b]$ by $\mathbb{I}_{[a,b]}$. 
For $k,N\in\mathbb{I}_{\geq 0}$, the modulo operator is denoted by $\mathrm{mod}(k,N)\in\mathbb{I}_{[0,N-1]}$. 
The quadratic norm w.r.t. a positive definite matrix $Q=Q^\top$ is denoted by $\|x\|_Q^2:=x^\top Q x$. 
The identify matrix is denoted by $I_n\in\mathbb{R}^{n\times n}$. 
For $x\in\mathbb{R}^{n_1}$, $y\in\mathbb{R}^{n_2}$, we abbreviate the stacked vector as $(x,y):=[x^\top,y^\top]^\top\in\mathbb{R}^{n_1+n_2}$. 
For a symmetric matrix $A=A^\top$, $A\succ 0$ ($A\succeq 0$) indicates that the matrix is positive definite (positive semidefinite). 
The interior of a set $\mathbb{X}\subseteq\mathbb{R}^n$ is denoted by $\mathrm{int}(\mathbb{X})$. 
By $\mathcal{K}_\infty$, we denote the class of functions $\alpha:\mathbb{R}_{\geq 0}\rightarrow\mathbb{R}_{\geq 0}$, which are continuous, strictly increasing, unbounded, and satisfy $\alpha(0)=0$. 
For a continuously differentiable function $F(x,y)$, $F:\mathbb{R}^{n_1}\times\mathbb{R}^{n_2}\rightarrow\mathbb{R}^m$, 
$\left.\left[\dfrac{\partial F}{\partial x}\right]\right|_{(\bar{x},\bar{y})}\in\mathbb{R}^{m\times n_1}$ denotes the Jacobian matrix of $F$ w.r.t. $x$ evaluated at $(\bar{x},\bar{y})$. 
 
\section{Preliminaries: Recursive feasibility \& stability in MPC}
\label{sec:prelim}
Before addressing the challenges intrinsic to dynamic operation~\ref{enum_motivation_1}--\ref{enum_motivation_3}, we first provide preliminaries regarding stabilizing MPC designs, analogous to the textbooks by
\cite{rawlings2017model} and \cite{grune2017nonlinear}. 
We consider a nonlinear discrete-time system 
\begin{align}
\label{eq:sys}
x(t+1)=f(x(t),u(t)),\quad x(0)=x_0, 
\end{align}
with the state $x(t)\in X=\mathbb{R}^n$, the control input $u(t)\in\mathbb{U}\subseteq\mathbb{R}^m$, and the time step $ t\in\mathbb{I}_{\geq 0}$. 
For a sequence $\mathbf{u}\in\mathbb{U}^N$, we denote the $k$-th element by $\mathbf{u}_k\in\mathbb{U}$ with $k\in\mathbb{I}_{[0,N-1]}$. 
Given an initial state $x\in X$ and an input sequence $\mathbf{u}\in\mathbb{U}^N$, we denote the solution to~\eqref{eq:sys} after $k\in\mathbb{I}_{\geq 0}$ steps by $x_{\mathbf{u}}(k,x)\in X$, $k\in\mathbb{I}_{[0,N]}$ with $x_{\mathbf{u}}(0,x)=x$. 
The system is subject to pointwise-in-time constraints
\begin{align}
\label{eq:constraints}
(x(t),u(t))\in\mathbb{Z}\subseteq X\times\mathbb{U},\quad \forall t\in\mathbb{I}_{\geq 0},
\end{align}
which, e.g., model actuator limitations or safety critical limits on the state.
The goal is to minimize a given stage cost $\ell:X\times\mathbb{U}\rightarrow\mathbb{R}$, resulting in the following optimal control problem
\begin{align}
\label{eq:optimal_control}
\mathcal{J}^\star_\infty(x):=&\limsup_{N\rightarrow\infty}\inf_{\mathbf{u}\in\mathbb{U}^N}\sum_{k=0}^{N-1}\ell(x_{\mathbf{u}}(k,x),\mathbf{u}_k)\\
\text{s.t. }&(x_{\mathbf{u}}(k,x),\mathbf{u}_k)\in\mathbb{Z},\quad k\in\mathbb{I}_{\geq 0}. \nonumber
\end{align}

\subsection{Stabilizing MPC design}
Solving Problem~\eqref{eq:optimal_control} is typically computationally intractable.
MPC approximately solves Problem~\eqref{eq:optimal_control} by repeatedly solving a finite-horizon open-loop optimal control problem:
\begin{align}
\label{problem:prelim}
\mathcal{J}_N^\star(x):=&\inf_{\mathbf{u}\in\mathbb{U}^N}\sum_{k=0}^{N-1}\ell(x_{\mathbf{u}}(k,x),\mathbf{u}_k)+V_{\mathrm{f}}(x_{\mathbf{u}}(N,x))\\
\text{s.t. }&(x_{\mathbf{u}}(k,x),\mathbf{u}_k)\in\mathbb{Z},~ k\in\mathbb{I}_{[0,N-1]},~ x_{\mathbf{u}}(N,x)\in\mathbb{X}_{\mathrm{f}},\nonumber
\end{align}
with prediction horizon $N\in\mathbb{I}_{\geq 1}$, a terminal cost $V_{\mathrm{f}}:X\rightarrow\mathbb{R}$, and a terminal set $\mathbb{X}_{\mathrm{f}}\subseteq X$. 
To simplify the theoretical exposition, we pose the following standing assumption.
\begin{assumption}(Continuity and compactness)
\label{asm:continuity}
The input constraint set $\mathbb{U}$ is compact. 
The dynamics $f$ and the cost functions $\ell,V_{\mathrm{f}}$ are continuous. 
\end{assumption}
These conditions imply that Problem~\eqref{problem:prelim} has a minimizer~\citep[Prop.~2.4]{rawlings2017model}, which is denoted by $\mathbf{u}^\star(x)\in\mathbb{U}^N$.\footnote{%
In case the minimizer is not unique, one minimizer can be selected. 
Although we assume that a minimizer is computed, most closed-loop guarantees also hold if a suboptimal feasible solution is computed~\citep{scokaert1999suboptimal,mcallister2023suboptimal}.} 
The closed-loop operation is defined by
\begin{align}
\label{eq:sys_closedloop}
x(t+1)=f(x(t),\mathbf{u}^\star_{0}(x(t))),\quad t\in\mathbb{I}_{\geq 0},
\end{align}
i.e., at each time $t$ we apply the first part of the optimal open-loop input sequence $\mathbf{u}^\star(x(t))\in\mathbb{U}^N$ computed based on the measured state $x(t)$.

The following result recaps standard design conditions and resulting closed-loop properties of the MPC, assuming a feasible steady state $(x_{\mathrm{r}},u_{\mathrm{r}})\in\mathbb{Z}$, $f(x_{\mathrm{r}},u_{\mathrm{r}})=x_{\mathrm{r}}$ should be stabilized.
\begin{assumption}
\label{asm:prelim_stabilizing} (Stabilizing stage cost)
There exist functions $\underline{\alpha}_\ell$, $\overline{\alpha}_\ell$, ${\alpha}_{\mathrm{f}}\in\mathcal{K}_{\infty}$ such that $\underline{\alpha}_{\ell}(\|x-x_{\mathrm{r}}\|)\leq \ell_{\min}(x)\leq \overline{\alpha}_{\ell}(\|x-x_{\mathrm{r}}\|)$ for all $(x,u)\in\mathbb{Z}$, $V_{\mathrm{f}}(x)\leq {\alpha}_{\mathrm{f}}(\|x-x_{\mathrm{r}}\|)$ for all $x\in\mathbb{X}_{\mathrm{f}}$, and $\ell(x_{\mathrm{r}},u_{\mathrm{r}})=0$, $V_{\mathrm{f}}(x_{\mathrm{r}})=0$, with $\ell_{\min}(x):=\min_{u\in\mathbb{U}}\ell(x,u)$. 
\end{assumption}
\begin{assumption}
\label{asm:prelim_terminal} (Terminal ingredients) 
There exists a terminal control law $k_{\mathrm{f}}:\mathbb{X}_{\mathrm{f}}\rightarrow\mathbb{U}$ such that for all $x\in\mathbb{X}_{\mathrm{f}}$:
\begin{enumerate}[label=(T.\arabic*)]
\item Constraint satisfaction: $(x,k_{\mathrm{f}}(x))\in\mathbb{Z}$;
\label{enum_term_1}
\item Positive invariance: $f(x,k_{\mathrm{f}}(x))\in\mathbb{X}_{\mathrm{f}}$;
\label{enum_term_2}
\item Local CLF: $V_{\mathrm{f}}(f(x,k_{\mathrm{f}}(x)))-V_{\mathrm{f}}(x)\leq -\ell(x,k_{\mathrm{f}}(x))$. 
\label{enum_term_3}
\end{enumerate}
Furthermore, there exists a function $\alpha_{\mathrm{V}}\in\mathcal{K}_\infty$ such that for any state $x\in{X}$ such that Problem~\eqref{problem:prelim} is feasible, it holds: 
\begin{enumerate}[label=(T.\arabic*), resume]
\item Weak controllability: $\mathcal{J}_N^\star(x)\leq \alpha_{\mathrm{V}}(\|x-x_{\mathrm{r}}\|)$. 
\label{enum_term_4}
\end{enumerate}
\end{assumption}
\begin{theorem}\cite[Thm.~2.19]{rawlings2017model} 
\label{thm:prelim_MPC_term}
Let Assumptions~\ref{asm:continuity}, \ref{asm:prelim_stabilizing}, and \ref{asm:prelim_terminal} hold. 
Suppose Problem~\eqref{problem:prelim} is feasible with $x=x_0$. 
Then, Problem~\eqref{problem:prelim} is feasible for all $t\in\mathbb{I}_{\geq 0}$, the constraints~\eqref{eq:constraints} are satisfied, $x_{\mathrm{r}}$ is asymptotically stable, and the following performance bound holds for the resulting closed-loop system~\eqref{eq:sys_closedloop}: 
 \begin{align}
\label{eq:prelim_performance}
\mathcal{J}^{\mathrm{cl}}_{\infty}(x_0):=\sum_{t=0}^{\infty}\ell(x(t),u(t))\leq \mathcal{J}_N^\star(x_0).
\end{align}
\end{theorem}
The intuition behind the terminal set and terminal cost is to approximate the infinite-horizon tail for $k\in\mathbb{I}_{\geq N}$~\citep{chen1998quasi}, i.e., a feasible solution to Problem~\eqref{eq:optimal_control} is given by appending $\mathbf{u}^\star(x)\in\mathbb{U}^N$ with the terminal control law $k_{\mathrm{f}}(x)$ for $k\in\mathbb{I}_{\geq N}$ due to~\ref{enum_term_1}, \ref{enum_term_2}. 
Stability is ensured by showing that the value function $\mathcal{J}_N^\star$ is a Lyapunov function, i.e., 
Condition~\ref{enum_term_3} implies $\mathcal{J}_N^\star(x(t+1))-\mathcal{J}_N^\star(x(t))\leq -\ell(x(t),u(t))$, $t\in\mathbb{I}_{\geq 0}$. 
Condition~\ref{enum_term_4} is a technical condition to ensure that $\mathcal{J}_N^\star$ is a Lyapunov function, which holds trivially if $x_{\mathrm{r}}\in\mathrm{int}(\mathbb{X}_{\mathrm{f}})$~\citep[Sec.~2.4.2]{rawlings2017model}. 
Under additional technical conditions, the performance bound~\eqref{eq:prelim_performance} also yields a suboptimality/regret bound w.r.t. the optimal performance $\mathcal{J}^\star_\infty$ (cf., \citet[App.~A]{kohler2021dynamic} and \citet[Thm.~5.22]{grune2017nonlinear}). 
Overall, Theorem~\ref{thm:prelim_MPC_term} provides all the desired closed-loop properties and the posed conditions (Asm.~\ref{asm:prelim_stabilizing}--\ref{asm:prelim_terminal}) can be constructively satisfied (cf. Sec.~\ref{sec:terminal}). 

\subsection{Challenges in dynamic operation}
\label{sec:challenges}
In the following, we explain how the challenges related to dynamic operation~\ref{enum_motivation_1}--\ref{enum_motivation_3} complicate the application of this design and how the different frameworks discussed in this article approach these problems. 
The design of the terminal ingredients (Asm.~\ref{asm:prelim_terminal}) is centred around the CLF $V_{\mathrm{f}}$ and a control law $k_{\mathrm{f}}$ that locally stabilizes the steady state $x_{\mathrm{r}}$. 
This offline design becomes challenging if the setpoint $x_{\mathrm{r}}$ can change online~\ref{enum_motivation_2} and non-stationary references are considered ~\ref{enum_motivation_1}. 
Constructive designs addressing this issue are presented in Section~\ref{sec:terminal}. 

The desired setpoint $x_{\mathrm{r}}$ can change arbitrary online~\ref{enum_motivation_2} and may even be physically infeasible~\ref{enum_motivation_3}. 
This can lead to infeasibility of Problem~\eqref{problem:prelim} due to the terminal set constraint $\mathbb{X}_{\mathrm{f}}$ and hence invalidate all closed-loop guarantees. 
In Section~\ref{sec:artificial}, artificial references are included in the MPC formulation to avoid these complications. 

All of these designs try to stabilize some given reference using a positive definite stage cost $\ell$ (Asm.~\ref{asm:prelim_stabilizing}). 
Section~\ref{sec:economic} shows how to directly minimize an economic (indefinite) cost $\ell$~\ref{enum_motivation_3}. 

Lastly, Section~\ref{sec:UCON} explores an alternative approach: deriving system theoretic conditions and a sufficiently long prediction horizon $N$ to ensure closed-loop properties for simpler MPC designs, which do not require $V_{\mathrm{f}}$, $\mathbb{X}_{\mathrm{f}}$ satisfying Assumption~\ref{asm:prelim_terminal}.

The summary (Sec.~\ref{sec:discussion_summary}) provides a detailed discussion how these different methods address the challenges~~\ref{enum_motivation_1}--\ref{enum_motivation_3}.

\section{Terminal ingredients for nonlinear tracking MPC}
\label{sec:terminal}
In this section, we focus on constructing a terminal cost $V_{\mathrm{f}}$ and terminal set $\mathbb{X}_{\mathrm{f}}$ (Asm.~\ref{asm:prelim_terminal}) for dynamic tracking problems, i.e., references that are non-stationary~\ref{enum_motivation_1} and subject to unpredictable online changes~\ref{enum_motivation_2}.  
We first summarize the standard linearization-based design for the regulation problem (Sec.~\ref{sec:terminal_1}) and discuss extensions to track online changing setpoints (Sec.~\ref{sec:terminal_2}). 
Then, we consider tracking of dynamic reference trajectories (Sec.~\ref{sec:terminal_3}), including the case where the full reference may change online (Sec.~\ref{sec:terminal_4}). 
Lastly, we provide some discussion (Sec.~\ref{sec:terminal_5}), an illustrative numerical example (Sec.~\ref{sec:terminal_example}), and mention open issues (Sec.~\ref{sec:terminal_openIssue}).

\subsection{Regulation problem}
\label{sec:terminal_1} 
We consider the basic stabilizing MPC formulation introduced in Section~\ref{sec:prelim} and provide a constructive design to satisfy Assumptions~\ref{asm:prelim_stabilizing}--\ref{asm:prelim_terminal}. 
The following assumption enables a local LQR design. 
\begin{assumption}
(Local LQR design)
\label{asm:quadCost_InteriorRef_DifferentialDynamics}
\begin{enumerate}[label=\alph*)]
\item Quadratic stage cost $\ell=\|x-x_{\mathrm{r}}\|_Q^2+\|u-u_{\mathrm{r}}\|_R^2$, $Q,R\succ 0$.
\label{enum_prelim_terminal_1}
\item Reference strictly feasible: $r=(x_{\mathrm{r}},u_{\mathrm{r}})\in \mathbb{Z}_{\mathrm{r}}\subseteq\mathrm{int}(\mathbb{Z})$, $f(x_{\mathrm{r}},u_{\mathrm{r}})=x_{\mathrm{r}}$. 
\label{enum_prelim_terminal_2}
\item The dynamics $f$ are twice continuously differentiable. 
\label{enum_prelim_terminal_3}
\end{enumerate}
\end{assumption}
The following method was first derived by~\cite{chen1998quasi}, cf. \citet[Sec.~2.5.5]{rawlings2017model} for the considered discrete-time variant. 
We denote the Jacobians by
\begin{align}
\label{eq:design_A_r}
A(r):=\left.\left[\dfrac{\partial f}{\partial x}\right]\right|_r,\quad B(r):=\left.\left[\dfrac{\partial f}{\partial u}\right]\right|_r
\end{align}
and abbreviate the following matrix expression related to the linear quadratic regulator (LQR):
\begin{align}
\label{eq:LQR_matrix}
&\mathrm{LQR}(A,B,K,P,P^+,Q,R)\\
:=&(A+BK)^\top P^+ (A+BK)-P+Q+K^\top R K.\nonumber
\end{align}
The condition $LQR\succeq 0$ can enforce Condition~\ref{enum_term_3} from Assumption~\ref{asm:prelim_terminal} for the linearization around the setpoint $r\in\mathbb{Z}_{\mathrm{r}}$. 
We account for the linearization error by imposing a stronger condition on the linearization with a tuning variable $\epsilon>0$. 
\begin{proposition}
\label{prop:terminal_LQR}
Let Assumption~\ref{asm:quadCost_InteriorRef_DifferentialDynamics} hold. 
Suppose $(A(r),B(r))$ is stabilizable and choose matrices $P,K$ satisfying $LQR(A(r),B(r),K,P,P,Q+\epsilon I_n,R)\succeq 0$, e.g., using the algebraic Riccati equation.  
Then, there exists a sufficiently small $\alpha>0$, such that the quadratic terminal cost $V_{\mathrm{f}}(x)=\|x-x_{\mathrm{r}}\|_{P}^2$, the linear terminal controller $k_{\mathrm{f}}(x)=u_{\mathrm{r}}+K(x-x_{\mathrm{r}})$ and the terminal set $\mathbb{X}_{\mathrm{f}}=\{x\in X|V_{\mathrm{f}}(x)\leq \alpha\}$ satisfy Assumptions~\ref{asm:prelim_stabilizing}--\ref{asm:prelim_terminal}. 
\end{proposition}
Twice continuous differentiability of $f$ ensures that Condition~\ref{enum_term_3} holds for the nonlinear system for all $\alpha\leq \alpha_1$, with some $\alpha_1>0$. 
Analytic formulas for $\alpha_1$ can be derived using bounds on the Hessian~\citep{chen1998quasi,rawlings2017model} and less conservative constants can be obtained using sampling-based approaches, cf. \citet[Rem.~3.1]{chen1998quasi}, \citet[Alg.~1]{koehler2020nonlinearTAC}, \citet{rajhans2019terminal}. 
Condition~\ref{enum_term_1} holds for all $\alpha\leq \alpha_2$ with some $\alpha_2>0$ since $(x_{\mathrm{r}},u_{\mathrm{r}})\in\mathbb{Z}_{\mathrm{r}}\subseteq\mathrm{int}(\mathbb{Z})$. 
In case $\mathbb{Z}$ is a polytope, $\alpha_2$ can be exactly determined using a linear program~\cite[Eq.~(10)]{conte2016distributed} and \citet[Lemma~3.37]{kohler2021dynamic} provide a similar procedure if $\mathbb{Z}$ is given by Lipschitz continuous inequality constraints.

\subsection{Tracking of online changing setpoints}
\label{sec:terminal_2}
In practice, the setpoint $x_{\mathrm{r}}$ to be stabilized is often subject to unpredictable online changes~\ref{enum_motivation_2}. 
However, the provided design requires some offline computation to determine a terminal cost $V_{\mathrm{f}}$ and a terminal set $\mathbb{X}_{\mathrm{f}}$ to stabilize a specific setpoint $x_{\mathrm{r}}$. 
This offline procedure needs to be repeated if we wish to stabilize a different setpoint $x_{\mathrm{r}}$, which is undesirable from a practical perspective. 
Due to its practical relevance, many approaches have been suggested to overcome this issue. 
\cite{findeisen2000nonlinear} suggest a fixed matrix $P$ for the quadratic terminal cost $V_{\mathrm{f}}$ for different setpoints $x_{\mathrm{r}}$ using a pseudo linearization, however, practical application of this theory is difficult. 
\cite{wan2003efficient,wan2004efficient} locally describe the nonlinear system as a linear difference inclusion (LDI) and compute constant matrices $P,K$, which are valid for a local set of steady states, i.e., enforcing $\mathrm{LQR}(A(r),B(r),K,P,P,Q,R)\succeq 0$ for all $r$ in some region. 
\citet[App.~B]{limon2018nonlinear} partition the steady-state manifold and compute different matrices $P,K$ for each partition using this LDI description, resulting in a piece-wise quadratic terminal cost. 
However, the manual partitioning can result in a cumbersome design process, especially if the system is strongly nonlinear or the steady-state manifold is high-dimensional, and the resulting discontinuity can bring additional complications.

\cite{koehler2020nonlinearTAC} alleviate these shortcomings with a continuous parametrization: 
 $V_{\mathrm{f}}(x,r)=\|x-x_{\mathrm{r}}\|_{P(r)}^2$, $k_{\mathrm{f}}(x,r)=u_{\mathrm{r}}+K(r)(x-x_{\mathrm{r}})$, $\mathbb{X}_{\mathrm{f}}(r)=\{x\in X|~V_{\mathrm{f}}(x,r)\leq \alpha(r)\}$ with $P(r),K(r),\alpha(r)$ continuous in $r$. 
The local CLF condition~\ref{enum_term_3} then holds with $\alpha(r)$ chosen sufficiently small (cf. Sec.~\ref{sec:artificial_alpha}) if
\begin{align}
\label{eq:terminal_setpoint_param}
\mathrm{LQR}(A(r),B(r),K(r),P(r),P(r),Q+\epsilon I_n,R)\succeq 0
\end{align}
holds for all feasible reference setpoints $r$. 
By interpreting the reference $r$ as a parameter, this can be viewed as a special case of gain-scheduling synthesis for linear-parameter varying (LPV) systems, a classical robust control problem~\citep{rugh2000research}. 
The computation of parametrized matrices $P,K$ satisfying Condition~\eqref{eq:terminal_setpoint_param} can be reformulated as linear matrix inequalities (LMIs), cf.~\citet{koehler2020nonlinearTAC} for details. 
As a result, suitable terminal ingredients can be obtained during runtime by simply evaluating the parametrized terminal ingredients around a new setpoint $x_{\mathrm{r}}$.

\subsection{Trajectory tracking}
\label{sec:terminal_3}
In the following, we address non-stationary operation~\ref{enum_motivation_1} in terms of tracking a time-varying reference trajectory $r(t)$, $t\in\mathbb{I}_{\geq 0}$. 
We assume that the reference is feasible, i.e., $x_{\mathrm{r}}(t+1)=f(x_{\mathrm{r}}(t),u_{\mathrm{r}}(t))$, $(x_{\mathrm{r}}(t),u_{\mathrm{r}}(t))\in\mathbb{Z}_{\mathrm{r}}$, $t\in\mathbb{I}_{\geq 0}$. 
For a given state $x$ and time $t$, the trajectory tracking MPC is characterized by 
\begin{align}
\label{problem:traj_track}
\min_{\mathbf{u}\in\mathbb{U}^N}&\sum_{k=0}^{N-1}\ell(x_{\mathbf{u}}(k,x),\mathbf{u}_k,t+k)+V_{\mathrm{f}}(x_{\mathbf{u}}(N,x),t+N)\\
\text{s.t. }&(x_{\mathbf{u}}(k,x),\mathbf{u}_k)\in\mathbb{Z},~ k\in\mathbb{I}_{[0,N-1]},~ x_{\mathbf{u}}(N,x)\in\mathbb{X}_{\mathrm{f}}(t+N),\nonumber
\end{align}
with $\ell(x,u,t)=\|x-x_{\mathrm{r}}(t)\|_Q^2+\|u-u_{\mathrm{r}}(t)\|_R^2$, minimizer $\mathbf{u}^\star(x,t)\in\mathbb{U}^N$, and value function $\mathcal{J}_N^\star(x,t)$.
The closed-loop system is given by 
\begin{align}
\label{eq:sys_closedloop_trajectory}
x(t+1)=f(x(t),\mathbf{u}^\star_{0}(x(t),t)),\quad t\in\mathbb{I}_{\geq 0}.
\end{align}
The analysis and closed-loop properties of such a trajectory tracking MPC are analogous to Section~\ref{sec:prelim}. 
\begin{assumption}
\label{asm:terminal_trajectory}
There exists a terminal control law $k_{\mathrm{f}}:X\times\mathbb{I}_{\geq 0}\rightarrow\mathbb{U}$ such that for all $t\in\mathbb{I}_{\geq 0}$ and all $x\in\mathbb{X}_{\mathrm{f}}(t)$:
\begin{enumerate}[label=(T.\arabic*)]
\item Constraint satisfaction: $(x,k_{\mathrm{f}}(x,t))\in\mathbb{Z}$;
\label{enum_term_traj_1}
\item Positive invariance: $f(x,k_{\mathrm{f}}(x,t))\in\mathbb{X}_{\mathrm{f}}(t+1)$;
\label{enum_term_traj_2}
\item Local CLF: \\
$V_{\mathrm{f}}(f(x,k_{\mathrm{f}}(x,t)),t+1)-V_{\mathrm{f}}(x,t)\leq -\ell(x,k_{\mathrm{f}}(x,t),t)$. 
\label{enum_term_traj_3}
\end{enumerate}
Furthermore, there exists a function $\alpha_{\mathrm{V}}\in\mathcal{K}_\infty$, such that for any $(x,t)$ such that Problem~\eqref{problem:traj_track} is feasible, it holds
\begin{enumerate}[label=(T.\arabic*), resume]
\item Weak controllability: $\mathcal{J}_N^\star(x,t)\leq \alpha_{\mathrm{V}}(\|x-x_{\mathrm{r}}(t)\|)$. 
\label{enum_term_traj_4}
\end{enumerate}
\end{assumption}
\begin{theorem}
\label{thm:MPC_traj_term}
Let Assumptions~\ref{asm:continuity}, \ref{asm:quadCost_InteriorRef_DifferentialDynamics}\ref{enum_prelim_terminal_1}, and \ref{asm:terminal_trajectory} hold. 
Suppose Problem~\eqref{problem:traj_track} is feasible with $(x,t)=(x_0,0)$. 
Then, Problem~\eqref{problem:traj_track} is feasible for all $t\in\mathbb{I}_{\geq 0}$, the constraints~\eqref{eq:constraints} are satisfied, and $x_{\mathrm{r}}(t)$ is uniformly asymptotically stable for the resulting closed-loop system~\eqref{eq:sys_closedloop_trajectory}. 
\end{theorem}
The proof and closed-loop properties are analogous to Theorem~\ref{thm:prelim_MPC_term} using $\mathcal{J}_N^\star(x,t)$ as a \textit{uniform} time-varying Lyapunov function~\cite[Thm.~2.22]{grune2017nonlinear}. 
Next, we focus on the constructive design of terminal ingredients for such time-varying reference trajectories $r(t)$ (Asm.~\ref{asm:terminal_trajectory}). 
The following design is largely based on the work by~\cite{faulwasser2011model,faulwasser2012optimization}. 
Analogous to the stabilization problem, we pick the linear-quadratic \textit{time-varying} parametrization 
$V_{\mathrm{f}}(x,t)=\|x-x_{\mathrm{r}}(t)\|_{P(t)}^2$, $k_{\mathrm{f}}(x,t)=u_{\mathrm{r}}(t)+K(t)(x-x_{\mathrm{r}}(t))$, $\mathbb{X}_{\mathrm{f}}=\{x\in X|V_{\mathrm{f}}(x,t)\leq \alpha(t)\}$. 
By linearizing Condition~\ref{enum_term_traj_3} and adding a slack $\epsilon>0$, we obtain the design conditions
\begin{align}
\label{eq:terminal_LTV}
\mathrm{LQR}(A(r(t)),B(r(t)),K(t),P(t),P(t+1),Q+\epsilon I,R)\succeq 0, 
\end{align}
which need to hold for all $t\in\mathbb{I}_{\geq 0}$. 
These conditions correspond to an LQR problem for the linear time-varying (LTV) system representing the dynamics of the error $x-x_{\mathrm{r}}$ locally around the reference $x_{\mathrm{r}}$. 
Tractable designs can, e.g., be obtained by assuming that the reference $r(t)$ becomes constant after some finite time. 
In this case, a stationary LQR is solved at the final state and then the time-varying LQR is solved backwards to obtain $P(t),K(t)$~\citep{faulwasser2011model}. 
The terminal set scaling $\alpha(t)$ can be chosen similar to the setpoint stabilization problem with the difference that $\alpha(t+1)$ and $\alpha(t)$ are coupled through the time-varying invariance condition~\ref{enum_term_traj_2}, cf.~\citet[Thm.~2]{faulwasser2011model}. 
\cite{aydiner2016periodic} propose a design for periodic reference trajectories, i.e., $r(t)=r(t+T)$, $\forall t\in\mathbb{I}_{\geq 0}$ with some $T\in\mathbb{I}_{\geq 1}$. 
By imposing the same periodicity in $P(t),K(t)$, the design can be cast as a periodic LQR problem or a sequence of $T$ coupled LMIs.

\subsection{Trajectory tracking with online changing trajectories}
\label{sec:terminal_4}
Next, we simultaneously address non-stationary operation~\ref{enum_motivation_1} with unpredictable online changes~\ref{enum_motivation_2} by considering time-varying reference trajectories $r$ that may change unpredictably during online operation. 
The design of a trajectory tracking MPC scheme (Sec.~\ref{sec:terminal_3}) requires computing the Jacobian of the nonlinear dynamics around the full reference trajectory $r(t)$ and solving a set of LMIs or Riccati equations to determine the LQR-based terminal ingredients. 
This can become computationally very expensive.
As a result, if the reference trajectory changes during runtime, re-computing terminal ingredients for a new reference $r(t)$ becomes a bottleneck in the practical application. 
To circumvent this issue, \cite{koehler2020nonlinearTAC} propose a \textit{reference generic offline computation}. 
The goal is to have one offline computation to obtain continuously parametrized terminal ingredients of the form $P(r)$, $K(r)$, such that 
\begin{align}
\label{eq:terminal_LPV}
\mathrm{LQR}(A(r),B(r),K(r),P(r),P(r^+),Q+\epsilon I,R)\succeq 0
\end{align}
for all $r,r^+\in\mathbb{Z}_{\mathrm{r}}$ with $x_{\mathrm{r}}^+=f(x_{\mathrm{r}},u_{\mathrm{r}})$. 
This problem can be cast as a gain-scheduling problem for LPV systems by viewing the reference $r$ as a parameter that is slowly time-varying, where a bound on the change can be deduced from the dynamics. 
There exists much literature on the synthesis of gain-scheduled controllers, which are used by~\citet[Prop.~1]{koehler2020nonlinearTAC} to derive a finite-dimensional semidefinite program (SDP) to compute $K(r),P(r)$ satisfying~\eqref{eq:terminal_LPV}. 
By defining $P(t)=P(r(t))$, $K(t)=K(r(t))$, these parametrized terminal ingredients also satisfy the LQR conditions~\eqref{eq:terminal_LTV} for any possible reference. 
Hence, the trajectory tracking MPC (Thm.~\ref{thm:MPC_traj_term}) can be implemented without a priori knowledge of the full reference and requires no repeated offline computations in case the reference changes. 
 
\subsection{Discussion}
\label{sec:terminal_5}
In the following, we discuss some variants. 
The efficient update of the scaling $\alpha>0$ characterizing the terminal set $\mathbb{X}_{\mathrm{f}}$ and general feasibility and stability questions under online changes of the reference are revisited in Section~\ref{sec:artificial}. 

\subsubsection{Terminal equality constraint} 
\label{sec:terminal_TEC}
A simpler design for the terminal ingredients is given by a terminal equality constraint, historically also called zero-terminal constraint~\citep{mayne1990receding}, with $\mathbb{X}_{\mathrm{f}}=\{x_{\mathrm{r}}\}$, $k_{\mathrm{f}}=u_{\mathrm{r}}$, $V_{\mathrm{f}}=0$. 
This design directly satisfies Conditions~\ref{enum_term_1}--\ref{enum_term_3} from Assumption~\ref{asm:prelim_terminal}. 
Condition~\ref{enum_term_4} holds if an additional local controllability condition is satisfied and the prediction horizon $N$ is larger than the controllability index~\cite[Prop. 3.10]{kohler2021dynamic}. 
This approach is easy to apply, which makes it particularly attractive for application with online changing setpoints or dynamic trajectories (Sec.~\ref{sec:terminal_2}, \ref{sec:terminal_4}, \ref{sec:artificial}), compare 
\citet[Sec.~III.A]{limon2018nonlinear}; \citet{fagiano2013generalized,muller2013economic,berberich2020tracking} and \citet{limon2016mpc,limon2014single}. 
However, there are significant drawbacks to this design: 
\cite{yu2014inherent} show that nonlinear MPC schemes with "proper" terminal ingredients are \textit{inherently robust}\footnote{%
Inherent robustness implies that recursive feasibility and some form of stability are preserved for sufficiently small model mismatch. 
These results also require that state constraints are relaxed using penalties.},
however, terminal equality constraints require a multi-step implementation\footnote{%
In a multi-step implementation, the first $\nu\in\mathbb{I}_{>1}$ elements of the optimal input sequence $\mathbf{u}^\star$ are applied to the system in open loop and the optimization problem is only solved every $\nu$ steps. 
\cite{berberich2022stability} choose $\nu$ larger than the controllability index to derive robustness guarantees.} to retain inherent robusness~\citep[Prop.~IV.1]{berberich2022stability}. 
Additional practical drawbacks include a small region of attraction and in general worse control performance, cf. the comparisons by \citet[Sec.~5]{chen1998quasi}, \citet[Sec.~V]{raff2006nonlinear}, \citet[Sec.~4.1]{koehler2020nonlinearAutomatica}, and \citet[Sec.~5.2]{koehler2020nonlinearTAC}.

\subsubsection{More general stage cost}
\label{sec:terminal_semidefinite}
Considering the design of the terminal cost $V_{\mathrm{f}}$ satisfying~\ref{enum_term_3} using $\mathrm{LQR}$: 
The same procedure can be applied for non-quadratic (twice cont. differentiable) stage costs $\ell$ by replacing $Q+K^\top R K$ by the Hessian of $\ell$, see \citet[Sec.~4.1]{amrit2011economic} and \citet[App. D]{koehler2020nonlinearTAC} for details.

\subsubsection{LPV \& incremental system properties}
\label{sec:terminal_LPV}
The design conditions~\eqref{eq:terminal_setpoint_param}/\eqref{eq:terminal_LPV} utilize LPV theory. 
There exists a rich history on addressing nonlinearity in MPC using LPV embeddings, compare the survey by~\cite{morato2020model}. 
Condition~\eqref{eq:terminal_LPV} ensures that any dynamically feasible reference $r$ can be stabilized, which is also referred to as \textit{incremental stability} with the \textit{contraction metric} $P$.\footnote{The presented quadratic terminal cost $V_{\mathrm{f}}$ is only a \textit{local} CLF. 
The Riemannian energy based on the metric $P(r)$ provides a "global" incremental Lyapunov function $V_{\mathrm{f}}$~\citep{manchester2017control}. 
Evaluating this function is computationally more expensive compared to the simple quadratic expression since it involves an integral. 
Details regarding the exact region where Condition~\ref{enum_term_traj_3} holds for $\mathbb{Z}_{\mathrm{r}}\neq \mathbb{R}^{n+m}$ can be found in~\cite[Prop.~5]{sasfi2022robust}.}
This relates to a long history on the interplay between LPV systems, incremental Lyapunov functions, and contraction metrics, compare \cite{angeli2002lyapunov,fromion2003theoretical,wang2019comparison,koelewijn2019linear} and \citet[Appendix C]{kohler2021dynamic}.

\subsection{Illustrative example}
\label{sec:terminal_example}
\begin{figure}
\begin{center} 
\includegraphics[scale=0.5]{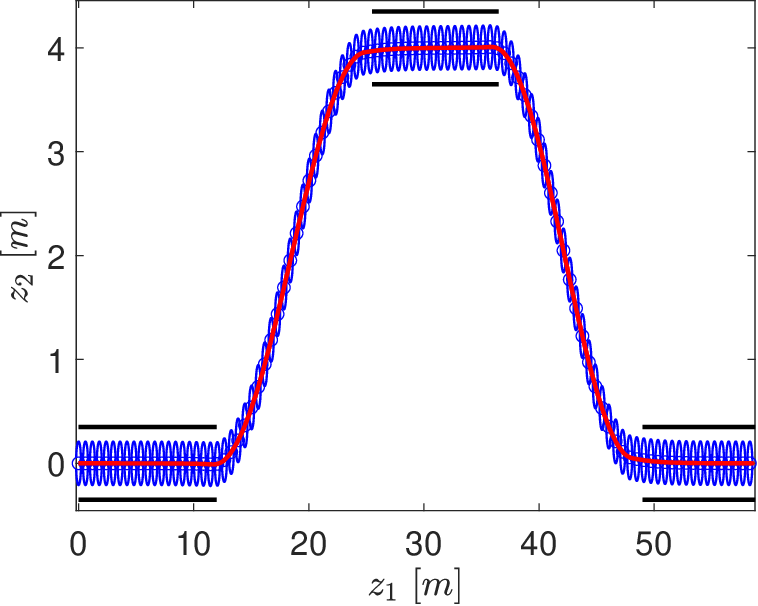} 
\caption{Evasion manoeuvrer of a car with reference trajectory $r$ (red), (projected) terminal sets (blue ellipses) and state constraints (black), adapted from~\cite{koehler2020nonlinearTAC}. 
The terminal ingredients are optimized offline before knowing the exact reference trajectory $r$ (cf. Sec.~\ref{sec:terminal_4}).}
\label{fig:trajtrack}
\end{center}
\end{figure}

The example by~\citet[Sec. IV]{koehler2020nonlinearTAC} considers a kinematic bicycle model of a car with $n=5$ states and $m=2$ inputs. 
The design~\eqref{eq:terminal_LPV} is utilized to compute terminal ingredients that are valid for any (dynamically feasible) reference trajectory $r(\cdot)$ in a specified constraint set $\mathbb{Z}_{\mathrm{r}}$. 
This offline optimization required 14 minutes using heuristic gridding. 
During online operation, an unexpected evasion manoeuvrer is required and valid terminal ingredients ensuring exponential stability are readily available, which are illustrated in Figure~\ref{fig:trajtrack}. 
This example demonstrates how to design nonlinear MPC schemes with guaranteed stability properties induced by suitable terminal ingredients, even for time-varying reference trajectories~\ref{enum_motivation_1} which are not known before hand~\ref{enum_motivation_2}.

\subsection{Open issues}
\label{sec:terminal_openIssue}
Limiting factors for practical deployment are complexity of the offline design for high-dimensional systems and performance limitations in case the terminal set $\mathbb{X}_{\mathrm{f}}$ is small. 
Methods to address scalability are outlined in Section~\ref{sec:discussion_distributed}.
Section~\ref{sec:discussion_UCON} provides a broad discussion on utilization of terminal ingredients in MPC and alternatives for easier deployment. 

 \section{Tracking MPC formulations using artificial references}
\label{sec:artificial}
In this section, we extend the tracking MPC formulations studied in Section~\ref{sec:terminal} to ensure closed-loop properties under online changing references~\ref{enum_motivation_2}, even if the desired target value is not achievable~\ref{enum_motivation_3}. 
Specifically, this is achieved by jointly optimizing \textit{artificial references}. 
We first present a setpoint tracking MPC with corresponding closed-loop properties (Sec.~\ref{sec:artificial_1}) and discuss practical implications (Sec.~\ref{sec:artificial_2}). 
Then, we present extensions to periodic reference tracking (Sec.~\ref{sec:artificial_3}) and discuss how partially decoupling tracking and planning can reduce the computational demand (Sec.~\ref{sec:artificial_4}). 
Lastly, we provide a numerical example (Sec.~\ref{sec:artificial_example}) and mention open issues (Sec.~\ref{sec:artificial_openIssue}). 

\subsection{Setpoint tracking MPC}
\label{sec:artificial_1}
We consider a system output $y=h(x,u)\in Y=\mathbb{R}^p$, with $h$ Lipschitz continuous, and some online supplied exogenous target value $y_{\mathrm{d}}(t)\in Y$, which may vary unpredictably during online operation~\ref{enum_motivation_2}. 
The controller should drive the system to a steady state that achieves this target output, i.e., $\lim_{t\rightarrow\infty}\|y_{\mathrm{d}}(t)-y(t)\|=0$. 
Since this is in general not feasible~\ref{enum_motivation_3}, we consider the more general problem of minimizing a continuous offset cost $V_{\mathrm{o}}(y,y_{\mathrm{d}})$, that quantifies the error between the system output $y$ and the target output $y_{\mathrm{d}}$.

\subsubsection{Setpoint tracking MPC - historical developments} 
The main difficulty stems from two factors: (i) $y_{\mathrm{d}}$ can arbitrarily change online~\ref{enum_motivation_2} and hence terminal set constraints centred around $y_{\mathrm{d}}$ are not recursively feasible; (ii) tracking $y_{\mathrm{d}}$ is in general ill-posed~\ref{enum_motivation_3} as it may not be feasible due to the constraints~\eqref{eq:constraints}. 
These challenges are related to the design of reference governors~\citep{garone2017reference}, a control unit that adjusts an externally provided reference $y_{\mathrm{d}}$ before passing it to a lower level controller in order to ensure constraint satisfaction. 
\cite{chisci2003dual,chisci2005predictive} propose a \textit{feasibility recovery mode}, which similarly adjusts the reference before passing it to an MPC to ensure feasibility. 
External loops to adjust the reference in MPC schemes are also used by~\cite{mayne2016generalized} and~\cite{skibik2021feasibility,skibik2022terminal}. 
The following exposition utilizes the framework of \textit{artificial references}, which includes this reference computation in the MPC formulation. 
This idea was initially developed by \cite{limon2008mpc}\footnote{%
For the special case of integrating processes, a similar idea was proposed earlier by~\cite{carrapicco2005stable}.} for linear systems and has led to many further developments in the literature concerning: 
\begin{itemize}
\item setpoint tracking for linear systems~\citep{Ferramosca2009,ferramosca2010mpc,simon2014reference,zeilinger2014real,limon2015tracking,berberich2020tracking,aboudonia2021distributed}, 
\item extensions to nonlinear systems~\citep{limon2018nonlinear,koehler2020nonlinearAutomatica,cotorruelo2020tracking,berberich2022linear_II,soloperto2022nonlinear,kohler2022distributed_setpoint,cunha2022robust,kohler2023transient,rickenbach2023active,galuppini2023nonlinear} 
\item and periodic reference trajectories~\citep{limon2014single,limon2016mpc,koehler2020nonlinearAutomatica,yang2021nonlinear,kohler2023distributed_periodic}. 
\end{itemize}

\subsubsection{Setpoint tracking MPC using artificial references}
For a given state $x\in X$ and a target value $y_{\mathrm{d}}\in Y$, the setpoint tracking MPC formulation is given by
\begin{align}
\label{problem:setpointTracking}
\min_{\mathbf{u}\in\mathbb{U}^N, r\in\mathbb{Z}_{\mathrm{r}}}&\sum_{k=0}^{N-1}\ell(x_{\mathbf{u}}(k,x),\mathbf{u}_k,r)+V_{\mathrm{f}}(x_{\mathbf{u}}(N,x),r)+V_{\mathrm{o}}(h(r),y_{\mathrm{d}})\nonumber\\
\text{s.t. }&(x_{\mathbf{u}}(k,x),\mathbf{u}_k)\in\mathbb{Z},~ k\in\mathbb{I}_{[0,N-1]},~ x_{\mathbf{u}}(N,x)\in\mathbb{X}_{\mathrm{f}}(r),\\
& f(x_{\mathrm{r}},u_{\mathrm{r}})=x_{\mathrm{r}}.\nonumber
\end{align} 
A minimizer is denoted by $r^\star(x,y_{\mathrm{d}})$, $\mathbf{u}^\star(x,y_{\mathrm{d}})$ with the value function $\mathcal{J}^\star_{\mathrm{tr}}(x,y_{\mathrm{d}})$. The closed-loop system is given by
\begin{align}
\label{eq:sys_closedloop_limon}
x(t+1)=f(x(t),\mathbf{u}^\star_{0}(x(t),y_{\mathrm{d}}(t))),\quad t\in\mathbb{I}_{\geq 0}.
\end{align}
The main difference to a standard MPC formulation is the joint optimization over the artificial reference $r=(x_{\mathrm{r}},u_{\mathrm{r}})\in\mathbb{Z}_{\mathrm{r}}$. 
The offset cost $V_{\mathrm{o}}$ penalizes the distance between the output $y=h(r)$ at the artificial reference $r$ and the desired target $y_{\mathrm{d}}$. 
For a fixed artificial reference $r$, Problem~\eqref{problem:setpointTracking} corresponds to a standard stabilizing MPC scheme as studied in Theorem~\ref{thm:prelim_MPC_term}. 
The external target $y_{\mathrm{d}}$ only appears in the cost $V_{\mathrm{o}}$ and hence feasibility can be guaranteed independent of $y_{\mathrm{d}}$. 
\begin{theorem}
\label{thm:limon_1} 
Let Assumptions~\ref{asm:continuity} and \ref{asm:quadCost_InteriorRef_DifferentialDynamics}\ref{enum_prelim_terminal_1} hold. 
Suppose the terminal ingredients $V_{\mathrm{f}}$, $\mathbb{X}_{\mathrm{f}}$ are designed according to Section~\ref{sec:terminal_2} or Section~\ref{sec:terminal_TEC}, ensuring that the conditions in Assumption~\ref{asm:prelim_terminal} hold for any $r\in\mathbb{Z}_{\mathrm{r}}$.\footnote{These designs also ensure feasibility of Problem~\eqref{problem:prelim} and a quadratic upper bound on the value function $\mathcal{J}_N^\star(x)$ in a neighbourhood of the reference $x_{\mathrm{r}}$, cf.~\citet[Prop. 4, Lemma~5]{koehler2020nonlinearAutomatica}. 
This guarantee holds uniformly for all  $x_r\in\mathbb{Z}_{\mathrm{r}}$ and the neighbourhood $\|x-x_{\mathrm{r}}\|\leq \epsilon$ with some uniform $\epsilon>0$.}
Suppose further Problem~\eqref{problem:setpointTracking} is feasible for $x=x_0$. 
Then, for any time-varying target sequence $y_{\mathrm{d}}(t)$, Problem~\eqref{problem:setpointTracking} is feasible for all $t\in\mathbb{I}_{\geq 0}$ and the constraints~\eqref{eq:constraints} are satisfied for the resulting closed-loop system~\eqref{eq:sys_closedloop_limon}. 
Furthermore, if $y_{\mathrm{d}}$ is constant, then $\lim_{t\rightarrow\infty}\|x_{\mathrm{r}}^\star(x(t),y_{\mathrm{d}})-x(t)\|=0$.
\end{theorem}
The proof is analogous to Theorem~\ref{thm:prelim_MPC_term} by using the previously optimal artificial reference $r^\star$ as a candidate solution. 
\subsubsection{Stability analysis} 
Theorem~\ref{thm:limon_1} only ensures convergence to the \textit{artificial} steady state $x_{\mathrm{r}}^\star$. 
Convergence and stability w.r.t. the target $y_{\mathrm{d}}$ require additional conditions. 
We consider a quadratic offset cost $V_{\mathrm{o}}(y,y_{\mathrm{d}})=\|y-y_{\mathrm{d}}\|_{S}^2$ with $S\succ 0$, but much of the presented theory generalizes to continuous (sub)differentiable, strictly convex functions~\cite[Asm.~2]{limon2018nonlinear}. 
The set of feasible output references is given by
\begin{align}
\label{eq:artificial_output_set}
\mathbb{Y}_{\mathrm{s}}:=\{y\in Y| ~\exists (x,u)\in\mathbb{Z}_{\mathrm{r}}, h(x,u)=y,~f(x,u)=x\}
\end{align}
and a minimizing setpoint is given by $y_{\mathrm{rd}}^\star(y_{\mathrm{d}})\in\arg\min_{y\in\mathbb{Y}_{\mathrm{s}}}V_{\mathrm{o}}(y,y_{\mathrm{d}})$. Note that $y_{\mathrm{rd}}^\star(y_{\mathrm{d}})=y_{\mathrm{d}}$ if $y_{\mathrm{d}}\in \mathbb{Y}_{\mathrm{s}}$. 
\begin{assumption}
\label{asm:limon_convex_unique}
\cite[Asm.~1--2]{limon2018nonlinear}
\begin{enumerate}[label=(Tr.\arabic*)]
\item Convexity: The set $\mathbb{Y}_{\mathrm{s}}$ is convex.
\label{enum_convex}
\item Uniqueness: There exists a unique Lipschitz continuous function $g:\mathbb{Y}_{\mathrm{s}}\rightarrow\mathbb{Z}_{\mathrm{r}}$, such that $g(y)=(x,u)$ for any steady state $(x,u)\in\mathbb{Z}_{\mathrm{r}}$, $f(x,u)=x$, $y=h(x,u)$. 
\label{enum_unique}
\end{enumerate}
\end{assumption}
Assumption~\ref{asm:limon_convex_unique} and $V_{\mathrm{o}}$ (strictly) convex imply that there exist a unique optimal steady state $(x_{\mathrm{rd}}^\star,u_{\mathrm{rd}}^\star)=g(y^\star_{\mathrm{rd}})$ and the following theorem ensures stability of this steady state.
\begin{theorem}
\label{thm:limon_2}
\cite[Thm.~8]{koehler2020nonlinearAutomatica} 
Let the conditions in Theorem~\ref{thm:limon_1} hold. Suppose further that  Assumption~\ref{asm:limon_convex_unique} holds and that  $y_{\mathrm{d}}(t)$ is constant for all $t\in\mathbb{I}_{\geq 0}$. 
Then, the optimal steady state $x_{\mathrm{rd}}^\star$ is exponentially stable for the resulting closed-loop system~\eqref{eq:sys_closedloop_limon} with the Lyapunov function $\mathcal{J}^\star_{\mathrm{tr}}(x,y_{\mathrm{d}})-V_{\mathrm{o}}(y_{\mathrm{rd}}^\star(y_{\mathrm{d}}),y_{\mathrm{d}})$. 
\end{theorem}
The main idea behind this result is that if $x$ is close to the artificial reference $x_{\mathrm{r}}^\star$, then the optimizer can move the artificial reference $y_{\mathrm{r}}$ towards the minimizer $y_{\mathrm{rd}}^\star$ (using convexity), thus also decreasing the offset cost $V_{\mathrm{o}}$. 
\cite{limon2008mpc,limon2018nonlinear} showed convergence with a case distinction, however, more recent results directly derive suitable Lyapunov inequalities~\citep{zeilinger2014real,limon2014single,limon2016mpc,koehler2020nonlinearAutomatica,soloperto2022nonlinear,cunha2022robust,kohler2023transient}. 
The following intermediate lemma provides further intuition for this stability result. 
\begin{lemma}
\label{lemma:setpoint_bound}
(Adapted from \cite[Prop.~2]{soloperto2022nonlinear})
Suppose the conditions in Theorem~\ref{thm:limon_2} hold. 
There exists a constant $c>0$, such that for any $(x,y_{\mathrm{d}})$ with Problem~\eqref{problem:setpointTracking} feasible it holds that $\|x-x_{\mathrm{r}}^\star(x,y_{\mathrm{d}})\|_Q^2\geq c\|r^\star(x,y_{\mathrm{d}})-g(y_{\mathrm{rd}}^\star(y_{\mathrm{d}}))\|^2$.
\end{lemma}
A similar bound is derived by~\citet[Lemma~2]{limon2014single} assuming continuity of minimizers w.r.t. $x$ and results comparable to Lemma~\ref{lemma:setpoint_bound} are also recently derived by \citet[Lemma~4]{cunha2022robust} and \citet[Lemma~2]{kohler2023transient}. 
Lemma~\ref{lemma:setpoint_bound} ensures that the distance between the artificial reference $r^\star$ and the optimal setpoint $r_{\mathrm{rd}}^\star=g(y_{\mathrm{rd}}^\star)$ is upper bounded by the distance of the system state $x$ to the artificial reference $x_{\mathrm{r}}^\star$, which converges to $0$ according to Theorem~\ref{thm:limon_1}. 

\subsection{Discussion}
\label{sec:artificial_2}
In the following, we discuss some of the properties of this setpoint tracking MPC formulation and different variations. 

\subsubsection{General properties}
The introduction of the artificial setpoint $r$ in Problem~\eqref{problem:setpointTracking} does not increase the design complexity and, depending on the prediction 
horizon $N$, has a small impact on the overall computational complexity. 
However, there are a number of practical benefits.  
Feasibility is completely independent of the external target $y_{\mathrm{d}}$, thus efficiently accounting for unpredictable online changes~\ref{enum_motivation_2}. 
Having a unified formulation that jointly solves planning and tracking simplifies the design and can effectively deal with infeasible targets~\ref{enum_motivation_3}. 
Furthermore, the offset cost $V_{\mathrm{o}}$ allows for intuitive tuning, e.g., \cite{Ferramosca2009} ensure local optimality using a large (exact) penalty while \cite{berberich2022linear_II} provide strong robustness properties using a small offset cost.
Notably, artificial references can also significantly increase the region of attraction, enabling large setpoint changes with arbitrary small prediction horizons. 
Artificial setpoints also provide a simple tool to enable coordination in a distributed setting~\citep{kohler2022distributed_setpoint,rickenbach2023active} and can easily accommodate online model updates~\citep{berberich2022linear_II,sasfi2022robust,peschke2023RAMPC_track}. 

\subsubsection{Convexity}
\label{sec:artificial_2_convex}
The presented analysis (Thm.~\ref{thm:limon_2}/Lemma~\ref{lemma:setpoint_bound}) uses convexity of the set of feasible steady-state outputs $\mathbb{Y}_{\mathrm{s}}$. 
This condition does \textit{not} require convexity of the steady-state manifold $(x,u)$, which can be significantly more restrictive for nonlinear systems (cf. \citet[Fig.~1]{koehler2020nonlinearAutomatica}). 
If $\mathbb{Y}_{\mathrm{s}}$ is not convex, the closed-loop system converges to some steady state, which is in general a \textit{local} minimum.
\cite{soloperto2022safe} show that this issue can sometimes be circumvented by simply increasing the offset cost $V_{\mathrm{o}}$ and using a large horizon $N$. 
More systematically, \cite{cotorruelo2020tracking} derive a transformation of the output to ensure the convexity condition holds for the special case of non-convex sets in normal form (e.g., star-shaped). 
\citet[Prop.~4]{soloperto2022nonlinear} show that convexity is not needed if the offset cost $V_{\mathrm{o}}$ is chosen as the path-distance restricted to the feasible steady-state manifold. 
The design by \cite{cotorruelo2020tracking} allows for an efficient implementation, but application for more general environments with multiple obstacles seems challenging. 
The approach by \cite{soloperto2022nonlinear} is applicable to complex non-convex environments. However, the implicit characterization of $V_{\mathrm{o}}$ essentially requires the solution to a (simplified) planning problem, thus increasing computational complexity. 

\subsubsection{Uniqueness and zone tracking}
\label{sec:artificial_zone}
Condition~\ref{enum_unique} in Assumption~\ref{asm:limon_convex_unique} ensures that the optimal steady state $x_{\mathrm{rd}}$ is unique. 
\citet[Rem.~1]{limon2018nonlinear} derive a unique Lipschitz continuous function $g$ using a rank condition on the Jacobians of $f,h$ and the implicit function theorem. 
If this condition does not hold, one can also specify a unique map $g$ in Problem~\eqref{problem:setpointTracking}, cf.~\cite[Equ. (9)]{limon2018nonlinear}. 
More generally, if stability of a specific point is not required, a more flexible \textit{zone tracking} MPC can be considered~\citep{ferramosca2010mpc,soloperto2022nonlinear}. 
In this case, the offset cost $V_{\mathrm{o}}$ can, e.g., penalize deviations of the output $y$ from some general desired set $\mathbb{Y}_{\mathrm{d}}\subseteq Y$. 
Correspondingly, one can show that the closed-loop system stabilizes a set of optimal steady states instead of one unique state. 
Notably, while the analysis, e.g., to show Lemma~\ref{lemma:setpoint_bound}, can become more complex~\cite[Prop.~1]{soloperto2022nonlinear}), the implementation of such a zone-tracking MPC remains simple.\footnote{%
 \cite{ferramosca2010mpc} explain in detail how the offset cost $V_{\mathrm{o}}$ can be cast as a linear or quadratic cost using additional linear inequality constraints. 
Notably, \cite{liu2019model} provide a more general zone-tracking MPC formulation that directly uses invariant sets instead of steady states $x_{\mathrm{r}}$, however, this requires a significantly different (set-based) MPC formulation.}
 
\subsubsection{Semidefinite input-output cost $\ell$}
For the intermediate result (Lemma~\ref{lemma:setpoint_bound}) and for the overall stability result (Thm.~\ref{thm:limon_2}), it is crucial that the stage cost $\ell$ is positive definite w.r.t. $x$. 
However, e.g., in case of input-output models, it is quite common to formulate the stage cost $\ell$ using the output $y$, which is only positive semidefinite in the state $x$. 
\cite{berberich2020tracking} show stability of the optimal steady state $x_{\mathrm{rd}}^\star$ with such stage costs for linear observable systems. 
\cite{galuppini2023nonlinear} provide initial results for more general nonlinear detectable systems. 
However, in contrast to~\cite{berberich2020tracking}, this implementation explicitly requires the storage function certifying detectability. 
This is due to a difference in the considered Lyapunov function, with the storage function evaluated at the online optimized steady state $x_{\mathrm{r}}$ or the (unknown) optimal steady state $x_{\mathrm{rd}}^\star$.

\subsubsection{Terminal set $\mathbb{X}_{\mathrm{f}}(r)$ for setpoint tracking}
\label{sec:artificial_alpha}
\cite{limon2008mpc} provide a polytopic terminal set $\mathbb{X}_{\mathrm{f}}(r)$ using the 
maximal admissible set of an augmented state for linear systems with polytopic constraints. 
However, computing this set can result in scalability issues and this approach does not transfer to nonlinear systems. 
Instead, a common approach is to scale a parametrized set: $\mathbb{X}_{\mathrm{f}}(r)=\{x\in X|~V_{\mathrm{f}}(x,r)\leq \alpha(r)\}$ with $V_{\mathrm{f}}$ quadratic and a scaling $\alpha(r)>0$. 
\citet[Lemma III.2]{zeilinger2014soft} derive a quadratic expression for $\alpha(r)$. 
\citet[App.~B]{limon2018nonlinear} compute a piece-wise constant scaling $\alpha(r)$ by partitioning the steady-state manifold offline. 
\cite{cotorruelo2020SOS} obtain a polynomial expression for $\alpha(r)$ by solving a sum-of-squares (SOS) problem offline. 
However, computing a parametrized function $\alpha(r)$ offline is conservative. 
 
\cite{koehler2020nonlinearAutomatica} address this issue by including the scaling $\alpha\in[\alpha_{\min},\alpha_1]\subseteq\mathbb{R}_{\geq 0}$ as a decision variable in Problem~\eqref{problem:setpointTracking}. 
Specifically, the constraint $r\in\mathbb{Z}_{\mathrm{r}}$ in Problem~\eqref{problem:setpointTracking} is replaced by comparable constraints on $(r,\alpha)$ that corresponds to Condition~\ref{enum_term_1}. 
In the special case of $k_{\mathrm{f}}/V_{\mathrm{f}}$ linear/quadratic (cf. Sec.~\ref{sec:terminal}) and $\mathbb{Z}$ polytopic, this is a set of linear inequality constraints. 
However, the approach can also be applied if $\mathbb{Z}$ is given by Lipschitz continuous inequality constraints~\cite[Sec.~3.2.2]{kohler2021dynamic}. 
The lower bound $\alpha\geq \alpha_{\min}>0$ ensures that Lemma~\ref{lemma:setpoint_bound} holds with a uniform constant $c>0$ and hence the exponential stability result (Thm.~\ref{thm:limon_2}) remains valid~\cite[Prop.~12]{koehler2020nonlinearAutomatica}. 
The increase in computational demand due to the online optimization of the scaling $\alpha$ tends to be negligible, while it can have significant performance benefits~\cite[Sec.~4]{koehler2020nonlinearAutomatica}.
For linear systems with a polytopic terminal set $\mathbb{X}_{\mathrm{f}}$, a comparable online optimization of the scaling $\alpha$ is also suggested by~\cite{simon2014reference}. 
It is possible to further generalize the terminal set by online optimizing over matrices $P\in\mathbb{R}^{n\times n}$, $K\in\mathbb{R}^{m\times n}$ characterizing $V_{\mathrm{f}},k_{\mathrm{f}}$ \citep{yang2021nonlinear,aboudonia2021distributed}. 
However, this introduces LMI constraints in the MPC formulation, which can significantly increase the computational complexity.

\subsection{Dynamic reference tracking}
\label{sec:artificial_3}
Section~\ref{sec:artificial_2} ensures recursive feasibility for arbitrary time-varying targets $y_{\mathrm{d}}(t)$, but stability/performance guarantees are restricted to (piece-wise) constant targets $y_{\mathrm{d}}$. 
Intuitively, online optimized artificial setpoints $r$ provide reasonably good performance if the target $y_{\mathrm{d}}$ is only "slowly" changing with time.\footnote{%
A corresponding performance result is not yet available in the literature, however, we conjecture that performance bounds similar to \cite[Thm.~10]{nonhoff2022online} can be established.} 
However, in many applications the optimal operation is intrinsically dynamic~\ref{enum_motivation_1} and trying to steer the system to a steady state $r\in\mathbb{Z}_{\mathrm{r}}$ may not result in satisfactory performance. 
Instead, similar to Sections~\ref{sec:terminal_3}--\ref{sec:terminal_4}, dynamic reference trajectories $\mathbf{r}_k\in\mathbb{Z}_{\mathrm{r}}$, $k\in\mathbb{I}_{\geq 0}$ need to be considered. 

\subsubsection{Periodic reference tracking}
\label{sec:artificial_3_1}
The special case of periodic target signals, i.e., $y_{\mathrm{d}}(t+T)=y_{\mathrm{d}}(t)$ with some period length $T\in\mathbb{I}_{\geq 1}$, has been studied for linear systems by ~\cite{limon2014single,limon2016mpc} with extensions to nonlinear systems by \cite{koehler2020nonlinearAutomatica,yang2021nonlinear,kohler2023distributed_periodic}. 
Let us denote the set of $T$-periodic references by $\mathcal{R}_T\subseteq\mathbb{Z}_{\mathrm{r}}^T$. 
Given a target signal $\mathbf{y}_{\mathrm{d}}\in Y^T$, optimal periodic operation is characterized by
\begin{align}
\label{eq:best_periodic}
\min_{\mathbf{r}\in\mathcal{R}_T}V_{\mathrm{o},T}(\mathbf{r},\mathbf{y}_{\mathrm{d}}):=\min_{\mathbf{r}\in\mathcal{R}_T}\sum_{k=0}^{T-1}\|h(\mathbf{r}_k)-\mathbf{y}_{\mathrm{d},k}\|_S^2,
\end{align}
with a minimizer $\mathbf{r}^\star_{\mathrm{rd}}$. 
Given the state $x\in X$ and target signal $\mathbf{y}_{\mathrm{d}}\in Y^T$, the periodic tracking MPC formulation is given by: 
\begin{align}
\label{problem:PeriodicTracking} 
\min_{\mathbf{u}\in\mathbb{U}^N, \mathbf{r}\in\mathbb{Z}_{\mathrm{r}}^T}&\sum_{k=0}^{N-1}\ell(x_{\mathbf{u}}(k,x),\mathbf{u}_k,\mathbf{r}_k)+V_{\mathrm{f}}(x_{\mathbf{u}}(N,x),\mathbf{r}_N)+V_{\mathrm{o},T}(\mathbf{r},\mathbf{y}_{\mathrm{d}})\nonumber\\
\text{s.t. }&(x_{\mathbf{u}}(k,x),\mathbf{u}_k)\in\mathbb{Z},~ k\in\mathbb{I}_{[0,N-1]},~ x_{\mathbf{u}}(N,x)\in\mathbb{X}_{\mathrm{f}}(\mathbf{r}_N),\nonumber\\
&\mathbf{r}\in\mathcal{R}_{T}, ~\mathbf{r}_{k}:=\mathbf{r}_{\mathrm{mod}(k,T)},~k\in\mathbb{I}_{[T,N]}. 
\end{align}
Problem~\eqref{problem:PeriodicTracking} jointly optimizes a periodic artificial reference $\mathbf{r}\in\mathbb{Z}_{\mathrm{r}}^T$ to match the target signal $\mathbf{y}_{\mathrm{d}}\in Y^T$ and computes an input sequence $\mathbf{u}\in\mathbb{U}^N$ to drive the system state $x$ to this artificial reference $\mathbf{r}\in\mathbb{Z}_{\mathrm{r}}^T$. 
For $T=1$, this reduces to the setpoint tracking MPC in Problem~\eqref{problem:setpointTracking}. 
For a fixed artificial reference $\mathbf{r}$, Problem~\eqref{problem:PeriodicTracking} reduces to the trajectory tracking MPC scheme studied in Theorem~\ref{thm:MPC_traj_term}. 
Assuming the terminal set $\mathbb{X}_{\mathrm{f}}$ and the terminal cost $V_{\mathrm{f}}$ are properly designed (cf. Sec.~\ref{sec:terminal_3}, \ref{sec:terminal_TEC}), this scheme inherits the same properties of Theorem~\ref{thm:limon_1}: (i) Recursive feasibility and constraint satisfaction holds for any $\mathbf{y}_{\mathrm{d}}$;
(ii) if $\mathbf{y}_{\mathrm{d}}$ is consistent/periodic, i.e., $\mathbf{y}_{\mathrm{d},k}(t+1)=\mathbf{y}_{\mathrm{d},\mathrm{mod}(k+1,T)}(t)$, then $\lim_{t\rightarrow\infty}\|\mathbf{x}_{\mathrm{r},0}^\star(x(t),\mathbf{y}_{\mathrm{d}}(t))-x(t)\|=0$, i.e., the system converges to the artificial periodic reference $\mathbf{r}$. 
Similar to Theorem~\ref{thm:limon_2}, stability of the optimal periodic reference trajectory can be guaranteed with a suitable uniqueness and convexity condition for $\mathbf{r}\in\mathcal{R}_T$, see~\citet[Thm.~8]{koehler2020nonlinearAutomatica} for details. 

\subsubsection{Discussion}
This periodic tracking MPC scheme shares all the theoretical properties of the setpoint tracking MPC (Sec.~\ref{sec:artificial_1}), enabling a seamless extension to dynamic/periodic problems. 
However, there are some practical concerns to be considered: 
(i) Computational complexity is increased, especially if $T\gg N$. This issue will be addressed in Section~\ref{sec:artificial_4}. 
(ii) The design of terminal ingredients $V_{\mathrm{f}},\mathbb{X}_{\mathrm{f}}$ is more challenging. 
One solution is to combine the parametrized terminal ingredients (Sec.~\ref{sec:terminal_4}) with an online optimized scaling $\alpha>0$ (Sec.~\ref{sec:artificial_alpha}), cf. \cite[Prop.~12]{koehler2020nonlinearAutomatica}. 
Other possibilities include simple terminal equality constraint~\cite[Prop.~4]{koehler2020nonlinearAutomatica} or online optimizing terminal ingredients~\citep{yang2021nonlinear}. 
(iii) 
There are numerous control problems which cannot be addressed using \textit{periodic} references $\mathbf{r}\in\mathbb{Z}_{\mathrm{r}}^T$ and target signals $\mathbf{y}_{\mathrm{d}}\in Y^T$.

\subsection{Partially decoupled tracking and planning} 
\label{sec:artificial_4}
The MPC formulations with artificial references provide a general framework to simultaneously address challenges~\ref{enum_motivation_1}--\ref{enum_motivation_3} by combining a \textit{tracking} MPC (Problem~\eqref{problem:traj_track}) with \textit{reference planning} (Problem~\eqref{eq:best_periodic}). 
However, there are different cases in which the reference planning problem becomes computationally expensive: periodic references $\mathbf{r}\in\mathbb{Z}^T_{\mathrm{r}}$ with $T\gg N$ (Sec.~\ref{sec:artificial_3}); non-trivial offset cost $V_{\mathrm{o}}$~\cite[Prop.~4]{soloperto2022nonlinear}, \cite[Sec.~5]{rickenbach2023active}; joint optimization of terminal ingredients~\citep{yang2021nonlinear,aboudonia2021distributed}. 
This increases the computational complexity of the MPC scheme, which can be a bottleneck for practical application.

\subsubsection{Decoupled tracking \& planning}
A natural solution is to split Problem~\eqref{problem:setpointTracking} in two problems: a tracking MPC and a reference planner that can be solved independently. 
This corresponds to a classical tracking--planning decomposition as often applied with some tracking controller (not necessarily MPC), see the overview by~\cite{schweidel2022safe}. 
Corresponding approaches where the tracking controller is given by an MPC scheme (Sec.~\ref{sec:terminal_3}) are also routinely applied, see the architecture by~\cite{mayne2011tube} or experimental results by~\cite{liniger2015optimization,romero2022model}. 
Online changes in the target $y_{\mathrm{d}}$ require online re-planning of the reference $\mathbf{r}$. 
This may yield feasibility issues in the tracking MPC scheme if the reference planning is completely decoupled from the tracking. 

\subsubsection{Constrained reference planning} 
\begin{figure}
\begin{center} 
\includegraphics[width=0.48\textwidth]{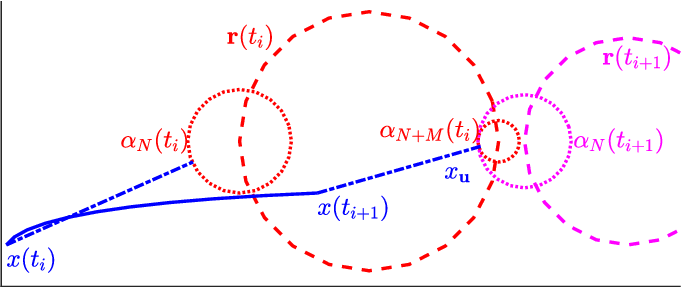} 
\caption{Illustration how the constrained reference planning~\eqref{problem:limon_partial} ensures recursive feasibility of the tracking MPC at time $t_{i+1}$. 
Closed-loop state $x(t)$, $t\in\mathbb{I}_{[t_i,t_{i+1}]}$ (blue, solid), predicted state sequence $x_{\mathbf{u}}$ of the tracker (blue, dash-dotted); artificial reference $\mathbf{r}$ at time $t_i$ (red, dashed) with terminal set scaling $\alpha$ (red-dotted);
artificial reference $\mathbf{r}$ at time $t_{i+1}$ (magenta, dashed) with terminal set scaling $\alpha$ (magenta, dotted). 
The predicted state sequence $x_{\mathbf{u}}$ of the tracker (blue, dash-dotted) satisfies the new terminal set constraint (magenta), since it contains the previous terminal set constraint (red), cf.~\eqref{problem:limon_partial_3}. } \label{fig:limon_partially_coupled}
\end{center}
\end{figure}

\citet[Sec.~3.4]{koehler2020nonlinearAutomatica} address this issue by updating the reference every $M\in\mathbb{I}_{\geq 1}$ steps and imposing additional constraints on the planner to ensure feasibility of the tracker. 
For simplicity, we consider again the special case of $T$-periodic reference trajectories (Sec.~\ref{sec:artificial_3}). 
At time $t_i=M\cdot i$, $i\in\mathbb{I}_{\geq 0}$, the reference planner computes a periodic reference $\mathbf{r}_k(t_i)$ and a terminal set scaling $\boldsymbol{\alpha}_k(t_i)$, $k\in\mathbb{I}_{[0,T-1]}$ corresponding to a terminal set $\mathbb{X}_{\mathrm{f}}(r,\alpha)=\{x\in X|~V_{\mathrm{f}}(x,r)\leq \alpha\}$, where we define $\mathbf{r}_k:=\mathbf{r}_{\mathrm{mod}(k,T)}$, $\boldsymbol{\alpha}_k:=\boldsymbol{\alpha}_{\mathrm{mod}(k,T)}$, $k\in\mathbb{I}_{\geq T}$. 
Then, for any $t\in[t_i,t_{i+1})$, the control input $u(t)$ is computed with a tracking MPC (Problem~\eqref{problem:traj_track}) using a terminal set $\mathbb{X}_{\mathrm{f}}(\mathbf{r}_{N+t-t_i}(t_i),\boldsymbol{\alpha}_{N+t-t_i}(t_i))$. 
The new reference for time $t_{i+1}$ is computed as:
\begin{subequations}
\label{problem:limon_partial}
\begin{align}
\label{problem:limon_partial_1}
\min_{\mathbf{r},\boldsymbol{\alpha}}&V_{\mathrm{o},T}(\mathbf{r},\mathbf{y}_{\mathrm{d}}(t_{i+1}))\\
\label{problem:limon_partial_2}
\text{s.t. }&(\mathbf{r},\boldsymbol{\alpha})\in\tilde{\mathcal{R}}_T,\\ 
\label{problem:limon_partial_3}
&\mathbb{X}_{\mathrm{f}}(\mathbf{r}_{N+M}(t_i),\boldsymbol{\alpha}_{N+M}(t_i))\subseteq \mathbb{X}_{\mathrm{f}}(\mathbf{r}_N,\boldsymbol{\alpha}_N). 
\end{align}
\end{subequations}
The set $\tilde{\mathcal{R}}_T$ characterizes the set of periodic reference trajectories $\mathbf{r}\in\mathbb{Z}_{\mathrm{r}}^T$ with scaling variables $\boldsymbol{\alpha}\in\mathbb{R}_{\geq 0}^T$ such that the periodic sequence of terminal sets $\mathbb{X}_{\mathrm{f}}(\mathbf{r}_k,\boldsymbol{\alpha}_k)$ satisfies Assumption~\ref{asm:terminal_trajectory}~\citep[Equ.~(29b)--(29e)]{koehler2020nonlinearAutomatica}.\footnote{%
Specifically, $(x,k_{\mathrm{f}}(x,r))\in\mathbb{Z}$ for all $x\in\mathbb{X}_{\mathrm{f}}(r,\alpha)$, which can be reformulated as discussed in Section~\ref{sec:artificial_alpha}. 
Furthermore, positive invariance requires a bound of the form $\boldsymbol{\alpha}_{k+1}\geq \rho \boldsymbol{\alpha}_k$, with some contraction rate $\rho<1$~\citep[Rem.~14]{koehler2020nonlinearAutomatica}, which trivially holds if $\alpha$ is constant.}
In the absence of constraint~\eqref{problem:limon_partial_3}, Problem~\eqref{problem:limon_partial} corresponds to a decoupled reference planning. 
Equation~\eqref{problem:limon_partial_3} constrains the planner such that feasibility of the terminal set constraint is preserved, see Figure~\ref{fig:limon_partially_coupled} for an illustration. 
Condition~\eqref{problem:limon_partial_3} is in general not computationally tractable, but simple sufficient conditions can be obtained using the quadratic parametrization of $V_{\mathrm{f}}$ from Section~\ref{sec:terminal_4}~\citep[Prop.~15]{koehler2020nonlinearAutomatica}. 
A feasible candidate solution to Problem~\eqref{problem:limon_partial} is given by the previous solution shifted by $M$ steps.

This design directly ensures recursive feasibility of the tracking MPC and the planner. 
However, Condition~\eqref{problem:limon_partial_3} may prevent convergence to the optimal periodic trajectory, i.e., in general~\eqref{problem:limon_partial_3} might only be feasible if $\mathbf{r}_N=\mathbf{r}_{N+M}(t_i)$. 
\citet[Prop.~16]{koehler2020nonlinearAutomatica} address this issue by using an exponential decay of the CLF $V_{\mathrm{f}}$ with some known factor $\rho<1$ to implement a contractive terminal set. 
This ensures that Constraint~\eqref{problem:limon_partial_3} is inactive for small enough reference updates and hence finite-time convergence of $\mathbf{r}$ to the optimal periodic reference can be ensured. 

\subsubsection{Partial coupling between tracking and planning}
Problem~\eqref{problem:limon_partial} uses no information of the current state of the tracker, which can result in slow convergence of the reference. 
\citet[Alg.~1]{koehler2020nonlinearAutomatica} propose a \textit{partial coupling}, which utilizes the most up to date information of the tracker to relax Condition~\eqref{problem:limon_partial_3}. 
Specifically, at time $t_{i}$ we know the predicted terminal cost in the tracking MPC, which satisfies $V_{\mathrm{f}}(x_{\mathbf{u}^\star}(N,x(t_{i})),\mathbf{r}_N(t_i))\leq \boldsymbol{\alpha}_{N}(t_{i})$. 
Thus, we shrink the scaling $\boldsymbol{\alpha}_{N+k}(t_i)$, $k\in\mathbb{I}_{\geq 0}$ to enforce more restrictive (but feasible) contractive constraints for the tracking MPC. 
As a result, Constraint~\eqref{problem:limon_partial_3} can be relaxed to 
\begin{align}
\label{eq:limon_partial_rho}
\mathbb{X}_{\mathrm{f}}(\mathbf{r}_{N+M}(t_i),\rho^M V_{\mathrm{f}}(x_{\mathbf{u}^\star}(N,x(t_i)),\mathbf{r}_{N}(t_i)))\subseteq \mathbb{X}_{\mathrm{f}}(\mathbf{r}_N,\boldsymbol{\alpha}_N). 
\end{align} 
The reduction in conservatism is due to the difference between the achieved convergence of the closed-loop tracking MPC and the a priori worst-case bound based on the local contraction rate $\rho<1$ of the terminal control law $k_{\mathrm{f}}$. 
\cite{skibik2022terminal} propose a similar strategy for the special case of linear systems, steady states ($T=1$), and synchronous updates ($M=1$).

\subsubsection{Discussion}
Overall, this partially decoupled strategy retains the theoretical properties of the MPC formulations with artificial references: 
recursive feasibility and convergence to the optimal feasible reference. 
A standard tracking MPC scheme is implemented and reference updates are computed in parallel on a different time scale ($M\gg 1$). 
Notably, the additional constraint~\eqref{eq:limon_partial_rho} to recompute the reference $\mathbf{r}(t_{i+1})$ in the planner (Problem~\eqref{problem:limon_partial}) depends on the reference $\mathbf{r}(t_i)$ and the state $x(t_i)$, which are available at time $t_i$.\footnote{%
Problem~\eqref{problem:limon_partial} also depends on the target signal $\mathbf{y}_{\mathrm{d}}(t_{i+1})$, which is in general only available at time $t_{i+1}$. However, all theoretical properties remain valid if it is replaced by the periodic continuation of the available target signal $\mathbf{y}_{\mathrm{d}}(t_{i})$.}
Hence, this more complex planning problem can be solved in the time interval $[t_i,t_{i+1}]$. 
This significantly reduces the computational requirements, allowing for fast feedback and complex planning problems in practical application. 
However, an increase in $M$ also slows down convergence to the optimal periodic reference $\mathbf{y}_{\mathrm{rd}}^\star$, resulting in a trade-off, see the numerical comparison by~\citet[Sec.~4]{koehler2020nonlinearAutomatica}.

\subsection{Illustrative example}
\label{sec:artificial_example}
\begin{figure}
\begin{center} 
\includegraphics[scale=0.5]{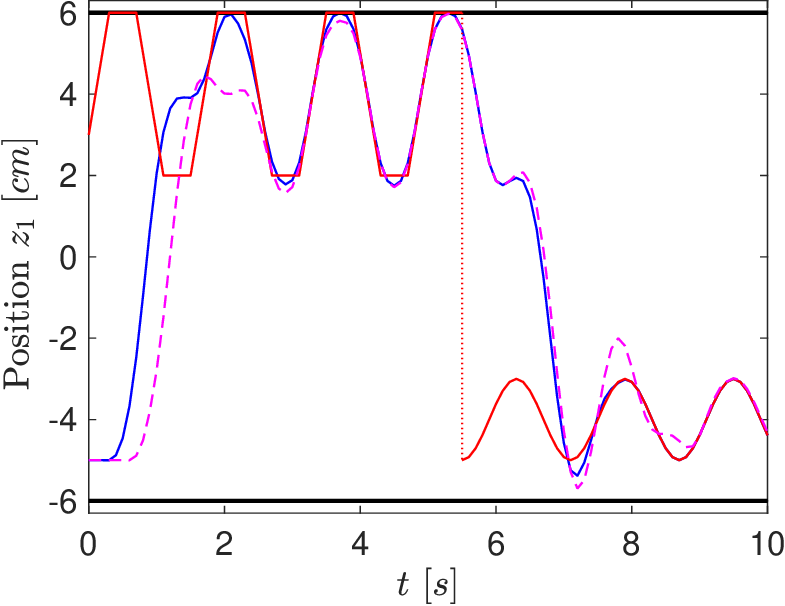} 
\end{center}
\caption{
Periodic tracking with a ball-and-plate system, adapted from~\cite{koehler2020nonlinearAutomatica}. Closed-loop state $x$ (blue, solid) resulting from periodic tracking MPC (Sec.~\ref{sec:artificial_3_1}) with a time-varying target signal $y_{\mathrm{d}}$ (red) and
state constraints $\mathbb{X}$ (black). 
Closed-loop state resulting from partially decoupled tracking and planning (Sec.~\ref{sec:artificial_4}) with $M=2$ shown dashed in magenta.}
\label{fig:artificial_example}
\end{figure}

Figure~\ref{fig:artificial_example} illustrates application of the tracking MPC formulations with a periodic artificial reference using a ball-and-plate system with $n=8$ states and $m=2$ inputs by~\cite{koehler2020nonlinearAutomatica}. 
This demonstrates optimal periodic operation, even if a time-varying infeasible target $y_{\mathrm{d}}$ is specified~\ref{enum_motivation_1},\ref{enum_motivation_3}, and effective dealing with online changes in the optimal mode of operation~\ref{enum_motivation_2}. 
Furthermore, a comparison with the partially decoupled tracking and planning (Sec.~\ref{sec:artificial_4}) clarifies that reduced computational complexity also reduces convergence speed. 
Experimental results demonstrating the practicality of the setpoint tracking MPC formulation (Sec.~\ref{sec:artificial_1}) are provided by~\citet{limon2018nonlinear,Nubert2020Robot,rickenbach2023active} with a four-tank, a robotic manipulator, and coordination of miniature cars, respectively. 

\subsection{Open issues}
\label{sec:artificial_openIssue}
Further research is required regarding convexity issues (Sec.~\ref{sec:artificial_2_convex}) to enhance applicability for cluttered obstacle domains in robotics. 
Practicality of the partially decoupled tracking and planning framework (Sec.~\ref{sec:artificial_4}) to complex real world problems remains to be demonstrated. 
The problem of non-periodic problems remains largely unexplored. 
We expect that further developments regarding flexible time parametrization (Sec.~\ref{sec:discussion_time}) can yield significant performance benefits.
Section~\ref{sec:discussion_artificial} provides a more detailed discussion regarding benefits and limitations of utilizing artificial references.

\section{Economic MPC}
\label{sec:economic}
In this section, we extend the problem formulation to \textit{economic} MPC~\citep{muller2017economic,ellis2017economic,faulwasser2018economic}. 
In control theory, one of the primary goals is to achieve \textit{stability} of the closed-loop system and steer the system state/output to some desired setpoint or trajectory. 
In Sections~\ref{sec:prelim}--\ref{sec:artificial}, we accomplished such tracking objectives with MPC schemes by using a stage cost $\ell$ which is positive definite w.r.t. a desired mode of operation (cf. Asm.~\ref{asm:prelim_stabilizing}). 
In economic MPC this condition is removed and instead an economic stage cost $\ell_{\mathrm{e}}$ is used 
that directly reflects the desired objective~\ref{enum_motivation_3}, e.g., minimizing energy consumption in HVAC or maximizing the yield in a chemical plant. 
Correspondingly, in this section our primary goal is \textit{not} to ensure some form of stability. 
Rather, we derive bounds on the closed-loop performance in terms of the economic stage cost $\ell_{\mathrm{e}}$. 
We first discuss how the basic stabilizing MPC can be extended to economic MPC (Sec.~\ref{sec:economic_1}), including non-stationary operation~\ref{enum_motivation_1} using a periodic reference (Sec.~\ref{sec:economic_2}). 
Then, we derive an economic MPC scheme for online changing operating conditions~\ref{enum_motivation_2} using artificial setpoints (Sec.~\ref{sec:economic_3}) and periodic artificial references (Sec.~\ref{sec:economic_4}). 
Furthermore, we discuss convergence/stability in economic MPC (Sec.~\ref{sec:economic_5}) and the design of an economic terminal cost (Sec.~\ref{sec:economic_6}). 
Finally, we provide a numerical example (Sec.~\ref{sec:eco_example}) and mention open issues (Sec.~\ref{sec:eco_openIssue}).

\subsection{Performance guarantees in economic MPC}
\label{sec:economic_1}
We extend the standard stabilizing MPC scheme from Section~\ref{sec:prelim} to directly consider an economic stage cost $\ell_{\mathrm{e}}$. 
\begin{assumption}(Continuity and compactness)
\label{asm:eco_continuity_compactness}
The economic stage cost $\ell_{\mathrm{e}}$ and the dynamics $f$ are continuous. 
The constraint set $\mathbb{Z}$ is compact. 
\end{assumption}
This assumption ensures that $\ell_{\mathrm{e}}$ is bounded. 
We define an optimal steady state $(x_{\mathrm{r}},u_{\mathrm{r}})\in\mathbb{Z}$ by
\begin{align}
\label{eq:eco_opt_steadyState}
\bar{\ell}_{\mathrm{e}}=\min_{(x_{\mathrm{r}},u_{\mathrm{r}})\in\mathbb{Z}}\ell_{\mathrm{e}}(x_{\mathrm{r}},u_{\mathrm{r}})\quad 
\text{s.t. } x_{\mathrm{r}}=f(x_{\mathrm{r}},u_{\mathrm{r}}).
\end{align}
We consider an economic MPC scheme which directly minimizes the economic stage cost $\ell_{\mathrm{e}}$ over the prediction horizon:
\begin{align}
\label{problem:eco}
\min_{\mathbf{u}\in\mathbb{U}^N}&\sum_{k=0}^{N-1}\ell_{\mathrm{e}}(x_{\mathbf{u}}(k,x),\mathbf{u}_k)+V_{\mathrm{f,e}}(x_{\mathbf{u}}(N,x))\\
\text{s.t. }&(x_{\mathbf{u}}(k,x),\mathbf{u}_k)\in\mathbb{Z},~ k\in\mathbb{I}_{[0,N-1]},~ x_{\mathbf{u}}(N,x)\in\mathbb{X}_{\mathrm{f}}.\nonumber
\end{align}
The closed-loop operation is defined by
\begin{align}
\label{eq:sys_closedloop_eco}
x(t+1)=f(x(t),\mathbf{u}^\star_{0}(x(t))),\quad t\in\mathbb{I}_{\geq 0},
\end{align}
where $\mathbf{u}^\star(x)$ is a minimizer to Problem~\eqref{problem:eco}. 
The conditions on the terminal ingredients are adjusted as follows. 
\begin{assumption}
\label{asm:terminal_eco}
Conditions~\ref{enum_term_1}--\ref{enum_term_2} from Assumption~\ref{asm:prelim_terminal} hold. 
Furthermore, the terminal cost $V_{\mathrm{f,e}}:X\rightarrow\mathbb{R}$ is continuous and satisfies
\begin{align}
\label{eq:eco_terminal}
V_{\mathrm{f,e}}(f(x,k_{\mathrm{f}}(x)))-V_{\mathrm{f,e}}(x)\leq -\ell_{\mathrm{e}}(x,k_{\mathrm{f}}(x))+\bar{\ell_{\mathrm{e}}}.
\end{align}
\end{assumption}
This condition can be trivially satisfied using a terminal equality constraint w.r.t. the steady state $x_{\mathrm{r}}$, i.e., $V_{\mathrm{f,e}}=0$, $\mathbb{X}_{\mathrm{f}}=\{x_{\mathrm{r}}\}$, $k_{\mathrm{f}}=u_{\mathrm{r}}$. More general designs are discussed in Section~\ref{sec:economic_6}. 
\begin{theorem}
\label{thm:eco_performance}
Let Assumptions~\ref{asm:eco_continuity_compactness} and \ref{asm:terminal_eco} hold. 
Suppose Problem~\eqref{problem:eco} is feasible with $x=x_0$. 
Then, Problem~\eqref{problem:eco} is feasible for all $t\in\mathbb{I}_{\geq 0}$, the constraints~\eqref{eq:constraints} are satisfied, and the following performance bound holds for the resulting closed-loop system~\eqref{eq:sys_closedloop_eco}:
\begin{align}
\label{eq:eco_performance}
\limsup_{K\rightarrow\infty}\dfrac{1}{K}\sum_{t=0}^{K-1}\ell_{\mathrm{e}}(x(t),u(t))\leq \bar{\ell_{\mathrm{e}}}.
\end{align}
\end{theorem} 
This result was initially developed by~\citet[Thm.~1]{angeli2012average} with a terminal equality constraint and then generalized by~\citet[Thm.~18]{amrit2011economic}. 
Theorem~\ref{thm:eco_performance} ensures that the closed-loop performance with an economic MPC is (on average) no worse than simply steering the system to a steady state, as done in a standard tracking MPC scheme. 
On a theoretical level, the only difference to Problem~\eqref{problem:prelim} is that Assumptions~\ref{asm:prelim_stabilizing}--\ref{asm:prelim_terminal} are relaxed to Assumption~\ref{asm:terminal_eco}. 
From a conceptual level, the more important difference is that the MPC directly minimizes the cost $\ell_{\mathrm{e}}$ instead of indirectly considering it by steering the system to some optimal setpoint $x_{\mathrm{r}}$. 
Such a direct minimization of the economic criterion $\ell_{\mathrm{e}}$ can significantly enhance performance, cf. the numerical investigations by~\citet{rawlings2008unreachable,rawlings2012fundamentals,ellis2017economic,Koehler2020Economic} and the discussion in Section~\ref{sec:discussion_eco}.

\subsection{Periodic economic MPC}
\label{sec:economic_2}
The design in Section~\ref{sec:economic_1} first determines an optimal steady state $x_{\mathrm{r}}$ and then Theorem~\ref{thm:eco_performance} ensures that the closed-loop performance is no worse than operation at this optimal steady state. 
In the following, we generalize this to provide performance guarantees w.r.t. optimal periodic operation~\ref{enum_motivation_1}, similar to~\cite{zanon2017periodic} and \cite{alessandretti2016convergence}. 
We consider an economic stage cost $\ell_{\mathrm{e}}(x,u,t)$ which is time-varying\footnote{%
Although we consider a time-varying cost $\ell_{\mathrm{e}}$ for generality, periodic operation can also outperform stationary operation in case of a time-invariant setting, see, e.g.,~\cite{zanon2017periodic,koehler2018periodic,Koehler2020Economic}.} and $T$-periodic with some period length $T\in\mathbb{I}_{\geq 1}$. 
We compute an optimal $T$-periodic reference $r(t+T)=r(t)\in\mathbb{Z}$, $t\in\mathbb{I}_{\geq 0}$: 
\begin{align}
\label{eq:periodic_eco_opt}
\bar{\ell}_{\mathrm{e}}:=\min_{\mathbf{r}\in\mathcal{R}_T}\dfrac{1}{T}\sum_{k=0}^{T-1}\ell_{\mathrm{e}}(\mathbf{r}_k,k),
\end{align}
where $\mathcal{R}_T\subseteq\mathbb{Z}^T$ is the set of $T$-periodic references (cf. Sec.~\ref{sec:artificial_3_1}). 
The goal is to design an economic MPC scheme which outperforms operation at this periodic reference $r(t)$. 
For a given state $x$ and time $t$, the periodic economic MPC formulation is given by: 
\begin{align}
\label{problem:eco_periodic}
\min_{\mathbf{u}\in\mathbb{U}^N}&\sum_{k=0}^{N-1}\ell_{\mathrm{e}}(x_{\mathbf{u}}(k,x),\mathbf{u}_k,t+k)+V_{\mathrm{f,e}}(x_{\mathbf{u}}(N,x),t+N)\\
\text{s.t. }&(x_{\mathbf{u}}(k,x),\mathbf{u}_k)\in\mathbb{Z},~ k\in\mathbb{I}_{[0,N-1]},~ x_{\mathbf{u}}(N,x)\in\mathbb{X}_{\mathrm{f}}(t+N),\nonumber
\end{align}
with a minimizer $\mathbf{u}^\star(x,t)\in\mathbb{U}^N$ and $\mathbb{X}_{\mathrm{f}}$, $V_{\mathrm{f,e}}$ are $T$-periodic. 
The closed-loop system is given by
\begin{align}
\label{eq:sys_closedloop_eco_periodic}
x(t+1)=f(x(t),\mathbf{u}^\star_{0}(x(t),t)),\quad t\in\mathbb{I}_{\geq 0}.
\end{align}
\begin{assumption}
\label{asm:terminal_eco_periodic}
There exists a $T$-periodic terminal control law $k_{\mathrm{f}}:X\times\mathbb{I}_{\geq 0}\rightarrow\mathbb{U}$ such that for all $t\in\mathbb{I}_{\geq 0}$ and all $x\in\mathbb{X}_{\mathrm{f}}(t)$ Conditions~\ref{enum_term_traj_1}--\ref{enum_term_traj_2} from Assumption~\ref{asm:terminal_trajectory} hold, $V_{\mathrm{f,e}}$ is continuous, and 
\begin{align}
\label{eq:eco_terminal_periodic}
&V_{\mathrm{f,e}}(f(x,k_{\mathrm{f}}(x,t)),t+1)-V_{\mathrm{f,e}}(x,t)\\
\leq& -\ell_{\mathrm{e}}(x,k_{\mathrm{f}}(x,t),t)+\ell_{\mathrm{e}}(x_{\mathrm{r}}(t),u_{\mathrm{r}}(t),t).\nonumber
\end{align}
\end{assumption}
\begin{theorem}
\label{thm:MPC_eco_periodic} 
Let Assumptions~\ref{asm:eco_continuity_compactness} and \ref{asm:terminal_eco_periodic} hold. 
Suppose Problem~\eqref{problem:eco_periodic} is feasible with $(x,t)=(x_0,0)$. 
Then, Problem~\eqref{problem:eco_periodic} is feasible for all $t\in\mathbb{I}_{\geq 0}$, the constraints~\eqref{eq:constraints} are satisfied, and the following performance bound holds for the resulting closed-loop system~\eqref{eq:sys_closedloop_eco_periodic}: 
\begin{align}
\label{eq:eco_performance_periodic}
\limsup_{K\rightarrow\infty}\dfrac{1}{K}\sum_{t=0}^{K-1}\ell_{\mathrm{e}}(x(t),u(t),t)\leq \bar{\ell}_{\mathrm{e}}. 
\end{align}
\end{theorem} 
Theorem~\ref{thm:MPC_eco_periodic} ensures that closed-loop performance with the economic MPC is (on average) no worse than using a trajectory tracking MPC scheme (Sec.~\ref{sec:terminal_3}) to drive the system to the optimal periodic reference $r$. 
The proof of this result is analogous to Theorem~\ref{thm:MPC_eco_periodic}~\cite[Rem.~5.8]{zanon2017periodic}. 
Assumption~\ref{asm:terminal_eco_periodic} can be satisfied with a simple terminal equality constraint. 

\subsubsection{Shifted terminal cost}
\label{sec:eco_shift_terminal}
The terminal cost decrease~\eqref{eq:eco_terminal_periodic} is w.r.t. the cost $\ell_{\mathrm{e}}$ at a specific reference point $x_{\mathrm{r}}(t)$, while the performance guarantee~\eqref{eq:eco_performance_periodic} is w.r.t. $\bar{\ell}_{\mathrm{e}}$, i.e., the average performance over one period length. 
The following proposition shows how we can modify the characterization of the terminal cost $V_{\mathrm{f,e}}$ to directly consider $\bar{\ell}_{\mathrm{e}}$. 
\begin{proposition}
\label{prop:eco_periodic_terminal}
\cite[Lemma~5]{Koehler2020Economic}
Let Assumptions~\ref{asm:eco_continuity_compactness} and \ref{asm:terminal_eco_periodic} hold. 
For any $t\in\mathbb{I}_{\geq 0}$ and any $x\in\mathbb{X}_{\mathrm{f}}(t)$, the shifted terminal cost 
\begin{align*} 
\tilde{V}_{\mathrm{f,e}}(x,t):=V_{\mathrm{f,e}}(x,t)+\sum_{k=0}^{T-2}\dfrac{T-1-k}{T}\ell_{\mathrm{e}}(x_{\mathrm{r}}(t+k),u_{\mathrm{r}}(t+k),t+k)
\end{align*} 
satisfies
\begin{align}
\label{eq:eco_term_shifted_result}
&\tilde{V}_{\mathrm{f,e}}(f(x,k_{\mathrm{f}}(x,t)),t+1)-\tilde{V}_{\mathrm{f,e}}(x,t)+\ell_{\mathrm{e}}(x,k_{\mathrm{f}}(x,t),t)
\leq \bar{\ell}_{\mathrm{e}}. 
\end{align}
\end{proposition}
Inequality~\eqref{eq:eco_term_shifted_result} direct relates the decrease in the terminal cost with $\bar{\ell}_{\mathrm{e}}$, the desired average cost of the periodic reference $r$, which becomes crucial for the design later (Sec.~\ref{sec:economic_4}). 
The same design condition is required to show that closed-loop performance is better than performance at some general set~\citep{dong2018analysis}.

\subsection{Online changing costs using artificial setpoints} 
\label{sec:economic_3}
In the following, we consider an economic stage cost $\ell_{\mathrm{e}}(x,u,y_{\mathrm{e}})$ that depends on some time-varying external parameters $y_{\mathrm{e}}(t)\in Y\subseteq\mathbb{R}^p$ with $Y$ compact, e.g., reflecting changes in online prices or user preference. 
Online variations in $y_{\mathrm{e}}(t)$ result in an online change of the optimal setpoint $x_{\mathrm{r}}$~\ref{enum_motivation_2}. 
Ideally, the economic MPC should provide a performance that is is no worse than this new optimal setpoint.
\cite{muller2013economic,muller2014performance,fagiano2013generalized} and~\cite{ferramosca2014economic} propose economic MPC formulations using artificial setpoints (cf. Sec.~\ref{sec:artificial_1}) to address this issue. 
We first present the approach by~\cite{muller2013economic,muller2014performance} and then explain the relation and difference to the approaches by \cite{fagiano2013generalized} and \cite{ferramosca2014economic}. 
For a given state $x$, parameters $y_{\mathrm{e}}\in {Y}\subseteq\mathbb{R}^p$, a weight $\beta\geq 0$, and a later specified variable $\kappa\in\mathbb{R}$, the economic MPC formulation is given by
\begin{align}
\label{problem:eco_selfTuning} 
\min_{\mathbf{u}\in\mathbb{U}^N, r\in\mathbb{Z}_{\mathrm{r}}}&\sum_{k=0}^{N-1}\ell_{\mathrm{e}}(x_{\mathbf{u}}(k,x),\mathbf{u}_k,y_{\mathrm{e}})+V_{\mathrm{f,e}}(x_{\mathbf{u}}(N,x),r,y_{\mathrm{e}})\nonumber\\
&+\beta \ell_{\mathrm{e}}(x_{\mathrm{r}},u_{\mathrm{r}},y_{\mathrm{e}})\nonumber\\
\text{s.t. }&(x_{\mathbf{u}}(k,x),\mathbf{u}_k)\in\mathbb{Z},~ k\in\mathbb{I}_{[0,N-1]},~ x_{\mathbf{u}}(N,x)\in\mathbb{X}_{\mathrm{f}}(r),\\
& f(x_{\mathrm{r}},u_{\mathrm{r}})=x_{\mathrm{r}}, ~\ell_{\mathrm{e}}(x_{\mathrm{r}},u_{\mathrm{r}},y_{\mathrm{e}})\leq \kappa. \nonumber
\end{align}
A minimizer is denoted by $r^\star(x,y_{\mathrm{e}},\kappa,\beta)$, $\mathbf{u}^\star(x,y_{\mathrm{e}},\kappa,\beta)$ and the closed-loop system is given by 
\begin{align}
\label{eq:sys_closedloop_selfTuning}
x(t+1)=&f(x(t),\mathbf{u}^\star_{0}(x(t),y_{\mathrm{e}}(t)),\kappa(t),\beta(t)),\quad t\in\mathbb{I}_{\geq 0},\\
\kappa(t+1)=&\ell_{\mathrm{e}}(r^\star(x(t),y_{\mathrm{e}}(t),\kappa(t),\beta(t)),y_{\mathrm{e}}(t+1)),\nonumber 
\end{align}
with some later specified weight $\beta(t)$, and initialization $\beta(0)=\beta_0\geq 0$ and $\kappa(0)=\kappa_0=\infty$. 
Compared to the setpoint tracking MPC (Problem~\eqref{problem:setpointTracking}), Problem~\eqref{problem:eco_selfTuning} directly minimizes the economic cost $\ell_{\mathrm{e}}$ instead of utilizing a positive definite tracking stage cost $\ell$. 
The added constraint involving $\kappa$ ensures that the cost at the artificial reference $r$ does not deteriorate compared to the previous solution. 
The weighting $\beta$ is used to ensure convergence of the artificial reference to a (local) minimum (Sec.~\ref{sec:economic_3_2}/\ref{sec:economic_3_3}). 

\subsubsection{Relative performance guarantees}
\label{sec:economic_3_1}
The following conditions on the self-tuning weight $\beta$ are trivially satisfied with a constant weight $\beta\geq 0$.
\begin{assumption}\citep[Asm.~1]{muller2013economic}
\label{asm:sefltuning_weight}
$\beta(t)$ is non-negative, $\beta(t+1)-\beta(t)$ admits a uniform upper bound, and $\limsup_{t\rightarrow\infty}\beta(t+1)-\beta(t)\leq 0$.
\end{assumption}
\begin{theorem}
\label{thm:selfTuning_1}%
\citep[Thm.~1]{muller2013economic} 
Let Assumptions~\ref{asm:eco_continuity_compactness} and \ref{asm:sefltuning_weight} hold. 
Suppose that $V_{\mathrm{f,e}}$, $k_{\mathrm{f}}$, $\mathbb{X}_{\mathrm{f}}$ satisfy the conditions in Assumption~\ref{asm:terminal_eco} for any (constant) feasible setpoint $r\in\mathbb{Z}_{\mathrm{r}}\subseteq\mathbb{Z}$ and any (constant) $y_{\mathrm{e}}\in Y$. 
Suppose further that Problem~\eqref{problem:eco_selfTuning} is feasible for $x=x_0$. 
Then, for any time-varying parameters $y_{\mathrm{e}}(t)$, Problem~\eqref{problem:eco_selfTuning} is feasible for all $t\in\mathbb{I}_{\geq 0}$ and the constraints~\eqref{eq:constraints} are satisfied for the resulting closed-loop system~\eqref{eq:sys_closedloop_selfTuning}. 
Furthermore, if $y_{\mathrm{e}}$ is constant, then the following performance bound holds
\begin{align}
\label{eq:eco_performance_selfTuning_1}
\limsup_{K\rightarrow\infty}\dfrac{1}{K}\sum_{t=0}^{K-1}\ell_{\mathrm{e}}(x(t),u(t),y_{\mathrm{e}})\leq \kappa_\infty:=\lim_{t\rightarrow\infty}\kappa(t). 
\end{align}
\end{theorem} 
Theorem~\ref{thm:selfTuning_1} ensures that the closed-loop performance is no worse than the performance at the artificial reference, which is given by $\kappa$. 
This result does \textit{not} imply that the artificial reference $r$ converges to some (local) optimum. 
In Theorem~\ref{thm:limon_2} we address a similar issue for a setpoint tracking MPC scheme, where we use the fact that the system converges close to the artificial reference and hence the artificial reference can be incrementally moved w.r.t. the candidate solution (see also Lemma~\ref{lemma:setpoint_bound}). 
However, the closed-loop system with the economic MPC will in general not converge close to the artificial setpoint $r$. 

\subsubsection{Self-tuning weight and local optimality}
\label{sec:economic_3_2}
\cite{muller2014performance} propose the following solution to this problem: 
The terminal set constraint $\mathbb{X}_{\mathrm{f}}$ should have a non-empty interior and be exponentially contractive, e.g., using a terminal set $\mathbb{X}_{\mathrm{f}}$ designed according to Section~\ref{sec:terminal_2}. 
This ensures that at any point in time, the artificial reference $r$ can be incrementally moved~\citep[Lemma~1]{muller2014performance}. 
Then, by suitably increasing the weighting $\beta(t)$ online if needed, \citet[Thm. ~3]{muller2014performance} ensure that the closed-loop performance is no worse than operating at a locally optimal steady state.\footnote{%
\cite{muller2013economic} provide explicit update rules for this \textit{self-tuning weight} $\beta$, which require that $\beta$ increases unbounded if $r$ is not a local minimizer.}
 
\subsubsection{Constant weight $\beta$ and suboptimality bound}
\label{sec:economic_3_3}
The above discussed work by~\cite{muller2013economic,muller2014performance} is inspired by \cite{fagiano2013generalized}. 
Instead of ensuring convergence of the artificial reference $r$ by increasing the weight $\beta$, \citet[Prop.~2]{fagiano2013generalized}
use a large constant weight $\beta$ and derive a suboptimality estimate $\epsilon(\beta)$ w.r.t. a local minimizer. 
Furthermore, \citet[Alg.~3]{fagiano2013generalized} use simple terminal equality constraint in combination with a multi-step implementation.\footnote{%
This analysis uses condition~\citep[Asm.~7]{fagiano2013generalized}, which is difficult to verify a priori. 
\citet[Lemma~4]{Koehler2020Economic} show that instead it suffices to assume local controllability, a prediction horizon $N$ longer than the controllability index $\nu$, and a multi-step implementation, cf. also the discussion regarding terminal equality constraints in Section~\ref{sec:terminal_TEC}.}

\subsubsection{Linear systems and strong duality}
\cite{ferramosca2014economic} address the problem in the special case of linear dynamics using a strong duality assumption. 
In particular, a shifted economic stage cost is minimized and a sufficiently large positive definite offset cost $V_{\mathrm{o}}$ (Sec.~\ref{sec:artificial_1}) is utilized to ensure asymptotic stability of the optimal steady state. 
However, these techniques cannot be transferred to the nonlinear setting and the minimized cost does in general not have an intuitive "economic" interpretation.

\subsection{Periodic optimal operation using artificial references}
\label{sec:economic_4}
In the following, we combine the designs and problem setups from Sections~\ref{sec:economic_2} and \ref{sec:economic_3}, following the work by~\cite{Koehler2020Economic}. 
Specifically, we consider a stage cost $\ell_{\mathrm{e}}(x,u,y_{\mathrm{e}})$, which depends on some parameters $y_{\mathrm{e}}(t)\in Y$. 
If $y_{\mathrm{e}}$ is $T$-periodic, then we wish to provide performance guarantees w.r.t. the optimal $T$-periodic operation. 
However, the parameters $y_{\mathrm{e}}$ may also be subject to additional unpredictable fluctuations and we need to guarantee reliable operation during such transient phases. Such problems naturally arise, e.g., in water distribution networks, electrical networks or HVAC systems, where the optimal mode of operation depends largely on external variables (e.g., the weather). These variables have a strong periodic component~\ref{enum_motivation_1} (e.g., due to the day--night cycle) but are also subject to additional variations that are difficult to predict~\ref{enum_motivation_2}.

\subsubsection{Economic MPC using periodic artificial references}
A natural solution to extend the economic MPC formulation in Section~\ref{sec:economic_3} to periodic problems is to replace the artificial setpoint $r\in\mathbb{Z}_{\mathrm{r}}$ by a periodic reference trajectory $\mathbf{r}\in\mathcal{R}_T\subseteq\mathbb{Z}_{\mathrm{r}}^T$ and consider a periodic prediction $\mathbf{y}_{\mathrm{e}}\in Y^T$ for the parameters with $\mathbf{y}_{\mathrm{e},0}(t)=y_{\mathrm{e}}(t)$. 
For a given state $x$, time $t$, parameters $\mathbf{y}_{\mathrm{e}}\in Y^T$, weight $\beta\geq 0$, and variable $\kappa\in \mathbb{R}$, an economic MPC formulation with periodic artificial references is given by:
\begin{align}
\label{problem:eco_selfTuning_periodic}
\min_{\mathbf{u}\in\mathbb{U}^N, \mathbf{r}\in\mathbb{Z}_{\mathrm{r}}^T}&\sum_{k=0}^{N-1}\ell_{\mathrm{e}}(x_{\mathbf{u}}(k,x),\mathbf{u}_k,\mathbf{y}_{\mathrm{e},k})+V_{\mathrm{f,e}}(x_{\mathbf{u}}(N,x),\mathbf{r},\mathbf{y}_{\mathrm{e}})\nonumber\\
&+\beta \sum_{k=0}^{T-1}\ell_{\mathrm{e}}(\mathbf{r}_k,\mathbf{y}_{\mathrm{e},k})\\
\text{s.t. }&(x_{\mathbf{u}}(k,x),\mathbf{u}_k)\in\mathbb{Z},~ k\in\mathbb{I}_{[0,N-1]},~ x_{\mathbf{u}}(N,x)\in\mathbb{X}_{\mathrm{f}}(\mathbf{r}),\nonumber\\
& \mathbf{r}\in\mathcal{R}_k,~\sum_{k=0}^{T-1}\ell_{\mathrm{e}}(\mathbf{r}_k,\mathbf{y}_{\mathrm{e},k})\leq \kappa,\nonumber
\end{align}
where we define $\mathbf{r}_{k}:=\mathbf{r}_{\mathrm{mod}(k,T)}$, $\mathbf{y}_{\mathrm{e},k}:=\mathbf{y}_{\mathrm{e},\mathrm{mod}(k,T)}$ for $k\in\mathbb{I}_{\geq T}$. 
A minimizer is denoted by $\mathbf{r}^\star(x,\mathbf{y}_{\mathrm{e}},\kappa,\beta)$, $\mathbf{u}^\star(x,\mathbf{y}_{\mathrm{e}},\kappa,\beta)$ and the closed-loop system is given by
\begin{align}
\label{eq:sys_closedloop_selfTuning_periodic}
x(t+1)=&f(x(t),\mathbf{u}^\star_{0}(x(t),\mathbf{y}_{\mathrm{e}}(t),\kappa(t),\beta(t))),\quad t\in\mathbb{I}_{\geq 0},\\
\kappa(t+1)=&\sum_{k=0}^{T-1}\ell_{\mathrm{e}}(\mathbf{r}_{k+1}^\star(x(t),\mathbf{y}_{\mathrm{e}}(t),\kappa(t),\beta(t)),\mathbf{y}_{\mathrm{e},k}(t+1)),\nonumber
\end{align}
with initialization $\kappa(0)=\kappa_0=\infty$ and a self-tuning weight $\beta(t)\geq 0$. 
This is a direct extension of the approach in Section~\ref{sec:economic_3}, replacing the artificial setpoint $r\in\mathbb{Z}_{\mathrm{r}}$ by an artificial periodic reference $\mathbf{r}\in\mathbb{Z}_{\mathrm{r}}^T$. 

\subsubsection{Pitfalls - periodic artificial references}
Next, we provide a negative result: A direct/na{\"i}ve extension of the design in Section~\ref{sec:economic_3} to the periodic setting does not provide the desired performance guarantees. 
We consider the following standard conditions for the terminal ingredients.
\begin{assumption}
\label{asm:eco_artificial_periodic}
There exists a control law $k_{\mathrm{f}}: X\times\mathcal{R}_T$, such that for any periodic reference $\mathbf{r}\in\mathcal{R}_T$ and any state $x\in\mathbb{X}_{\mathrm{f}}(\mathbf{r})$ and any $y\in Y^T$:
\begin{enumerate}[label=(E.\arabic*)]
\item Constraint satisfaction: $(x,k_{\mathrm{f}}(x,\mathbf{r}))\in\mathbb{Z}$;
\label{enum_eco_art_per_1}
\item Positive invariance: $f(x,k_{\mathrm{f}}(x,\mathbf{r}))\in\mathbb{X}_{\mathrm{f}}(\tilde{\mathbf{r}})$;
\label{enum_eco_art_per_2}
\item Terminal cost: 
$V_{\mathrm{f,e}}(f(x,k_{\mathrm{f}}(x,\mathbf{r})),\tilde{\mathbf{r}},\tilde{\mathbf{y}}_{\mathrm{e}})-V_{\mathrm{f,e}}(x,\mathbf{r},\mathbf{y}_{\mathrm{e}})\leq -\ell_{\mathrm{e}}(x,k_{\mathrm{f}}(x,\mathbf{r}),\mathbf{y}_{\mathrm{e},N})+\ell_{\mathrm{e}}(\mathbf{r}_N,\mathbf{y}_{\mathrm{e},N})$,
\label{enum_eco_art_per_3}
\end{enumerate}
with the periodically shifted sequences $\tilde{\mathbf{r}}_k=\mathbf{r}_{k+1}$, $\tilde{\mathbf{y}}_{\mathrm{e},k}=\mathbf{y}_{\mathrm{e},k+1}$. 
\end{assumption}
For a given periodic trajectory $\mathbf{r}$ and parameters $\mathbf{y}_{\mathrm{e}}$, Assumption~\ref{asm:eco_artificial_periodic} corresponds to Assumption~\ref{asm:terminal_eco_periodic} and is trivially satisfied by a terminal equality constraint $\mathbb{X}_{\mathrm{f}}(\mathbf{r})=\{\mathbf{x}_{\mathrm{r},N}\}$, $V_{\mathrm{f,e}}=0$. 
\begin{theorem}
\label{thm:eco_artificial_periodic_rec}
\citep{Koehler2020Economic}
Let Assumptions~\ref{asm:eco_continuity_compactness}, \ref{asm:sefltuning_weight}, and \ref{asm:eco_artificial_periodic} hold. 
Suppose Problem~\eqref{problem:eco_selfTuning_periodic} is feasible for $x=x_0$. 
Then, for any parameters $\mathbf{y}_{\mathrm{e}}(t)$, Problem~\eqref{problem:eco_selfTuning_periodic} is feasible for all $t\in\mathbb{I}_{\geq 0}$ and the constraints~\eqref{eq:constraints} are satisfied for the resulting closed-loop system~\eqref{eq:sys_closedloop_selfTuning_periodic}. 
Furthermore, if $\mathbf{y}_{\mathrm{e}}$ is consistently periodic, i.e., $y_{\mathrm{e}}(t+k)=\mathbf{y}_{\mathrm{e},k}(t)$, $t,k\in\mathbb{I}_{\geq 0}$, then the following performance bound holds
\begin{align}
\label{eq:eco_periodic_bad_bound}
&\limsup_{K\rightarrow\infty}\dfrac{1}{K}\sum_{t=0}^{K-1}\ell_{\mathrm{e}}(x(t),u(t),y_{\mathrm{e}}(t))\\
\leq& \limsup_{t\rightarrow\infty}\dfrac{1}{T}\sum_{k=t}^{t+T-1}\ell_{\mathrm{e}}(\mathbf{r}_{0}^\star(x(k),\mathbf{y}_{\mathrm{e}}(k),\kappa(k),\beta(k)),\mathbf{y}_{\mathrm{e},0}(k)).\nonumber
\end{align}
\end{theorem}
Theorem~\ref{thm:eco_artificial_periodic_rec} ensures recursive feasibility and constraint satisfaction independent of the parameters $\mathbf{y}_{\mathrm{e}}$. 
The performance bound~\eqref{eq:eco_periodic_bad_bound} follows from Condition~\ref{enum_eco_art_per_3} of the terminal cost $V_{\mathrm{f,e}}$. 
However, this bound is not w.r.t. the average cost of the periodic reference $\mathbf{r}^\star$, but only w.r.t. $\mathbf{r}_0^\star$, the first point of the periodic reference. 
\citet[Sec.~II.C]{Koehler2020Economic} provide a simple example (inspired by the counterexample by~\cite{muller2016economic}) that demonstrates that the closed-loop performance can be arbitrarily bad compared to the periodic reference $\mathbf{r}^\star$. 
The problem is that the sequence of optimized trajectories $\mathbf{r}^\star_0(t)$, $t\in\mathbb{I}_{\geq 0}$ can differ completely from the periodic artificial reference $\mathbf{r}^\star\in\mathbb{Z}_{\mathrm{r}}^T$, even if the average cost of the periodic trajectories coincide. 

\subsubsection{Performance guarantees} 
\cite{Koehler2020Economic} provided a simple solution to this problem. 
Using Proposition~\ref{prop:eco_periodic_terminal}, we construct a shifted terminal cost
\begin{align*} 
\tilde{V}_{\mathrm{f,e}}(x,\mathbf{r},\mathbf{y}_{\mathrm{e}}):=V_{\mathrm{f,e}}(x,\mathbf{r},\mathbf{y}_{\mathrm{e}})+\sum_{k=0}^{T-2}\dfrac{T-1-k}{T}\ell_{\mathrm{e}}(\mathbf{r}_{k+N},\mathbf{y}_{\mathrm{e},k+N})
\end{align*}
which satisfies 
\begin{align*}
&\tilde{V}_{\mathrm{f,e}}(f(x,k_{\mathrm{f}}(x,\mathbf{r})),\tilde{\mathbf{r}},\tilde{\mathbf{y}}_{\mathrm{e}})-\tilde{V}_{\mathrm{f,e}}(x,\mathbf{r},\mathbf{y}_{\mathrm{e}})\\
\leq& -\ell_{\mathrm{e}}(x,k_{\mathrm{f}}(x,\mathbf{r}),\mathbf{y}_{\mathrm{e},N})+ \dfrac{1}{T}\sum_{k=0}^{T-1}\ell_{\mathrm{e}}(\mathbf{r}_k,\mathbf{y}_{\mathrm{e},k}).
\end{align*}
This shifted terminal cost directly provides a decrease condition w.r.t. the average performance at the periodic reference. 
The following theorem demonstrates that this yields the desired performance guarantees. 
\begin{theorem}
\label{thm:eco_periodic_artificial}
\citep[Prop.~3, Prop.~6, Cor.~1]{Koehler2020Economic}
Suppose the conditions in Theorem~\ref{thm:eco_artificial_periodic_rec} hold and we replace the terminal cost $V_{\mathrm{f,e}}$ in Problem~\eqref{problem:eco_selfTuning_periodic} by $\tilde{V}_{\mathrm{f,e}}$. 
Then, the closed-loop system~\eqref{eq:sys_closedloop_selfTuning_periodic} satisfies the following performance bound
\begin{align}
\label{eq:eco_performance_selfTuning_periodic}
\limsup_{K\rightarrow\infty}\dfrac{1}{K}\sum_{t=0}^{K-1}\ell_{\mathrm{e}}(x(t),u(t),y_{\mathrm{e}}(t))\leq \lim_{t\rightarrow\infty}\dfrac{\kappa(t)}{T}. 
\end{align} 
Furthermore, if $\beta(t)$ is updated using~\citep[Update scheme 2 or 6]{muller2013economic} and the terminal set $\mathbb{X}_{\mathrm{f}}$ is contractive~\citep[Asm.~3]{Koehler2020Economic}, then the closed-loop performance is on average no worse than the performance at a locally optimal periodic orbit. 
\end{theorem}
Theorem~\ref{thm:eco_periodic_artificial} shows that we obtain the desired performance guarantees, outperforming standard stabilization of some optimal periodic trajectory. 
The design of this shifted terminal cost $\tilde{V}_{\mathrm{f,e}}$ is motivated by the characterization of periodic optimality by~\citet[Prop.~1]{koehler2018periodic}, which is inspired by non-monotonic Lyapunov functions~\citep{ahmadi2008non}. 
Unfortunately, the shifted terminal cost $\tilde{V}_{\mathrm{f,e}}$ does not admit an intuitive interpretation. 
\citet[Prop.~2]{Koehler2020Economic} also report an alternative design ensuring the performance bound~\eqref{eq:eco_performance_selfTuning_periodic} by constraining $\ell_{\mathrm{e}}(\mathbf{r}_k,\mathbf{y}_{\mathrm{e},k})$, $k\in\mathbb{I}_{[0,T-1]}$. 
However, this requires an additional continuity condition to ensure local optimality~\citep[Asm.~5]{Koehler2020Economic}.

\subsubsection{Periodicity constrained economic MPC}
\label{sec:economic_periodicity}
\cite{houska2017cost,wang2018economic} develop a \textit{periodicity constrained} economic MPC scheme that intrinsically provides closed-loop performance bounds w.r.t. the optimized periodic reference. 
Instead of separately optimizing a predicted input sequence $\mathbf{u}\in\mathbb{U}^N$ and a periodic trajectory $\mathbf{r}\in\mathcal{R}_T$, this approach only optimizes over periodic trajectories that starts at the current state $x$: 
\begin{align}
\label{problem:periodicity_constrained}
\min_{\mathbf{r}\in\mathcal{R}_T}\sum_{k=0}^{T-1}\ell_{\mathrm{e}}(\mathbf{r}_k,\mathbf{y}_{\mathrm{e},k})~\text{s.t. } \mathbf{x}_{\mathrm{r},0}=x.
\end{align}
Problem~\eqref{problem:periodicity_constrained} can be recovered as a special case from Problem~\eqref{problem:eco_selfTuning_periodic} by setting the prediction horizon $N=0$ and using a simple terminal equality constraint $\mathbb{X}_{\mathrm{f}}=\{\mathbf{x}_{\mathrm{r},0}\}$, $V_{\mathrm{f,e}}=0$. 
This special case does not satisfy the technical conditions we posed on the terminal set $\mathbb{X}_{\mathrm{f}}$~\citep[Asm.~3]{Koehler2020Economic}. 
Hence, Theorem~\ref{thm:eco_periodic_artificial}, which shows convergence to a locally optimal reference, is not applicable. 
In fact, convergence to a locally optimal periodic orbit can only be established under rather restrictive conditions, cf.~\citet[Example~6, Thm.~1]{wang2018economic} and \citet[Lemma~3, Thm.~4]{houska2017cost}. 
\citet[Sec.~V.B, App.~A]{Koehler2020Economic} also provide a numerical comparison, demonstrating that the lack of a prediction horizon $N$ limits the performance of this approach.

\subsection{Stability and convergence in economic MPC}
\label{sec:economic_5}
While Sections~\ref{sec:economic_1}--\ref{sec:economic_4} focus on performance guarantees, the question of stability and convergence can also plays an important role when studying economic MPC schemes. 

\subsubsection{Dissipativity, optimality, stability}
\label{sec:economic_5_dissip}
Dissipativity plays a crucial role to study stability in economic MPC~\citep{faulwasser2018economic} and optimal control~\citep{grune2022dissipativity}. 
Neglecting some technical conditions, one can show equivalence between the following properties:
\begin{itemize}
\item the system is optimally operated at the steady state $x_{\mathrm{r}}$ (in a strict sense);
\item (strict) dissipativity w.r.t. the supply rate $s(x,u)=\ell(x,u)-\ell(x_{\mathrm{r}},u_{\mathrm{r}})$, i.e., 
$\exists \lambda:X\rightarrow\mathbb{R}_{\geq 0}$, s.t. $\lambda(f(x,u))-\lambda(x)\leq +s(x,u)-\alpha_{\ell}(\|x-x_{\mathrm{r}}\|)$, $\alpha_\ell\in\mathcal{K}_\infty$;
\item the closed-loop system~\eqref{eq:sys_closedloop_eco} (asymptotically) stabilizes $x_{\mathrm{r}}$;
\item the turnpike property holds, i.e., optimal solutions stay close to $x_{\mathrm{r}}$ most of the time;
\end{itemize}
see~\citet{angeli2012average,grune2016relation,muller2014necessity,faulwasser2017turnpike,faulwasser2018economic,muller2021dissipativity} for details. 
Loosely speaking, these results imply that the economic MPC scheme "does the right thing": If operation at the steady state is optimal, then the closed-loop will converge to this steady state. 
If the closed-loop system does not stabilize the steady state, then superior performance is achieved using some dynamic operation. 
Given strict dissipativity, transient performance bounds of economic MPC compared to the infinite horizon optimal performance can also be established~\citep{grune2015non}. 
Generalizations of this dissipativity concept are available for periodic orbits~\citep{zanon2017periodic,koehler2018periodic}, general sets~\citep{dong2018analysis,martin2019dissipativity}, and the constructive verification is studied by~\cite{pirkelmann2019approximate,berberich2020dissipativity}, cf.~\cite{muller2021dissipativity} and~\cite{grune2022dissipativity} for recent overviews. 
While dissipativity is not used for the designs in this section, the analysis in Section~\ref{sec:UCON} needs this property.

\subsubsection{Enforcing convergence/stability in economic MPC}
In various applications, we might wish to constructively enforce convergence/stability to some mode of operation, irrespective of its optimality in terms of the stage cost $\ell_{\mathrm{e}}$. 
One can simply use a tracking MPC formulation (Sec.~\ref{sec:artificial}) to stabilize the optimal steady state or periodic orbit~\citep{limon2014single}. 
Furthermore, it is possible to choose a tracking stage cost $\ell$, which locally approximates an economic MPC scheme~\citep{de2020tunempc}. 
In the special case of linear systems and (strictly) convex cost $\ell_{\mathrm{e}}$, an economic MPC behaves like a stabilizing MPC due to duality/dissipativity, and thus stability can be naturally established~\citep{ferramosca2014economic,broomhead2015robust}. 
\cite{alamir2020new} ensure asymptotic stability of the optimal steady state by adding a large enough penalty on non-stationary operation. 
\cite{gutekunst2020economic} achieve convergence to the optimal periodic trajectory by penalizing the non-periodicity in the economic stage cost $\ell_{\mathrm{e}}$. 
Convergence/stability can also be imposed with
Lyapunov-based constraints~\citep{heidarinejad2012economic}, \citep[Sec.~4]{ellis2017economic}, 
average constraints~\citep{muller2014transient,muller2014convergence,rosenfelder2020stability}, or by considering a multi-objective formulation~\citep{he2015stability}, \citep[Sec.~9]{faulwasser2018economic}, \citep{soloperto2020augmenting,eichfelder2022relaxed}.

\subsection{Design of terminal ingredients in economic MPC}
\label{sec:economic_6}
In the following, we discuss how to modify the design of the terminal ingredients from Section~\ref{sec:terminal} to satisfy the conditions in economic MPC (Asm.~\ref{asm:terminal_eco}, \ref{asm:terminal_eco_periodic}, \ref{asm:eco_artificial_periodic}). 
A recent overview for these design methods can be found in~\citet[Sec.~IV.A]{Koehler2020Economic}.

The conditions on the terminal set $\mathbb{X}_{\mathrm{f}}$, i.e., positive invariance and constraint satisfaction, are equivalent to the conditions needed in the tracking MPC schemes (Sec.~\ref{sec:terminal}--\ref{sec:artificial}) and hence the terminal set can be constructed using the design procedures in Section~\ref{sec:terminal}. 
Recall that the economic MPC schemes with artificial references (Sec.~\ref{sec:economic_3}--\ref{sec:economic_4}) require a contractive terminal set, while a simple terminal equality constraint requires modifications to ensure convergence to a local minimum. 

\subsubsection{Terminal cost - standard design}
The conditions on the terminal cost $V_{\mathrm{f,e}}$ involve the economic stage cost $\ell_{\mathrm{e}}$, which is in general neither quadratic nor positive definite. 
This necessitates modifications to the design procedures in Section~\ref{sec:terminal}. 
Let us first consider a fixed optimal steady state $x_{\mathrm{r}}$ (Sec.~\ref{sec:economic_1}). 
\cite{amrit2011economic} propose a simple design to compute an economic terminal cost of the form $V_{\mathrm{f,e}}(x)=\|x\|_P^2+p^\top x$ that satisfies~\eqref{eq:eco_terminal}. 
Specifically, one can derive a local linear-quadratic upper bound on the stage cost $\ell_{\mathrm{e}}$ using the Jacobian and Hessian of $\ell_{\mathrm{e}}$~\citep[Lemma~23]{amrit2011economic}. 
Then, the matrix $P$ is designed using a Lyapunov equation comparable to $\mathrm{LQR}\succeq 0$~\eqref{eq:LQR_matrix}, where the role of $Q,R$ is replaced by a positive semidefinite matrix that upper bounds the Hessian of $\ell_{\mathrm{e}}$. 
The vector $p\in\mathbb{R}^n$ is chosen such that it exactly compensates the Jacobian of the stage cost $\ell_{\mathrm{e}}$. 
Then, analogous to Proposition~\ref{prop:terminal_LQR}, Inequality~\eqref{eq:eco_terminal} holds in a sufficiently small terminal set~\citep[Lemma~24]{amrit2011economic}. 
Notably, this terminal cost $V_{\mathrm{f,e}}$ is \textit{not} positive definite w.r.t. $x_{\mathrm{r}}$. 
This captures the fact that, for some initial conditions in the terminal set, the transient cost is strictly smaller than the cost at the optimal steady state. 
Due to this, the terminal set $\mathbb{X}_{\mathrm{f}}$ cannot be chosen as a sublevel set of the economic terminal cost $V_{\mathrm{f,e}}$ (as done in tracking MPC, Sec.~\ref{sec:artificial_alpha}). 
Note that there is no degree of freedom in the choice of $p$ as it needs to \textit{exactly} cancel the Jacobian of $\ell_{\mathrm{e}}$. 

\subsubsection{Extensions}
Next, we discuss how to extend this design to periodic problems (Sec.~\ref{sec:economic_2}), artificial setpoints (Sec.~\ref{sec:economic_3}), and artificial periodic reference trajectories (Sec.~\ref{sec:economic_4}) using the design by~\citet[Sec.~IV.A]{Koehler2020Economic}.

For periodic reference trajectories (Sec.~\ref{sec:economic_2}), we consider $V_{\mathrm{f,e}}(x,t)=\|x-x_{\mathrm{r}}(t)\|_{P(t)}^2+(x-x_{\mathrm{r}}(t))^\top p(t)$. 
Time-varying matrices $P(t)$ can be computed using a time-varying (periodic) LQR, similar to~\citep{aydiner2016periodic}. 
\citet[Prop.~4]{Koehler2020Economic} provide a formula to compute $p(t)$ by solving a linear system of equations that includes the Jacobian of the dynamics $f$ and the stage cost $\ell_{\mathrm{e}}$ along the periodic reference $r(t)$. 

To account for artificial setpoints (Sec.~\ref{sec:economic_3}), the main challenge is related to the linear correction factor $p$. 
Specifically, while the approach from Section~\ref{sec:terminal_2} can be used to compute a parametrized matrix $P(r)$, this is in general not possible for $p(r)$. 
Instead, we include the vector $p\in X$ as a decision variable in the MPC that needs to satisfy an equality constraint involving the Jacobian of the dynamics $f$ and the stage cost $\ell_{\mathrm{e}}$, cf.~\citet[Rem.~7]{muller2014performance}.

For periodic artificial reference trajectories (Sec.~\ref{sec:economic_4}), we simply combine the previous two approaches. 
A parametrized matrix $P(r)$ is computed offline following the design in Section~\ref{sec:terminal_4}. 
A periodic correction term $\mathbf{p}\in X^T$ is included in the online optimization with suitable constraints involving the Jacobian of $f$ and $\ell_{\mathrm{e}}$, cf.~\citet[Sec.~IV.A]{Koehler2020Economic}.

\subsubsection{Positive definite terminal cost}
The computation of the correction factor $p$ can become complex, e.g., when considering artificial references. 
\cite{alessandretti2016convergence} propose a positive definite terminal cost $V_{\mathrm{f,e}}$ by suitably scaling a tracking terminal cost $V_{\mathrm{f}}$, which does not require this correction term $p$. 
The same approach can be applied to general artificial references using the parametrized terminal cost (Sec.~\ref{sec:terminal}) to provide a terminal cost for artificial references without the need to online compute the correction term $p$~\cite[Rem.~5]{Koehler2020Economic}.
However, this neglects the gradient information captured by $p$ and is hence more conservative, see also the comparison by~\citet[App.~A]{Koehler2020Economic}.

\subsection{Illustrative example}
\label{sec:eco_example}
\begin{figure}
\begin{center} 
\includegraphics[scale=0.5]{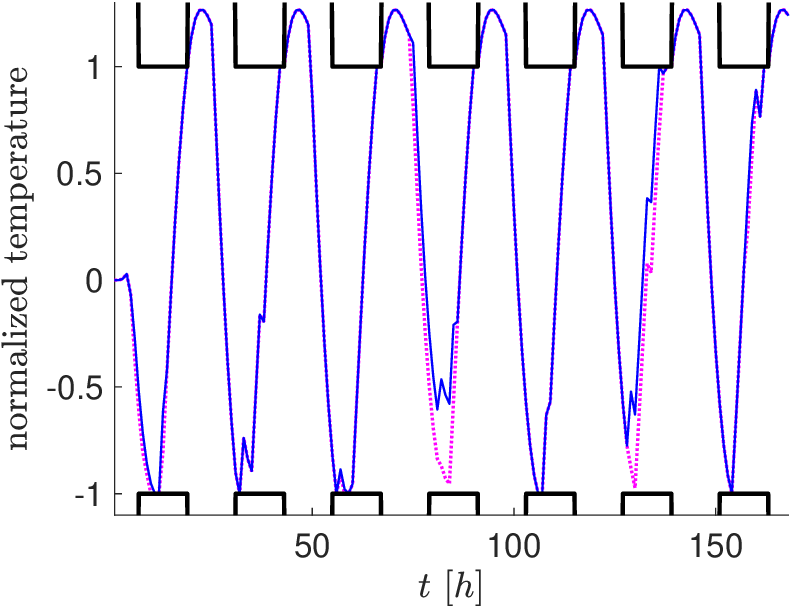} 
\end{center}
\caption{Temperature control in a building, adapted from~\cite{Koehler2020Economic}. Closed-loop state $x$ (blue, solid) resulting from periodic economic MPC (Thm.~\ref{thm:eco_periodic_artificial}) with time-varying ambient temperature and price signal, which also changes unpredictably during online operation. Time-varying state constraints (black, solid). 
Optimal trajectory computed in hindsight with known price signal $y_{\mathrm{e}}$ shown in magenta, dashed.}
\label{fig:eco_example}
\end{figure}

Figure~\ref{fig:eco_example} illustrates application of the economic MPC formulation with artificial periodic references (Thm.~\ref{thm:eco_periodic_artificial}) with a simple scalar system modelling temperature control in a building by~\cite{Koehler2020Economic}. 
The system is subject to time-varying temperature constraints and a disjoint input constraint modelling a discrete set of chillers. 
The economic objective $\ell_{\mathrm{e}}$ is minimizing energy cost. 
Ambient temperature and electricity cost are time-varying with period length $24$ hours~\ref{enum_motivation_1}. 
As the true external quantities are not (exactly) periodic, the closed-loop system needs to react to unexpected changes in the optimal mode of operation~\ref{enum_motivation_2}. 
This example highlights the performance of the economic MPC formulation, achieving close to optimal performance despite unpredictable online changes ~\ref{enum_motivation_2} in external time-varying quantities~\ref{enum_motivation_1}. 

\subsection{Open issues}
\label{sec:eco_openIssue}
Performance guarantees beyond periodic optimal operation remain largely an open issue, see, e.g., \cite{dong2018analysis,martin2019dissipativity} and \cite{grune2020economic} for some recent results. 
The utilization of artificial references inherits the limitations of tracking MPC formulations with artificial references (Sec.~\ref{sec:artificial_openIssue}) and hence may also benefit from the extensions in Section~\ref{sec:artificial_4} and Section~\ref{sec:discussion_time}. 
Transient performance/regret bounds relative to the optimal performance are a practically relevant and theoretically challenging issue, especially considering 
unpredictable online changes in the cost~\ref{enum_motivation_2} or lack of knowledge of the optimal mode of operation~\ref{enum_motivation_3}.
   
\section{MPC without terminal constraints/cost}
\label{sec:UCON}
In this section, we analyse simpler MPC formulations, which do not rely on a terminal cost $V_{\mathrm{f}}$ or a terminal set $\mathbb{X}_{\mathrm{f}}$ satisfying Assumption~\ref{asm:prelim_terminal}. 
First, we provide some motivation and historical context regarding the analysis of such MPC schemes (Sec.~\ref{sec:UCON_1}). 
Then, we derive qualitative conditions in terms of the stage cost $\ell$ and intrinsic system properties, such that a sufficiently long prediction horizon $N$ ensures desired closed-loop properties (Sec.~\ref{sec:UCON_2}). 
Furthermore, we provide quantitative bounds for a sufficiently long horizon $N$ and discuss how to use these analysis results constructively in the design (Sec.~\ref{sec:UCON_3}). 
Finally, we provide a numerical example (Sec.~\ref{sec:UCON_example}) and mention open issues (Sec.~\ref{sec:UCON_openIssue}).

\subsection{Motivation and historical context}
\label{sec:UCON_1}
In the following, we study an MPC formulation given by
\begin{align}
\label{problem:UCON}
\mathcal{J}_N^\star(x):=&\min_{\mathbf{u}\in\mathbb{U}^N}\sum_{k=0}^{N-1}\ell(x_{\mathbf{u}}(k,x),\mathbf{u}_k)\\
\text{s.t. }&(x_{\mathbf{u}}(k,x),\mathbf{u}_k)\in\mathbb{Z},~ k\in\mathbb{I}_{[0,N-1]},\nonumber
\end{align}
with a minimizer $\mathbf{u}^\star(x)$ and the closed-loop system 
\begin{align}
\label{eq:sys_closedloop_UCON}
x(t+1)=f(x(t),\mathbf{u}^\star_{0}(x(t))),\quad t\in\mathbb{I}_{\geq 0}. 
\end{align}
Compared to the standard formulation in Problem~\eqref{problem:prelim}, we simply remove the terminal cost $V_{\mathrm{f}}$ and the terminal set constraint $\mathbb{X}_{\mathrm{f}}$ in Problem~\eqref{problem:UCON}.

The design and analysis of the MPC formulations in Sections~\ref{sec:prelim}--\ref{sec:economic} revolve around the terminal ingredients. 
However, the design of the terminal ingredients (Sec.~\ref{sec:terminal}) can be complicated and application to large scale systems brings additional challenges (cf. Sec.~\ref{sec:discussion_distributed}). 
This design also requires stabilizability of the linearization, which is not given for non-holonomic systems like a car~\citep{worthmann2015model}. 
Furthermore, the terminal set constraint $\mathbb{X}_{\mathrm{f}}$ can lead to feasibility issues, e.g., if the design is done incorrectly or the model mismatch is neglected. 
In contrast, Problem~\eqref{problem:UCON} is strikingly simple from an application perspective. 
Due to these reasons, many MPC implementations do not utilize properly designed terminal ingredients, cf. Section~\ref{sec:discussion_UCON} for a balanced discussion regarding drawbacks and merits of terminal ingredients in MPC.

\subsubsection{What can go wrong?}
For $N\rightarrow\infty$, Problem~\eqref{problem:UCON} results in the infinite horizon optimal controller, which is known to be stabilizing under standard conditions. 
Hence, one may expect that the stability properties of the closed-loop system~\eqref{eq:sys_closedloop_UCON} improve if we increase the prediction horizon $N$. 
However, stability is in general \textit{not} monotone w.r.t. the horizon $N$, i.e., the closed-loop system~\eqref{eq:sys_closedloop_UCON} may become unstable if the prediction horizon $N$ increases, see, e.g., the numerical examples by~\citet[Sec.~4.2]{primbs1999nonlinear} and \citet[Sec.~VII.A]{kohler2023stability}. 
Instability due to a finite horizon $N$ can even be observed for simple stable linear systems, e.g., water
tanks~\citep{raff2006nonlinear} or mass--spring--damper systems~\citep{kohler2023stability}. 
Hence, a theoretical analysis is required to a priori judge the stability properties of such MPC formulations.

\subsubsection{Stability conditions - historical developments}
Historically, MPC emerged from the process control industry without any system theoretic guarantees and terminal ingredients are (mostly) not used, cf.~\citet[Sec.~V5]{qin2003survey}, \citet[Sec.~3 (i)]{mayne2013apologia}. 
Such MPC formulations are also called \textit{unconstrained} MPC~\citep{grune2009analysis,grune2010analysis,reble2012unconstrained} due to the absence of a stabilizing terminal constraint (despite the presence of input or state constraints). 
Given that such MPC formulations are routinely applied, there is a significant interest in understanding the theoretical properties. 
Early works use regularity properties to ensure stability under some sufficiently long horizon $N$, cf., e.g., the works by \cite{alamir1995stability} and \cite{jadbabaie2005stability}. 
However, it is difficult to obtain quantitative bounds regarding the horizon $N$ using this analysis. 

Over the last two decades, there has been a tremendous advancement on this problem with a rich theory deriving quantitative bounds, see, e.g., the overview and the textbook by~\cite{grune2012nmpc,grune2017nonlinear}. 
Specifically, the works by \cite{grimm2005model,tuna2006shorter} and \cite{grune2008infinite,grune2009analysis,grune2010analysis} provide a foundation to determine stability properties and a sufficiently long prediction horizon $N$. 
This has lead to multiple subsequent advancements, addressing among others: 
\begin{itemize}
\item tight\footnote{%
Tight in the sense that no smaller bound can be derived given the posed assumptions and Lyapunov function.} estimates for the stabilizing horizon $N$~\citep{grune2009analysis,grune2010analysis,reble2012unconstrained,kohler2021dynamic,kohler2023stability};
\item characterizing the region of attraction~\citep{boccia2014stability,kohlernonlinear19,esterhuizen2020recursive,kohler2021stability,pan2021first,kohler2021constrained};
\item semidefinite cost~\citep{grimm2005model,kohler2021constrained,westenbroek2022computational,kohler2023stability};
\item economic cost~\citep{grune2013economic,grune2014asymptotic,faulwasser2015design,muller2016economic,kohlernonlinear19,grune2020economic,long2021unconstrained}; \item terminal weights~\citep{tuna2006shorter,grune2010analysis,reble2012improved,kohler2021stability,beckenbach2022approximate,moreno2022predictive,kohler2023stability};
\item continuous-time problems~\citep{reble2012unconstrained,reble2012improved,esterhuizen2020recursive,faulwasser2021predictive,pan2021first,long2021unconstrained,westenbroek2022computational};
\item non-holonomic systems~\citep{worthmann2015model,coron2020model,rosenfelder2022model}.
\end{itemize}
In the following, we summarize the main results in these directions. 

\subsection{Qualitative conditions}
\label{sec:UCON_2}
In the following, we discuss qualitative conditions on the system and the stage cost $\ell$, which ensure desired closed-loop properties with a sufficiently long prediction horizon $N$. 
We first focus on the standard problem of driving the system to the origin using a positive definite stage cost $\ell$ (Sec.~\ref{sec:UCON_2_reg}). Then, we extend the problem to a positive semidefinite input-output stage cost $\ell$ (Sec.~\ref{sec:UCON_2_detect}) and a singular output cost $\ell$ (Sec.~\ref{sec:UCON_2_output}). 
The generalizations to an economic stage cost $\ell_{\mathrm{e}}$ and dynamic problems are studied in Sections~\ref{sec:UCON_2_economic} and \ref{sec:UCON_2_dynamic}, respectively. 
This exposition considers global stability with pure input constraints, i.e., $\mathbb{Z}=X\times\mathbb{U}$. 
Section~\ref{sec:UCON_2_local} shows how these arguments naturally extend to state constraints with a suitable region of attraction.

\subsubsection{Cost controllability}
\label{sec:UCON_2_reg}
We first focus on the prototypical problem of stabilizing the origin with $0=f(0,0)$, $0\in\mathbb{U}$, and a positive definite stage cost $\ell$ (Asm.~\ref{asm:prelim_stabilizing}), with $r=0$ for simplicity.  
\begin{assumption}
\citep[Asm.~3.5]{grune2012nmpc}
\label{asm:CostControl} 
There exists a constant $\gamma\geq 1$, such that for any $x\in {X}$ and any horizon $N\in\mathbb{I}_{\geq 1}$, the value function satisfies
\begin{align}
\label{eq:exp_cost_control}
\mathcal{J}_N^\star(x)\leq \gamma\ell_{\min}(x).
\end{align}
\end{assumption}
A sufficient condition for~\eqref{eq:exp_cost_control} is \textit{exponential cost controllability}, which is characterized as follows: 
For any $x\in{X}$, there exists an input sequence $\mathbf{u}\in\mathbb{U}^\infty$ satisfying 
$\ell(x_{\mathbf{u}}(k,x),\mathbf{u}_k)\leq C\rho^k\ell_{\min}(x)$, $k\in\mathbb{I}_{\geq 0}$ with constants $C\geq 1$, $\rho\in[0,1)$.
If a system is exponentially stabilizable, then this condition holds with a quadratic stage cost $\ell(x,u)=\|x\|_Q^2+\|u\|_R^2$, $Q,R\succ 0$.
In general, Assumption~\ref{asm:CostControl} does not hold and there may not exist any finite horizon $N$ ensuring asymptotic stability when using a quadratic stage cost $\ell$ ~\citep{muller2017quadratic}. 
Systematic choices of $\ell$ can be derived using \textit{homogeneity}~\citep{coron2020model} and non-holonomic vehicles are explored by~\cite{worthmann2015model,rosenfelder2022model}. 
\begin{theorem}\citep[Thm.~4.5]{grune2008infinite}
\label{thm:UCON}
Let Assumptions~\ref{asm:continuity}, \ref{asm:prelim_stabilizing}, and \ref{asm:CostControl} hold. 
There exists a constant $\underline{N}\geq 0$, such that for all $N>\underline{N}$ and all $x_0\in{X}$, the closed-loop system~\eqref{eq:sys_closedloop_UCON} satisfies
\begin{align}
\label{eq:DP_relaxed}
\mathcal{J}_N^\star(x(t+1))-\mathcal{J}_N(x(t))\leq -\alpha_N\ell(x(t),u(t)),~t\in\mathbb{I}_{\geq 0},
\end{align}
with $\alpha_N\in(0,1]$, the origin is asymptotically stable with the Lyapunov function $\mathcal{J}_N^\star$, and the following performance bound holds
\begin{align}
\label{eq:UCON_performance}
\sum_{t=0}^{\infty}\ell(x(t),u(t))\leq \dfrac{1}{\alpha_N}\mathcal{J}_\infty^\star(x_0).
\end{align}
\end{theorem}
Theorem~\ref{thm:UCON} ensures stability and \eqref{eq:UCON_performance} provides a relative suboptimality/regret bound $1/\alpha_N\in[0,\infty)$ w.r.t. the infinite-horizon optimal solution $\mathcal{J}_\infty^\star$. 
Inequality~\eqref{eq:DP_relaxed} is also called a relaxed dynamic programming inequality. It holds that $\lim_{N\rightarrow\infty}\alpha_N=1$ and hence optimality is recovered in the limit.

To simplify the exposition, in the following we assume that the nonlinear system is exponentially stabilizable and hence a simple quadratic cost $\ell$ can be used. 

\subsubsection{Cost detectability}
\label{sec:UCON_2_detect}
In many applications, we encounter a cost of the form $\ell(x,u)=\|h(x)\|_Q^2+\|u\|_R^2$, where $y=h(x)$ is some lower dimensional output of the system (Sec.~\ref{sec:artificial_1}). 
This is particularly relevant if we identified an input-output model, e.g., an impulse response model, from data. 
Condition~\eqref{eq:exp_cost_control} is not applicable in this case and we require a \textit{detectability} condition. 
\begin{assumption}\citep[SA 3/4]{grimm2005model}
\label{asm:Costdetectability}
There exist constants $\gamma\geq 1$, $\gamma_{\mathrm{o}}\geq 0$, $\epsilon_{\mathrm{o}}>0$ and a storage function $W:X\rightarrow\mathbb{R}_{\geq 0}$, such that for any horizon $N\in\mathbb{I}_{\geq 1}$ and for any $x\in X$, $u\in \mathbb{U}$: 
\begin{enumerate}[label=\arabic*)]
\item Cost controllability: $\mathcal{J}_N^\star(x)\leq \gamma\|x\|^2$;
\label{enum_cost_control}
\item Cost detectability: $W(x)\leq \gamma_{\mathrm{o}}\|x\|^2$, \\
$W(f(x,u))-W(x)\leq -\epsilon_{\mathrm{o}}\|x\|^2+\ell(x,u)$.
\label{enum_cost_detect}
\end{enumerate}
\end{assumption}
Condition~\ref{enum_cost_control} holds if the system is exponentially stabilizable and $h$ is Lipschitz continuous~\citep[Prop.~2]{kohler2021constrained}. 
Condition~\ref{enum_cost_detect} ensures that $\lim_{t\rightarrow\infty}\ell(x(t),u(t))=0$ implies $\lim_{t\rightarrow\infty}\|x(t)\|=0$, i.e., the system is zero-state detectable with the output $y$.
This property is also called input-output-to-state stability (IOSS), $W(x)$ is an IOSS Lyapunov function, and for linear systems this reduces to the standard detectability condition~\citep{cai2008input}. 
In case of nonlinear autoregressive models (NARX), a simple quadratic IOSS Lyapunov function $W(x)$ can be constructed~\citep[Rem.~3]{kohler2023stability}. 
Assumption~\ref{asm:CostControl} is a special case of Assumption~\ref{asm:Costdetectability} considering $W(x)=0$. 
\begin{theorem}
\label{thm:UCON_detect}
(\citet[Thm.~1--2]{grimm2005model}, \citet[Thm.~1]{kohler2023stability})
Let Assumptions~\ref{asm:continuity} and \ref{asm:Costdetectability} hold. 
There exists a constant $\underline{N}\geq 0$, such that for all $N>\underline{N}$ and all $x_0\in{X}$, the origin is exponentially stable for the resulting closed-loop system~\eqref{eq:sys_closedloop_UCON} with the Lyapunov function $\mathcal{J}_N^\star(x)+W(x)$.
Furthermore, there exists a constant $\alpha_N\in(0,1]$, such that 
\begin{align}
\label{eq:UCON_performance_detectable}
\sum_{t=0}^{\infty}\ell(x(t),u(t))\leq \dfrac{1}{\alpha_N}\mathcal{J}_\infty^\star(x_0)+\dfrac{1-\alpha_N}{\alpha_N}W(x_0).
\end{align}
\end{theorem}
Theorem~\ref{thm:UCON_detect} yields closed-loop properties comparable to Theorem~\ref{thm:UCON} with a semidefinite stage cost under a detectability condition (Asm.~\ref{asm:Costdetectability}).

\subsubsection{Singular output cost}
\label{sec:UCON_2_output}
Next, we consider a pure output stage cost $\ell(x,u)=\|h(x)\|_Q^2$ without any input regularization. 
Such a cost appears naturally in case no input reference is available, e.g., in the classical output regulation setting~\citep{kohler2021constrained}. 
Furthermore, studying this singular cost provides insight for the implications of decreasing the input weighting $R$~\citep{westenbroek2022computational}. 
Due to the absence of an input regularization $R\succ 0$, the cost detectability condition~\ref{enum_cost_detect} is more restrictive. 
\citet[Prop. 4]{kohler2021constrained} show that this cost $\ell$ satisfies a comparable detectability condition if the system is \textit{minimum-phase}/\textit{output-input stable}~\citep{liberzon2002output}, i.e., the zero dynamics ($h(x)\equiv 0$) are asymptotically stable. 
\citet[Lemma~3,Thm.~3]{westenbroek2022computational} derive a converse result: If the system is \textit{non}-minimum-phase, then for any (finite) prediction horizon $N$, the closed-loop system is unstable if the input regularization $R$ is chosen small enough. 
Some intuition for these results can be found by relating the MPC with this singular cost $\ell$ to a high-gain controller, 
which is known to destabilize non-minimum-phase systems~\citep[Thm.~8]{davison1974properties}.

\subsubsection{Economic cost and dissipativity}
\label{sec:UCON_2_economic}
In the following, we discuss results for an economic stage cost $\ell_{\mathrm{e}}(x,u)$ (Sec.~\ref{sec:economic}), which is indefinite (Asm.~\ref{asm:prelim_stabilizing} does not hold). 
\cite{grune2013economic,grune2014asymptotic} show that a sufficiently long prediction horizon $N$ ensures (practical) stability of the optimal steady state $x_{\mathrm{r}}$, given strict dissipativity (cf. Sec.~\ref{sec:economic_5_dissip}) and a local continuity/controllability condition. 
The main difference between this strict dissipativity condition and the cost detectability condition~\ref{enum_cost_detect} is the fact that $W(x)$ in Assumption~\ref{asm:Costdetectability} needs to be positive semidefinite~\citep[Thm.~2]{hoger2019relation}. 
The implementation of this economic MPC scheme (Problem~\eqref{problem:UCON}) only requires the specification of the stage cost $\ell_{\mathrm{e}}$, but no knowledge of the optimal steady state $x_{\mathrm{r}}$, i.e., the closed-loop system "finds" the optimal steady state. 
Due to the absence of terminal ingredients, these economic MPC schemes also yield transient performance bounds relative to the infinite horizon optimal controller~\citep{grune2014asymptotic}. 

\subsubsection{Dynamic operation}
\label{sec:UCON_2_dynamic}
In the following, we explain how the results in Sections~\ref{sec:UCON_2_reg}--\ref{sec:UCON_2_economic} can be naturally extended to address the challenges related to dynamic operation~\ref{enum_motivation_1}--\ref{enum_motivation_3}. 
Most difficulties listed in Section~\ref{sec:challenges} are not present due to the simple design in Problem~\eqref{problem:UCON}. 
In particular, as no specific design is required for the implementation, this MPC formulation can directly cope with online changes in the mode of operation~\ref{enum_motivation_2}. 
Furthermore, Sections~\ref{sec:UCON_2_detect}--\ref{sec:UCON_2_economic} study output references or economic costs~\ref{enum_motivation_3}. 
Notably, the results for economic stage costs $\ell_{\mathrm{e}}$ also apply to unreachable references~\citep{rawlings2008unreachable}. 
Specifically, suppose the stage cost is positive definite w.r.t. some state $x_{\mathrm{r}}$, but this setpoint is not feasible (due to dynamics or constraints). Then, the closed-loop system (practically) converges to the steady state with the smallest distance to the unfeasible target $x_{\mathrm{r}}$~\citep{kohlernonlinear19,long2021unconstrained}.

Next, we discuss results for non-stationary operation~\ref{enum_motivation_1}. 
Generally speaking, the previously presented results extend to the time-varying case, assuming the corresponding conditions related to cost controllability, cost detectability or dissipativity are suitably adjusted, as discussed below in detail. 

Trajectory tracking MPC schemes without terminal ingredients are investigated by~\cite{kohlernonlinear19,faulwasser2021predictive,kohler2021constrained}. 
\citet[Prop.~2]{kohlernonlinear19} ensure cost controllability (Asm.~\ref{asm:CostControl}) for any dynamically feasible reference trajectory if the system is incrementally stabilizable (cf. Sec.~\ref{sec:terminal_LPV}). 
\citet[Prop.~3]{kohler2021constrained} show cost detectability (Asm.~\ref{asm:Costdetectability}) if additionally the output $h$ is Lipschitz continuous and the system is incrementally IOSS, which characterizes detectability for nonlinear systems~\citep{allan2021nonlinear}. 
\citet[Sec.~1.4]{faulwasser2021predictive} use differential flatness to verify cost controllability. 
 
Considering the singular output stage cost (Sec.~\ref{sec:UCON_2_output}), \cite{kohler2021constrained} apply this analysis to solve the \textit{output regulation problem}~\citep{isidori1990output}, i.e., an exogenous system generates a time-varying output reference $y_{\mathrm{r}}(t)$ that should be tracked. 
The classical solution to this problem requires the solution to a partial differential equation to determine the regulator manifold, where this tracking objective is achieved. 
\citet[Thm.~2]{kohler2021constrained} show that a simple MPC formulation (Problem~\eqref{problem:UCON}) with an output tracking cost $\ell=\|h(x)-y_{\mathrm{r}}\|_Q^2$ implicitly steers the system to this regulator manifold, given a minimum-phase condition. 
Since this manifold is unknown, regularization w.r.t. some reference input $u_{\mathrm{r}}$ cannot be directly applied. 
However, in case the exogenous signal is periodic, it is possible to penalize the non-periodicity in the input. In this case, the minimum-phase condition can be relaxed to a standard detectability/IOSS condition in combination with a technical non-resonance condition~\citep[Sec.~IV]{kohler2021constrained}. 

For the case of time-varying economic stage costs $\ell(x,u,t)$, 
\citet[Thm.~3, Lemma~4]{kohlernonlinear19} investigate periodic problems, show strict dissipativity, and prove that the closed-loop system implicitly tracks the optimal periodic trajectory, see also \cite{long2021unconstrained} for the continuous-time extension. 
\citet{grune2020economic} extend this result to general time-varying problems. 
These results assume that there exists a unique time-varying optimal trajectory $x_{\mathrm{rd}}^\star(t)$. 
However, in a time-invariant setup, the optimal mode of operation may also be non-stationary, cf., e.g., maximizing the yield of a continuous stirred-tank reactor~\citep{bailey1971cyclic}. 
In case the system is optimally operated at a (time-invariant) periodic orbit, \citet[Sec.~3]{muller2016economic} show that the closed-loop performance can be arbitrarily suboptimal, even with an arbitrarily large horizon $N$. 
Suitable stability and performance guarantees can be obtained using: a multi-step implementation~\citep{muller2016economic}; additional symmetry conditions~\citep{koehler2018periodic}; or discounting in the cost~\citep{schwenkel2022linearly}. 
Results for more general optimal operation in some (non-periodic) set are not yet available.

\subsubsection{State constraints and region of attraction} 
\label{sec:UCON_2_local}
The previous analysis uses a \textit{global} stabilizability condition (Asm.~\ref{asm:CostControl}/\ref{asm:Costdetectability}). 
However, this precludes hard state constraints and most unstable systems. 
Due to its global nature, it is in general also challenging to verify numerically. 
We address this issue by considering a relaxed \textit{local} condition and suitably characterize the region of attraction. 
In the following, we again consider general state and input constraints~\eqref{eq:constraints}. 

First, we investigate the regulation problem with $\ell(x,u)=\|x\|_Q^2+\|u\|_R^2$ (Sec.~\ref{sec:UCON_2_reg}). 
\begin{assumption}
\label{asm:CostControl_local} 
\citep[Asm.~1]{boccia2014stability}
There exist constants $\gamma\geq 1$, $c>0$, such that for any $x\in {X}$ satisfying $\ell_{\min}(x)\leq c$ 
 and any horizon $N\in\mathbb{I}_{\geq 1}$, Problem~\eqref{problem:UCON} is feasible and the value function satisfies~\eqref{eq:exp_cost_control}. 
\end{assumption}
Compared to Assumption~\ref{asm:CostControl}, this relaxes the requirement to hold only locally around the origin. 
Similar to Proposition~\ref{prop:terminal_LQR}, if the nonlinear dynamics are twice-continuously differentiable, the linearization at the origin is stabilizable, and $0\in\mathrm{int}(\mathbb{Z})$, then Assumption~\ref{asm:CostControl_local} holds with some $c>0,\gamma\geq 1$~\citep{darup2015missing}. 
Hence, Assumption~\ref{asm:CostControl_local} allows for hard state constraints and unstable systems. 
Due to the presence of hard state constraints, recursive feasibility of Problem~\eqref{problem:UCON} is not immediately obvious. 
\begin{theorem}
\label{thm:UCON_local}
\citep[Thm.~4]{boccia2014stability}
Let Assumptions~\ref{asm:continuity}, \ref{asm:prelim_stabilizing}, and \ref{asm:CostControl_local} hold. 
For any constant $\bar{V}>0$, there exists a horizon $N_{\bar{V}}\geq 0$, such that for any horizon $N> N_{\bar{V}}$ and any initial condition $x_0\in \mathbb{X}_{\bar{V}}:=\left\{x\in X|~\mathcal{J}_N^\star(x)\leq \bar{V}\right\}$, Problem~\eqref{problem:UCON} is feasible for all $t\in\mathbb{I}_{\geq 0}$, the constraints~\eqref{eq:constraints} are satisfied, the set $\mathbb{X}_{\bar{V}}$ is positively invariant, and the origin is exponentially stable with the Lyapunov function $\mathcal{J}_N^\star$ for the resulting closed-loop system~\eqref{eq:sys_closedloop_UCON}. 
\end{theorem}
Theorem~\ref{thm:UCON_local} reveals a connection between the region of attraction $\mathbb{X}_{\bar{V}}$ and the horizon bound $N>N_{\bar{V}}$. 
Compared to Theorem~\ref{thm:prelim_MPC_term}, recursive feasibility is \textit{not} ensured for any feasible initial condition, but only a subset $\mathbb{X}_{\bar{V}}$. 
The proof first establishes a decrease condition on the Lyapunov function $\mathcal{J}_N^\star$, which ensures positive invariance of the sublevel set $\mathbb{X}_{\bar{V}}$ and thus recursive feasibility of Problem~\eqref{problem:UCON}, i.e., "stability implies feasibility". 
This is in contrast to the standard paradigm for MPC with terminal ingredients, i.e., "feasibility implies stability"~\citep{scokaert1999suboptimal}, which first establishes recursive feasibility independent of any stability properties. 
Since the stability properties depend on a sufficiently long horizon $N$, also recursive feasibility can be lost here if the horizon is not sufficiently large. 
 
\citet[Thm.~1]{esterhuizen2020recursive} extend Theorem~\ref{thm:UCON_local} to continuous-time problems. 
For detectable and singular costs $\ell$, \citet[Thm.~1,Thm.~2]{kohler2021constrained} analyse the region of attraction as a sublevel set of the Lyapunov function. 
For economic stage costs, \citet[Lemma~3]{faulwasser2015design} show that the turnpike property (Sec.~\ref{sec:economic_5_dissip}) ensures recursive feasibility, cf. also~\citet[Sec.~IV.2]{faulwasser2018economic}. 
\citet[Thm.~3]{kohlernonlinear19} analyse the region of attraction of an economic MPC scheme using a sublevel set of the (practical) Lyapunov function, compare also the continuous-time extension by~\citet[Thm.~3.3]{long2021unconstrained}.

\subsection{Quantitative bounds}
\label{sec:UCON_3}
In the following, we discuss quantitative bounds on the sufficiently long horizon $\underline{N}$ ensuring desired closed-loop properties.
First, we present quantitative, partially tight, bounds on the prediction horizon for the different problem setups (Sec.~\ref{sec:UCON_3_1}). 
Then, we outline how the theoretical results can be constructively used in the design (Sec.~\ref{sec:UCON_3_2}). 
 
\subsubsection{Tight horizon bounds}
\label{sec:UCON_3_1}
We first consider the regulation problem (Sec.~\ref{sec:UCON_2_reg}) using Assumption~\ref{asm:CostControl} and discuss different techniques to obtain $\underline{N}$ for this problem, see also~\citep{grune2012nmpc}. 

Observe that there always exists a $k\in\mathbb{I}_{[0,N-1]}$, such that $\ell(x_{\mathbf{u}^\star}(k,x),\mathbf{u}_k)\leq \mathcal{J}_N^\star(x)/N$. 
Hence, a simple candidate solution is obtained by appending the finite-horizon optimal solution at this point $k$, which ensures~\eqref{eq:DP_relaxed} with $\alpha_N>0$ for $N>\underline{N}\approx\gamma^2$~\citep{grimm2005model}. 
By using the principle of optimality and recursively applying~\eqref{eq:exp_cost_control}, \cite{tuna2006shorter} and \cite{grune2008infinite} show an exponential decay 
\begin{align}
\label{eq:UCON_expDecay_openLoop}
\mathcal{J}^\star_{N-k}(x_{\mathbf{u}^\star}(k,x))\leq \rho_\gamma^k \mathcal{J}_N^\star(x), ~\rho_\gamma=(\gamma-1)/{\gamma}\in[0,1), 
\end{align}
 which ensures stability for $N>\underline{N}\approx 2\gamma\log(\gamma)$. 
\cite{grune2009analysis} determine a tight constant $\alpha_N$ satisfying~\eqref{eq:DP_relaxed} using a linear program (LP) and \cite{grune2010analysis} derive the analytical solution $\underline{N}\approx\gamma\log(\gamma)$, see also~\citet{reble2012unconstrained} for the continuous-time case. 
Given Assumption~\ref{asm:CostControl}, this is the least conservative\footnote{%
In practice we may still achieve stability with a shorter horizon $N\ll \underline{N}$. 
This has two reasons: a) This derivation uses a fixed Lyapunov function $\mathcal{J}_N^\star$.
b) The horizon bound $\underline{N}$ is valid for all systems satisfying condition~\eqref{eq:exp_cost_control}, not only the considered dynamics $f$ and cost $\ell$.} 
 bound s.t. $\mathcal{J}_N^\star$ is a Lyapunov function (cf.~\eqref{eq:DP_relaxed}).

\citet[Thm.~3]{kohler2023stability} extend this LP analysis to derive a tight estimate $\underline{N}$ for detectable stage costs (cf. Thm.~\ref{thm:UCON_detect}), which recovers the LP bounds by~\cite{grune2009analysis,grune2010analysis} as a special case. 
This improves bounds previously derived in \citet[Thm.~1--2]{grimm2005model} and \citet[App.~A]{kohler2021constrained}.

For economic stage costs, the results by~\cite{grune2013economic,grune2014asymptotic} are largely qualitative and convergence is only ensured up to a remainder term that vanishes as $N\rightarrow\infty$.

Considering the region of attraction $\mathbb{X}_{\bar{V}}$ based on local conditions (Thm.~\ref{thm:UCON_local}):
\citet[Rem.~4.32]{kohler2021dynamic} derives a tight estimate coupling the region of attraction $\mathbb{X}_{\bar{V}}$ and a sufficiently long horizon $N_{\bar{V}}$. 
This result shows that the required horizon $N_{\bar{V}}$ can be decomposed as a sum of two terms: a horizon $\underline{N}$ which only depends on the local bound $\gamma$ (Asm.~\ref{asm:CostControl_local}) and a horizon $N_0$ that depends linearly on the region of attraction $\bar{V}$. 
The derivation of this result incorporates the LP analysis by~\cite{grune2010analysis} and contradiction arguments similar to~\cite{limon2006stability}, which ensure $\ell_{\min}(x_{\mathbf{u}^\star(x)}(k,x))\leq c$ for all $k\in\mathbb{I}_{[N_0,N-1]}$. 
Related results include early work by~\cite{boccia2014stability}, continuous-time results~\citep{esterhuizen2020recursive}, a bound $N_{\bar{V}}$ linear in $\bar{V}$~\citep[Thm.~2]{kohlernonlinear19}, and recent bounds for semi-definite costs (cf. \citet[Thm.~1, Thm.~2]{kohler2021constrained}, \citet[Thm.~4.12, Thm.~4.50]{kohler2021dynamic}) and economic costs (cf. \citet[Thm.~3]{kohlernonlinear19} and \citet[Thm.~3.3]{long2021unconstrained}).

In the special case of open-loop stable systems, \citet[Rem.~4]{kohler2023stability} show that for any horizon $N\in\mathbb{I}_{\geq 1}$, asymptotic stability can be ensured by choosing a sufficiently large input penalty $R\succ0 $. 
This follows from Theorem~\ref{thm:UCON_detect} by picking $W(x)$ (Asm.~\ref{asm:Costdetectability}) as a Lyapunov function characterizing the open-loop (input-to state) stability. In case of hard state constraints, a large region of attraction requires a (proportional) increase in the prediction horizon $N$~\citep[Thm.~4.12]{kohler2021dynamic}.

For minimum-phase systems, \citet[Rem.~7]{kohler2021constrained} show asymptotic stability with a singular cost $\ell(x,u)=\|h(x)\|_Q^2$ and a very short horizon $N$ (independent of $\gamma$). 
In the continuous-time case with an input-output stage cost $\ell(x,u)=\|h(x)\|_Q^2+\|u\|_R^2$, \citet[Thm.~2]{westenbroek2022computational} show asymptotic stability for any horizon $N$ by choosing a sufficiently small input penality $R\succ 0$. 
Both results only hold \textit{locally} in case of compact input constraints $u\in\mathbb{U}$ and a longer horizon is required to ensure a large region of attraction~\citep[Thm.~4.50]{kohler2021dynamic}.

\subsubsection{From analysis to design}
\label{sec:UCON_3_2}
In contrast to Sections~\ref{sec:prelim}--\ref{sec:economic}, this section mainly provides \textit{analysis} results and hence we next discuss how to apply these theoretical results. 
First, we explain how these theoretical results can be used to inform the choice of the prediction horizon $N$ and stage cost $\ell$.
Then we mention some modifications to Problem~\eqref{problem:UCON} to enable theoretical guarantees with significantly shorter horizons $N$ while keeping a simple design. 

For a given stage cost $\ell$, the previous results provide a lower bound $\underline{N}$ on the prediction horizon such that desired closed-loop properties can be guaranteed for $N>\underline{N}$. 
Consider for example a trajectory tracking MPC scheme (Sec.~\ref{sec:terminal_3}) with some feasible reference trajectory $r(t)$ which is generated online~\ref{enum_motivation_2}. 
By verifying cost controllability (Asm.~\ref{asm:CostControl}) for any dynamic reference trajectory, we can derive a horizon $N$ ensuring asymptotic stability for any reference trajectory~\citep[Sec.~III]{kohlernonlinear19}.

Depending on the system dynamics and the chosen stage cost $\ell$, the derived horizon bound $\underline{N}$ may simply be too large for real-time implementation. 
For practical application, we may instead treat the prediction horizon $N$ as given (due to computational requirements) and consider the stage cost $\ell$ as a design parameter. 
The qualitative results (Sec.~\ref{sec:UCON_2}) provide some indication how the stage cost should be chosen and we mentioned special cases (open-loop stable, minimum-phase) for which we can derive arbitrarily small horizon bounds. 
Considering the horizon bounds (Sec.~\ref{sec:UCON_3_1}), the stage cost $\ell$ should be chosen s.t. the cost controllability constant $\gamma$ (Asm.~\ref{asm:CostControl}) is sufficiently small. 

In Section~\ref{sec:artificial}, tracking MPC formulations use artificial setpoints to effectively deal with infeasible references and provide a large region of attraction, even for short horizons $N$. 
For the simple MPC formulation (Problem~\eqref{problem:UCON}), infeasible references result in an economic formulation and closed-loop properties require a potentially very large prediction horizon $N$. 
Even for feasible references, a large region of attraction requires a large prediction horizon $N$. 
Hence, it is desirable to merge the benefits of the simple design of Problem~\eqref{problem:UCON} and the properties of artificial references (Sec.~\ref{sec:artificial}). 
\citet[Thm.~3]{limon2018nonlinear} show that we can forgo the design of a terminal set constraint $\mathbb{X}_{\mathrm{f}}$ for setpoint tracking MPC formulations by using sublevel set arguments (cf. Sec.~\ref{sec:UCON_2_local}). 
However, implementation still requires the design of a local CLF satisfying Condition~\ref{enum_term_2} in Assumption~\ref{asm:prelim_terminal}. 
\citet[Thm.~2]{soloperto2022nonlinear} directly integrate an artificial setpoint in Problem~\eqref{problem:UCON}, without any terminal cost $V_{\mathrm{f}}$. The corresponding analysis characterizes the region of attraction as a sublevel sets of the tracking cost (cf. Sec.~\ref{sec:UCON_2_local}) w.r.t. \textit{any} artificial reference $r\in\mathbb{Z}_{\mathrm{r}}$. 
This provides a large region of attraction and stability of the optimal setpoint with relatively short prediction horizons $N$, while the absence of terminal ingredients allows for a simple design.

Using a (local) CLF as a terminal cost $V_{\mathrm{f}}$ can ensure stability properties for any horizon $N\in\mathbb{I}_{\geq 1}$ (Sec.~\ref{sec:prelim}). 
However, computing a local CLF can be cumbersome and avoiding this is one of the main motivations to consider the simple MPC formulation in Problem~\eqref{problem:UCON}. 
In order merge the complementary advantages we investigate two questions: 
\begin{enumerate}[label=\arabic*)]
\item Can we ensure stability with a shorter horizon $N$ if we add a terminal cost $V_{\mathrm{f}}$ satisfying a relaxed CLF condition?
\label{enum_relax_CLF_1}
\item Is the design of such a relaxed CLF significantly simpler? 
\label{enum_relax_CLF_2}
\end{enumerate}
\ref{enum_relax_CLF_1}: \citet[A~3]{tuna2006shorter} characterize a relaxed CLF as
\begin{align}
\label{eq:CLF_approx}
\min_{u\in\mathbb{U}}V_{\mathrm{f}}(f(x,u))+\ell(x,u)\leq (1+\epsilon_{\mathrm{f}})V_{\mathrm{f}}(x),\quad \epsilon_{\mathrm{f}}\in[0,\infty),
\end{align}
and similar conditions are used by \cite{reble2012improved,kohler2021stability,kohler2023stability}. 
For $\epsilon_{\mathrm{f}}=0$, Condition~\eqref{eq:CLF_approx} corresponds to Assumption~\ref{asm:prelim_terminal}, however, \eqref{eq:CLF_approx} is also naturally satisfied for any positive definite terminal cost $V_{\mathrm{f}}$ with some finite $\epsilon_{\mathrm{f}}<\infty$. 
\citet[Thm. ~5--8]{kohler2023stability} provide a (tight) LP analysis ensuring stability for 
$N>\underline{N}\approx \gamma(\log(\gamma)-\log(1+1/\epsilon_{\mathrm{f}}))$, see also the continuous-time results by~\cite{reble2012improved}.
For $\epsilon_{\mathrm{f}}\rightarrow\infty$, this recovers the horizon bounds by~\cite{grune2010analysis}, while $\epsilon_{\mathrm{f}}\rightarrow 0$ enables asymptotic stability with arbitrary short horizons $N$. 
The analysis can be naturally extended to characterize the region of attraction (cf. Sec.~\ref{sec:UCON_2_local}), see \citet{kohler2021stability} and \citet[Thm.~4.37]{kohler2021dynamic}. 
\citet[Thm.~5]{kohler2023stability} and \citet[Thm.~5]{kohler2021stability} also discuss how the added terminal cost $V_{\mathrm{f}}$ changes the performance bounds~\eqref{eq:UCON_performance}/\eqref{eq:UCON_performance_detectable}. 

\ref{enum_relax_CLF_2}: For practical application, we require a simple way to obtain a terminal cost $V_{\mathrm{f}}$. 
\cite{beckenbach2022approximate} and \cite{moreno2022predictive} compute $V_{\mathrm{f}}$ using approximate dynamic programming and relate $\epsilon_{\mathrm{f}}$ to the stopping condition. 
\citet[Thm.~5.4]{grune2010analysis} suggest $V_{\mathrm{f}}(x)=\omega \ell_{\min}(x)$ with a simple weighting $\omega\geq 1$, which is especially attractive if $\min_{u\in\mathbb{U}}\ell_{\min}(f(x,u))\leq \ell_{\min}(x)$. 
In this case, asymptotic stability holds for any horizon $N\in\mathbb{I}_{\geq 1}$ if $\omega$ is chosen sufficiently large~\citep[Sec.~VI]{kohler2023stability}, see also \cite{reble2012improved} and \cite{alamir2018stability} regarding stability results by increasing the weighting over the horizon. 
\cite{magni2001stabilizing,kohler2021stability} define $V_{\mathrm{f}}(x)$ use a finite-horizon rollout, i.e., $V_{\mathrm{f}}(x)=\sum_{k=0}^{M-1}\ell(x_{\mathbf{u}}(k,x),\mathbf{u}_k)$ with the rollout horizon $M\in\mathbb{I}_{\geq 1}$ and some locally stabilizing control law $\mathbf{u}\in\mathbb{U}^M$. 
The resulting MPC formulation is equivalent to Problem~\eqref{problem:UCON} with a horizon $N+M$ with an additional constraint on the last $M$ inputs. 
\citet[Prop.~4]{kohler2021stability} provide an LP analysis to derive a tight constant $\epsilon_{\mathrm{f}}$ satisfying~\eqref{eq:CLF_approx} and show that asymptotic stability can be ensured with a significantly smaller horizon $N+M$. 
Such a combination of online optimization and a finite-horizon rollout is also key in various reinforcement learning algorithms~\citep{bertsekas2022lessons}. 
\citet[Thm.~1]{bonassi2024nonlinear} show that a scaled finite-horizon rollout cost can provides a local CLF satisfying Assumption~\ref{asm:prelim_terminal}. 
For the special case of linear systems with polytopic constraints, \citet[Sec.~V]{dutta2014certification} and \citet[Thm.~4]{rakovic2023model} show that this finite-horizon rollout can implicitly characterize the terminal set constraint $\mathbb{X}_{\mathrm{f}}$.  
The implicit characterization of the terminal cost in terms of a finite-horizon rollout ensures that no redesign is required for online changing setpoints (cf. Sec.~\ref{sec:terminal_2}/\ref{sec:artificial_1}) if a (locally) stabilizing feedback is known~\citep{magni2005solution}, which makes it very attractive for practical application.

For economic MPC without terminal ingredients, the theoretical bounds on the prediction horizon are rather qualitative. 
\citet[Thm.~5]{zanon2018economic} show that for any finite horizon $N$, the economic MPC scheme results in suboptimal operation. 
\citet[Thm.~5]{zanon2018economic} ensure (local) asymptotic stability by adding a linear gradient correction as a terminal cost. 
\citet[Thm.~4]{liu2016economic} use a finite-horizon rollout of the economic cost $\ell_{\mathrm{e}}$ as a terminal cost, which ensures exponential stability for a sufficiently large rollout horizon $M$.

\subsection{Illustrative example}
\label{sec:UCON_example}
\begin{figure}
\begin{center} 
\includegraphics[scale=0.5]{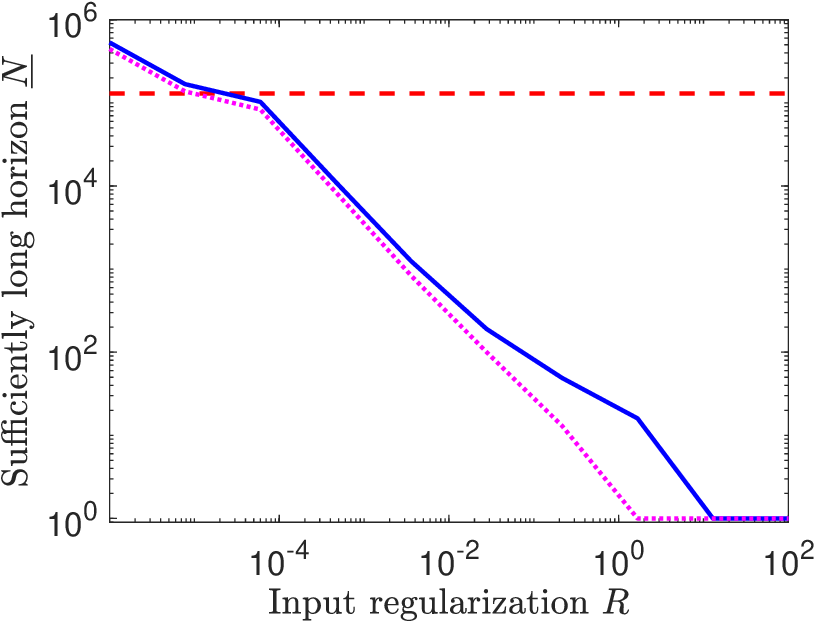} 
\end{center}
\caption{Chain of mass--spring--dampers, adapted from~\cite{kohler2023stability}. Sufficient horizon $\underline{N}$ for stability for varying $R$ based on Theorem~\ref{thm:UCON_detect} in blue, solid. 
Bound $\underline{N}$ utilizing an additional approximate terminal cost (Sec.~\ref{sec:UCON_3_2}) using a finite-horizon roll-out with $M=10$ shown in magenta, dotted.
Theoretical bounds $\underline{N}$ using Theorem~\ref{thm:UCON} shown in red, dashed.} 
\label{fig:UCON_example}
\end{figure}

We illustrate the application of these theoretical results at the example of a chain of linear mass--spring--damper systems with $n=12$ states and $m=1$ input by \citet[Sec.~VII.A]{kohler2023stability}. 
The stage cost is $\ell(x,u)=\|y\|^2+\|u\|_R^2+\|x\|_Q^2$ with controlled position $y$, a small weight $Q\succ 0$, and a tunable input regularization $R>0$. 
The system is open-loop stable, lightly damped, severely under actuated, and has non-minimum-phase behaviour w.r.t. $y$.
The MPC formulation~\eqref{problem:UCON} destabilizes the system if a short horizon $N$ or small regularization parameters $R$ is chosen. 
We utilize Theorem~\ref{thm:UCON_detect} to systematically determine sufficiently large prediction horizons $N$ ensuring global exponential stability for different choice of $R$, which is visualized in Figure~\ref{fig:UCON_example}. 
Numerically computing $\underline{N}$ for the different parameter combinations takes about $12$ seconds. 
This provides a constructive method to choose an input regularization $R$ ensuring exponential stability for a given prediction horizon $N$. 
Conservatism of the analysis in Theorem~\ref{thm:UCON} and significantly shorter horizons by including a simple approximate terminal cost are also visible in Figure~\ref{fig:UCON_example}. 
Given that this offline verification utilizes no specific requirements regarding the considered setpoint, the same design choices $N,R$ also ensure stability for arbitrary reference setpoints~\ref{enum_motivation_2} or (feasible) reference dynamic trajectories~\ref{enum_motivation_1}. 
Constructive utilization of the theory in Section~\ref{sec:UCON} is (currently) primarily considered for non-holonomic vehicles, see numerical studies by~\cite{worthmann2015model,coron2020model} and experimental results by~\citep{rosenfelder2022model,rickenbach2023active}. 

\subsection{Open issues}
\label{sec:UCON_openIssue}
The lack of tools to automatically verify the posed conditions (Asm.~\ref{asm:CostControl}--\ref{asm:CostControl_local}) limits application for high-dimensional nonlinear systems. 
The theoretical bound on a sufficiently long horizon $N$ is often too conservative to be applied. 
Incorporating a terminal penalty while keeping a simple design procedure is a promising research direction, see also discussion in Section~\ref{sec:discussion_UCON}.

\section{Discussion and Conclusions}
\label{sec:discuss}
In this section, we provide some high-level discussion and conclusions. 
First, we provide a concise summary of the different frameworks introduced in this article (Sec.~\ref{sec:discussion_summary}) and address some important extensions (Sec.~\ref{sec:discussion_extensions}). 
Lastly, we discuss complementary advantages of these frameworks and possible unifying approaches (Sec.~\ref{sec:discussion_complementary}). 
 
\subsection{Summary}
\label{sec:discussion_summary} 
In the following, we summarize the main tools and methods developed in Sections~\ref{sec:terminal}--\ref{sec:UCON} and discuss how they address the challenges of dynamic operation~\ref{enum_motivation_1}--\ref{enum_motivation_3}.

In Section~\ref{sec:terminal}, we provide design procedures for the terminal ingredients $V_{\mathrm{f}},\mathbb{X}_{\mathrm{f}}$. 
For a given steady state, this design uses the linearization to construct a local CLF with the LQR~\citep{chen1998quasi}.
For non-stationary operation~\ref{enum_motivation_1}, we use a local LTV model around a reference trajectory and construct a local CLF using a time-varying LQR~\citep{faulwasser2011model}. 
To account for online changes in the reference~\ref{enum_motivation_2}, we develop a reference generic offline computation using tools from LPV systems to yield parametrized terminal ingredients for any feasible reference setpoint/trajectory~\citep{koehler2020nonlinearTAC}. 
In addition, we discuss terminal equality constraints as a simple but conservative alternative (Sec.~\ref{sec:terminal_TEC}). 

In Section~\ref{sec:artificial}, we design tracking MPC formulations using artificial references.
These designs address online changing operating conditions~\ref{enum_motivation_2} by ensuring feasibility independent of online changes in the reference. 
Furthermore, the offset cost $V_{\mathrm{o}}$ enables indirect specification of the optimal mode of operation through a (possibly infeasible) output target $y_{\mathrm{d}}$~\ref{enum_motivation_3}, while stability of the implicitly defined optimal steady state is ensured~\citep{limon2018nonlinear}. 
To account for non-stationary operation~\ref{enum_motivation_1}, a periodic artificial reference trajectory is used and stability of the (unknown) optimal periodic trajectory is established~\citep{limon2016mpc,koehler2020nonlinearAutomatica}. 
The deployment of these nonlinear tracking MPC schemes hinges on the design of terminal ingredients for online changing references (Sec.~\ref{sec:terminal}). 
Important extensions include the design of an offset cost $V_{\mathrm{o}}$ addressing non-convexity~\citep{soloperto2022nonlinear} and the reduction of the online computational demand by operating a tracking MPC and a planner at different frequencies~\citep{koehler2020nonlinearAutomatica}. 

In Section~\ref{sec:economic}, we directly express the control goal in the cost $\ell_{\mathrm{e}}$~\ref{enum_motivation_3} using an economic MPC formulation~\citep{faulwasser2018economic}. 
The economic MPC scheme ensures that the closed-loop performance is no worse than tracking MPC schemes~\citep{angeli2012average}. 
Non-stationary operation~\ref{enum_motivation_1} in terms of periodic trajectories can also be directly addressed~\citep{zanon2017periodic}. 
To account for online changing operating conditions~\ref{enum_motivation_2}, we use artificial setpoints (Sec.~\ref{sec:artificial}). 
Optimality guarantees with economic stage costs require a self-tuning weight $\beta(t)$~\citep{muller2014performance}. 
We use artificial periodic references to account for non-stationary and online changing operating conditions \ref{enum_motivation_1}--\ref{enum_motivation_2}, however, this may result in suboptimal performance. 
We recover the desired performance guarantees by introducing a shifted terminal cost $\tilde{V}_{\mathrm{f}}$ that yields a monotonic decrease condition~\citep{Koehler2020Economic}. 
We also discuss stability and the design of terminal ingredients for economic MPC schemes. 

In Section~\ref{sec:UCON}, we analyse MPC formulations without terminal ingredients. 
Given a suitable stabilizability condition, asymptotic stability is ensured with a sufficiently long horizon $N$ and a tight bound for this horizon is computed using an LP analysis~\citep{grune2010analysis}. 
Since this framework requires no explicit offline design, the results can be directly applied to non-stationary operation~\ref{enum_motivation_1} with online changing operating conditions~\ref{enum_motivation_2}. 
We address input-output/output references~\ref{enum_motivation_3} assuming detectability/minimum-phase~\citep{grimm2005model,kohler2021constrained,westenbroek2022computational} and a tight horizon bound $N$ is obtained using an LP analysis~\citep{kohler2023stability}. 
The consideration of infeasible references complicates the analysis, however, 
 (practical) stability of the optimal feasible trajectory can be ensured with a sufficiently large horizon~\citep{kohlernonlinear19,long2021unconstrained}. 
General economic stage cost can be handled similarly~\citep{grune2013economic,grune2014asymptotic,grune2020economic}. 
Although most of these results consider \textit{global} properties, we also study the region of attraction~\citep{boccia2014stability}, including corresponding
bounds on the prediction horizon $N$~\citep{kohler2021dynamic}. 
By adding a terminal weight, stability can be ensured with a significantly shorter horizon $N$~\citep{kohler2023stability}.

\subsection{Extensions} 
\label{sec:discussion_extensions}
In the following, we discuss some important extensions w.r.t. the exposition in Sections~\ref{sec:prelim}--\ref{sec:UCON}. 
First, we provide more flexible MPC formulations for non-stationary operation~\ref{enum_motivation_1} by introducing a \textit{time-parametrization} (Sec.~\ref{sec:discussion_time}). 
Then, we discuss large scale systems and \textit{distributed} solutions (Sec.~\ref{sec:discussion_distributed}). 
Finally, we consider the presence of \textit{model mismatch} (Sec.~\ref{sec:discussion_robust}).

\subsubsection{Flexible time parametrization}
\label{sec:discussion_time}
In the following, we re-parametrize time-varying reference signals to increase the flexibility of the MPC approaches. 

The MPC formulations based on artificial periodic reference trajectories $\mathbf{r}\in\mathbb{Z}_{\mathrm{r}}^T$(Sec.~\ref{sec:artificial_3}/\ref{sec:economic_4}) fix a period length $T\in\mathbb{I}_{\geq 1}$ and then provide stability/optimality guarantees w.r.t. all $T$-periodic trajectories. 
By adopting a continuous-time formulation with a variable step-size, we can optimize over periodic trajectories without fixing a period length, cf.~\citet[Sec.~2]{houska2015enforcing}, \citet[Sec.~3]{gutekunst2020economic}, \citet[Rem.~2, App.~A]{Koehler2020Economic}. 

In Section~\ref{sec:terminal_2}, we introduce the trajectory tracking problem using a time-varying reference trajectory $x_{\mathrm{r}}(t)$, $t\in\mathbb{I}_{\geq 0}$. 
This problem can be generalized to a \textit{path-following} problem, where only the geometric curve of the reference is provided, i.e., $x_{\mathrm{r}}(\theta)$ with a scalar parametrization $\theta\in[0,1]$. 
The timing can be flexibly changed by optimizing $\dot{\theta}(t)>0$ as a free control input and a desired speed $\dot{\theta}$ can be set in the cost function. 
Corresponding path-following MPC formulations with stability and convergence guarantees are derived by~\cite{faulwasser2012optimization,faulwasser2015nonlinear,yu2015nonlinear}, including an MPC formulation without terminal ingredients~\citep{faulwasser2021predictive}. 
Such path-following implementations are also successfully applied in different experiments~\citep{faulwasser2016implementation,liniger2015optimization,romero2022model}. These approaches are also called \textit{model predictive contouring control}~\citep{lam2010model}. 

It is natural to combine a path-following formulation with an artificial reference path (Sec.~\ref{sec:artificial}). 
\cite{sanchez2023tracking} provide a first step in this direction, by optimizing a periodic artificial reference $\mathbf{r}\in\mathcal{R}_T$ and evaluating the offset cost $V_{\mathrm{o}}$ w.r.t. a periodic reference curve. 
By jointly optimizing the path-progress $\theta$, convergence to the optimal periodic curve is ensured. However, this approach still requires a fixed period length $T$ in the artificial reference. 
Further research is required in this direction.

In summary, most of the designs in this article that use a time-varying reference can be generalized to use a more flexible time parametrization, which can significantly improve performance.

\subsubsection{Large scale systems and distributed solutions}
\label{sec:discussion_distributed}
In the following, we discuss application of the presented approaches to large scale systems. 
Scalability for large systems is often achieved using \textit{distributed} approaches, see the overviews by \cite{christofides2013distributed,muller2017economic} on distributed MPC.
We discuss design procedures that are scalable and preserve sparsity of the optimization problems. 
Hence, efficient distributed optimization methods can be applied~\citep{engelmann2022aladin}, see also~\citet{kohler2019distributed} regarding closed-loop properties under inexact distributed optimization. 

Considering the design of terminal ingredients $V_{\mathrm{f}},k_{\mathrm{f}}$ (Sec.~\ref{sec:terminal}), computing the CLF based on the LQR destroys the sparsity structure in Problem~\eqref{problem:prelim}/\eqref{problem:traj_track}. 
Furthermore, the LMI based computation of parametrized terminal ingredients by~\cite{koehler2020nonlinearTAC} does not scale to high dimensional systems. 
\cite{conte2016distributed} propose an LMI-based design of the terminal ingredients for linear systems using a distributed parametrization of $V_{\mathrm{f}},k_{\mathrm{f}}$. 
This ensures that the offline design and the online optimization can be efficiently distributed. 
Similar distributed parametrizations can be utilized to compute parametrized terminal ingredients for nonlinear systems, cf.~\cite{wang2017distributed}. 
 
Artificial references (Sec.~\ref{sec:artificial}) can be naturally combined with distributed formulations. 
Distributed setpoint tracking MPC formulations (Sec.~\ref{sec:artificial_1}) for large scale systems are derived by~\cite{ferramosca2013cooperative,aboudonia2021distributed}. 
Artificial references can also facilitate complex coordination in multi-agent systems~\citep{kohler2023distributed_periodic,kohler2022distributed_setpoint,rickenbach2023active}. 
\cite{kohler2022distributed_setpoint} achieve consensus by only communicating artificial references with other agents. 
Similarly, \cite{rickenbach2023active} solve the coverage control problem and ensure collision avoidance by only exchanging artificial references. 
 
In general, few existing results extend economic MPC (Sec.~\ref{sec:economic}) to large scale distributed systems~\citep[Sec.~4.5]{muller2017economic}. 
The design of the terminal ingredients for large scale systems also utilizes a distributed parametrization~\citep[Sec.~2.2]{kohler2017distributed}. 
Coordination of multiple agents with economic cost functions is addressed by~\cite{kohler2018distributed} using a separate coordination algorithm to compute reference setpoints.

MPC formulations without terminal ingredients (Sec.~\ref{sec:UCON}) are simple to apply for large scale systems due to the absence of any offline designs. 
For example, \cite{giselsson2013feasibility} and \citet[App.~A]{kohler2019distributed} derive closed-loop guarantees for regulation and economic performance despite inexact distributed optimization.

\subsubsection{Model mismatch and robustness}
\label{sec:discussion_robust}
In this article, we assume that the system evolves exactly according to our model~\eqref{eq:sys}, which is rarely the case in practice. 
In general, recursive feasibility and closed-loop properties under model mismatch require a \textit{robust} MPC design~\citep{kouvaritakis2016model,mayne2016robust}. 
A simple robust MPC design uses a nominal prediction and confines the system state $x(t)$ in a \textit{tube} around this nominal prediction with an additional tube feedback, cf.~\citet{mayne2011tube,mayne2016robust,rakovic2022homothetic,sasfi2022robust}; \citet[Sec.~3.5]{kouvaritakis2016model}. 
By using a nominal cost, the nominal closed-loop properties we derive in this article also apply to this nominal trajectory and the system state $x(t)$ is confined to a neighbourhood of this nominal trajectory.

While the design of such tubes is beyond the scope of this article, we highlight a connection to the parametrized terminal cost $V_{\mathrm{f}}(x,r)$ introduced in Section~\ref{sec:terminal_4}. 
Specifically, this design results in an incremental Lyapunov function (cf. Sec.~\ref{sec:terminal_LPV}), which is used in the nonlinear robust MPC designs by~\cite{bayer2013discrete,singh2017robust,kohler2018novel,Koehler2020Robust,sasfi2022robust} to parametrize this tube.

In the following, we mention existing results that combine the nominal results in Sections~\ref{sec:terminal}--\ref{sec:UCON} with a robust MPC formulation. 
Robust trajectory tracking (Sec.~\ref{sec:terminal_3}) is studied by~\citet[App.~B]{koehler2020nonlinearTAC}. 
%
Robust setpoint tracking formulations (Sec.~\ref{sec:artificial_1}) for linear and nonlinear systems are developed by~\cite{limon2010robust,zeilinger2014soft} and 
\cite{Nubert2020Robot,cunha2022robust}, cf. also~\cite{sasfi2022robust,peschke2023RAMPC_track} 
regarding online model updates. 
For artificial periodic references (Sec.~\ref{sec:artificial_3}), \cite{pereira2017robust} provide a robust design for linear systems, cf. also~\cite{broomhead2015robust}.
Robust \textit{economic} MPC formulations (Sec.~\ref{sec:economic_1}) are developed by~\cite{bayer2014tube,bayer2016robust} and a periodic extension (Sec.~\ref{sec:economic_2}) is derived by~\cite{wabersich2018economic}. 
Robust MPC formulations without terminal ingredients (Sec.~\ref{sec:UCON}) are developed by~\cite{kohler2018novel} 
and results for output regulation and economic MPC are derived by \cite{kohler2021constrained} and \cite{schwenkel2020robust,kloppelt2021transient}.

\textit{Offset-free} tracking MPC formulations are routinely applied to ensure convergence to the desired reference~\citep{magni2005solution,limon2010robust,morari2012nonlinear,betti2013robust}.
In particular, asymptotically constant model mismatch can be compensated (asymptotically) using a disturbance observer or a velocity formulation~\citep{muske2002disturbance,pannocchia2003disturbance,cisneros2020velocity}. 
This does not ensure convergence to an optimal steady state~\eqref{eq:eco_opt_steadyState} in case 
of an economic stage cost $\ell_{\mathrm{e}}$. 
\textit{Modifier adaptation} estimates a correction for the steady-state optimization~\eqref{eq:eco_opt_steadyState} that ensures convergence to the (economically) optimal steady state~\citep{alamo2014gradient,faulwasser2019toward,vaccari2021offset}. 
Developing similar approaches for more dynamic problems~\ref{enum_motivation_1} is an open problem.

\subsection{Complementary benefits and limitations}
\label{sec:discussion_complementary}
In the following, we contrast the complementary advantages of the different MPC frameworks presented in Sections~\ref{sec:terminal}--\ref{sec:UCON} and highlight some unifying approaches. 
First, we analyse the role of artificial references (Sec.~\ref{sec:discussion_artificial}). 
Then, we discuss benefits of using economic MPC formulations (Sec.~\ref{sec:discussion_eco}). 
Finally, we contrast the benefits and limitations of using terminal ingredients in MPC (Sec.~\ref{sec:discussion_UCON}).

\subsubsection{Artificial references}
\label{sec:discussion_artificial} 
One of the key tools we use in Sections~\ref{sec:artificial}--\ref{sec:economic} to deal with online changing operating conditions~\ref{enum_motivation_2} are artificial references (Sec.~\ref{sec:artificial}). 
Artificial references ensure recursive feasibility and constraint satisfaction completely independent of the control goal, which is expressed by some target to be tracked or an economic cost, both of which may vary unpredictably online. 
The non-trivial part is that under nominal conditions, we obtain the same performance/stability guarantees that we would achieve with a standard MPC formulation with a fixed optimal reference (cf. Thm.~\ref{thm:limon_2}/\ref{thm:eco_periodic_artificial}).
From a design perspective, the addition of an artificial reference does not require complex offline operations and is hence easy to apply. 
Additional benefits of artificial references include the large region of attraction and flexibility, e.g., if multiple setpoints are optimal (cf. Sec.~\ref{sec:artificial_zone}). 
Main challenges and limitations in the application of artificial references include: (i) design of terminal ingredients; (ii) periodicity requirement for dynamic references; and (iii) computational complexity. 

(i) One drawback of this approach is arguably the fact that terminal ingredients need to be designed for all possible references, e.g., using~\cite{koehler2020nonlinearTAC}. 
The approach by \cite{soloperto2022nonlinear} circumvents this issue by using artificial references without any terminal ingredients. 
This combines many of the benefits of artificial references (Sec.~\ref{sec:artificial}) and MPC without terminal ingredients (Sec.~\ref{sec:UCON}), such as a simple design, a large region of attraction, and direct handling of online changing or infeasible references~\ref{enum_motivation_2}--\ref{enum_motivation_3}. 

(ii) We address optimal operation with non-stationary targets~\ref{enum_motivation_1} using periodic artificial reference trajectories $\mathbf{r}\in\mathbb{Z}_{\mathrm{r}}^T$ and provide guarantees w.r.t. optimal periodic operation with period length $T$. 
This requires a priori knowledge of the optimal periodic length $T$, which can be relaxed using a suitable parametrization (Sec.~\ref{sec:discussion_time}). 
More importantly, there are many control problems where optimal operation is completely non-periodic. 
For example, batch processes in process control or motion planning problems for autonomous robots are typically posed as finite-horizon control problems and searching for a periodic solution may be infeasible or highly suboptimal. 
Addressing such problems with artificial references remains an open issue. 

(iii) Considering the computational complexity, 
Problem~\eqref{problem:setpointTracking}/\eqref{problem:eco_selfTuning} uses an artificial setpoint, which slightly increases the computational complexity.\footnote{%
Solvers exploiting the structure in Problem~\eqref{problem:prelim} (cf.~\citep[Sec.~2.7.2]{verschueren2022acados}) are not directly applicable to Problem~\eqref{problem:setpointTracking}/\eqref{problem:eco_selfTuning}. 
A simple workaround by~\cite{rickenbach2023active} is to use an augmented state $(x,r)$.}  
However, the increase in computational complexity can be very large in case of artificial periodic reference trajectories $\mathbf{r}\in\mathbb{Z}_{\mathrm{r}}^T$ (cf. Problem~\eqref{problem:PeriodicTracking}/\eqref{problem:eco_selfTuning_periodic}) with $T\gg 1$.  
\cite{koehler2020nonlinearAutomatica} address this issue by decomposing Problem~\eqref{problem:PeriodicTracking} into a tracking MPC and a planning problem (Sec.~\ref{sec:artificial_4}). 
This circumvents the computational increase by running a standard tracking MPC scheme while the planning problem is solved in parallel on a longer time-scale. 
In general, decomposing complex control problems into a standard tracking MPC scheme combined with a more complex optimization problem solved on a slower time-scale is a promising approach, cf., e.g., tracker--planner hierarchy~\citep{schweidel2022safe} or asynchronous updates in robust MPC~\citep{sieber2022asynchronous}.\footnote{%
The idea by~\cite{sieber2022asynchronous} can also be adapted to the problem in Section~\ref{sec:artificial_4} by removing the feasibility preserving constraint~\eqref{problem:limon_partial_3} and instead optimizing the reference $\mathbf{y}_{\mathrm{r}}$ in the tracking MPC as a convex combination with a scalar interpolating variable (given convexity, Asm.~\ref{asm:limon_convex_unique}).}

\subsubsection{Directly minimizing economic costs} 
\label{sec:discussion_eco}
In Section~\ref{sec:economic}, we discuss economic MPC formulations that directly optimize the performance measure $\ell_{\mathrm{e}}$, e.g., reflecting production yield and energy consumption. 
This is in contrast to the tracking MPC formulations (Sec.~\ref{sec:prelim}--\ref{sec:artificial}), which use a positive definite stage cost $\ell$ to regulate the system to some desired mode of operation. In the following, we discuss the benefits of directly using an economic stage cost $\ell_{\mathrm{e}}$ vs. a regulation cost $\ell$. 

A benefit of tracking MPC formulations is simplicity, 
 i.e., we ensure stability of the optimal setpoint and hence closed-loop operation is no worse than operation at this setpoint. 
In economic MPC, we may require additional modifications to ensure the same performance bounds with artificial references and even more care is required for economic MPC formulations without terminal ingredients. 
The fact that economic MPC schemes do not necessarily guarantee stability may be a practical hindrance for deployment. 
For example, monitoring a process may be less trivial as significant fluctuations in the controlled variables could indicate some failure, even if these fluctuations do improve the economic performance. 
 
Theorem~\ref{thm:eco_performance} ensures that the average performance of an economic MPC scheme is no worse than any tracking MPC scheme. 
There are numerous examples demonstrating significant performance benefits using economic MPC schemes~\citep{rawlings2008unreachable,rawlings2012fundamentals,ellis2017economic}. 
The adoption of economic MPC is especially relevant in applications where the objective is largely independent of the stability of the system, e.g., 
in HVAC systems economic cost functions are common to reduce energy consumption~\citep[Table~3]{taheri2022model}. 
Even though similar economic considerations are paramount in various industries, the deployment of tracking MPC formulations remains the norm. 

The introduction of economic MPC schemes in practice may be facilitated by formulations that also enable a user to flexibly set desired stability properties. 
Approaches to address this issue include Lyapunov constraints~\citep{heidarinejad2012economic,ellis2017economic}, average constraints~\citep{muller2014transient,muller2014convergence}, multi-objective formulations~\citep{soloperto2020augmenting,eichfelder2022relaxed}, or using the external variable $y_{\mathrm{e}}$ to flexibly change the cost online (cf. \citet[Fig.~4]{Koehler2020Economic}, Sec.~\ref{sec:economic_3}). 
 
\subsubsection{On terminal ingredients in MPC} 
\label{sec:discussion_UCON}
In the following, we discuss advantages and limitations of using terminal ingredients in MPC, see also the discussions by \cite{mayne2013apologia} and \citet[Sec.~7.4]{grune2017nonlinear}. 
 First, we provide a general discussion contrasting the standard MPC formulation (Sec.~\ref{sec:prelim}) to MPC formulations without terminal ingredients (Sec.~\ref{sec:UCON}). Then, we focus on the challenges related to dynamic operation~\ref{enum_motivation_1}--\ref{enum_motivation_3}. 
Finally, we highlight some approaches that combine ideas from both frameworks. 

We first focus on stabilizing the origin with a quadratic stage cost $\ell(x,u)=\|x\|_Q^2+\|u\|_R^2$ and distinguish three MPC formulations:
\begin{itemize}
\item no terminal ingredients, also called "unconstrained" MPC (UCON, Sec.~\ref{sec:UCON}, \cite{grune2010analysis} ); 
\item a classical quadratic terminal cost $V_{\mathrm{f}}$, also called quasi-infinite horizon (QINF, Sec.~\ref{sec:terminal_1}, \cite{chen1998quasi});
\item a simple terminal equality constraint (TEC, Sec.~\ref{sec:terminal_TEC}). 
\end{itemize}
Considering the closed-loop performance and region of attraction, QINF is generally superior to TEC, see, e.g., comparisons by~\citet[Sec.~5]{chen1998quasi}, \citet[Sec.~V]{raff2006nonlinear}, \citet[Sec.~4.1]{koehler2020nonlinearAutomatica}, and \citet[Sec.~5.2]{koehler2020nonlinearTAC}. 
General statements regarding the performance of UC vs. QINF/TEC are difficult, however, two important extreme cases can be regarded. 
(i) The performance of QINF is locally close to optimal (cf. \citet[Thm.~5.22]{grune2017nonlinear}, \citet[App.~A]{kohler2021dynamic}), while the performance of UCON with a short horizon $N$ can be highly suboptimal. 
(ii) Theorem~\ref{thm:UCON} can ensure any desired performance bound $\alpha_N\in[0,1)$ for UCON with a sufficiently large (finite) horizon $N$. In contrast, the terminal set constraint in QINF/TEC typically yields a bounded feasible set, which may result in severe performance limitations. 

TEC requires a more restrictive local \textit{controllability} condition to be applied. 
TEC and UCON require no offline design, while QINF requires an LQR design based on the linearization and the determination of a suitable terminal set scaling $\alpha>0$. 
Given terminal ingredients $V_{\mathrm{f}}$, $\mathbb{X}_{\mathrm{f}}$ (Asm.~\ref{asm:prelim_terminal}), we can directly compute constants $\gamma,c$ satisfying cost controllability (Asm.~\ref{asm:CostControl_local}), cf.~\citep{darup2015missing}. 
However, there is no algorithm to determine an analytical formula for a local CLF $V_{\mathrm{f}}$ using given constants $c,\gamma$ that satisfy the cost controllability condition (Asm.~\ref{asm:CostControl_local}). 
Hence, the design of a CLF $V_{\mathrm{f}}$ and a positive invariant set $\mathbb{X}_{\mathrm{f}}$ for QINF is intrinsically more complex.

In the following, we focus on challenges related to dynamic operation~\ref{enum_motivation_1}--\ref{enum_motivation_3}. 
The design of a terminal cost $V_{\mathrm{f}}$ for time-varying references~\ref{enum_motivation_1} or online changing references~\ref{enum_motivation_2} becomes significantly more complex (Sec.~\ref{sec:terminal}). 
In addition, evaluation of the terminal cost/set $V_{\mathrm{f}}/\mathbb{X}_{\mathrm{f}}$ in Problem~\eqref{problem:traj_track} requires the state reference $x_{\mathrm{r}}(t)$, which is difficult to implement if only an output reference $y_{\mathrm{r}}(t)$ is available~\ref{enum_motivation_3}. 
Furthermore, all theoretical properties break down if the provided reference does not satisfy the constraints or dynamics~\ref{enum_motivation_3}. 
Although the deployment of artificial references can avoid some of these issues, this results in additional challenges (cf. Sec.~\ref{sec:discussion_artificial}). 
In contrast, MPC formulations without terminal ingredients can be directly applied to time-varying or online changing references~\ref{enum_motivation_1}--\ref{enum_motivation_2}. 
Even output references $y_{\mathrm{r}}(t)$ or infeasible references can be handled (cf. Sec.~\ref{sec:UCON_2_dynamic}). 
Hence, the advantages of a simple MPC formulation without terminal ingredients become more pronounced when considering such dynamic operation. 
However, derivation of a sufficiently long horizon $N$ also becomes more challenging and in general a long prediction horizon $N$ may be needed due to the absence of terminal ingredients. 
Such MPC formulations without terminal ingredients are also particularly popular in robotics experiments with dynamic operation, cf., e.g.,~\cite{liniger2015optimization,faulwasser2016implementation,rosenfelder2022model,romero2022model,rickenbach2023active}. 
This important factor is often neglected in discussions regarding benefits and drawbacks of terminal ingredients in MPC, cf.~\cite{mayne2013apologia} and \citet[Sec.~7.4]{grune2017nonlinear}. 
One important exception is economic MPC (Sec.~\ref{sec:economic}):
There exist results for time-varying economically optimal operation without terminal ingredients~\citep{kohlernonlinear19,grune2020economic,long2021unconstrained}, however, the corresponding conditions are very challenging to verify. 
On the other hand, Section~\ref{sec:economic} introduces simple design methods for economic MPC schemes with terminal ingredients that provide strong theoretical properties.

Lastly, we discuss approaches that unify the benefits of MPC frameworks with and without terminal ingredients. 
\cite{limon2006stability} show that we can drop the terminal set constraint $\mathbb{X}_{\mathrm{f}}$ in Problem~\eqref{problem:prelim} for any initial condition in a region of attraction $x\in\mathbb{X}_{\bar{V}}$. 
\citet[Sec.~III.B]{limon2018nonlinear} combine this with artificial setpoints to handle infeasible references. 
\cite{soloperto2022nonlinear} further remove the requirement of a local CLF $V_{\mathrm{f}}$, thus simplifying the design. 
\cite{magni2001stabilizing} and \cite{kohler2021stability} use a finite-horizon rollout cost $V_{\mathrm{f}}$ to approximate a CLF. 
\cite{kohler2023stability} provide a theoretical analysis to ensure stability with a short horizon $N$ using such an approximate CLF, see also the continuous-time results by~\cite{reble2012improved}. 
In combination, these approaches can be implemented with a short horizon $N$, without any complex offline design, and for infeasible references.  
\bibliographystyle{elsarticle-harv}
\biboptions{authoryear} 
\bibliography{bibfileDiss} 

\end{document}